\documentclass[twocolumn,aps,amssymb,superscriptaddress]{revtex4-1}
\usepackage[final]{changes}
\usepackage[colorlinks=true,citecolor=blue,linkcolor=blue,urlcolor=blue]{hyperref}
\usepackage{amsmath}
\usepackage{amssymb}
\usepackage{bbm}
\usepackage{xcolor}
\usepackage{graphicx}
\usepackage{dcolumn}
\usepackage{bm}
\usepackage{hyperref}
\usepackage{afterpage}
\usepackage{tabularx}
\graphicspath{{Latexpicture/}}

\begin{document}

\title{Multi-port few-photon diodes}

\author{Teng-Fei Xiao}
\altaffiliation{These authors contributed equally to this work.}
\affiliation{Hunan Key Laboratory for Micro-Nano Energy Materials and Devices\\ and School of Physics and Optoelectronics, Xiangtan University, Hunan 411105, China}

\author{Liang Li}
\altaffiliation{These authors contributed equally to this work.}
\affiliation{Hunan Key Laboratory for Micro-Nano Energy Materials and Devices\\ and School of Physics and Optoelectronics, Xiangtan University, Hunan 411105, China}
\author{Renli Xu}
\affiliation{School of Artificial Intelligence and Information Technology,
Jiangsu Province Engineering Research Center of TCM Intelligence Health Service,
Nanjing University of Chinese Medicine, Nanjing 210023, China}
\author{Tiantian Li}
\affiliation{Hunan Key Laboratory for Micro-Nano Energy Materials and Devices\\ and School of Physics and Optoelectronics, Xiangtan University, Hunan 411105, China}

\author{Junlong Tian}
\affiliation{Department of Electronic Science, College of Big Data and
Information Engineering, Guizhou University, Guiyang 550025, China}

\author{Jinlei Tan}
\email{jinleitan@jsu.edu.cn}
\affiliation{Department of Physics, College of Physics, Mechanical and Electrical Engineering, Jishou University, Jishou 416000, China}
	
\author{Jie Peng}
\email{jpeng@xtu.edu.cn}
\affiliation{Hunan Key Laboratory for Micro-Nano Energy Materials and Devices\\ and School of Physics and Optoelectronics, Xiangtan University, Hunan 411105, China}

\begin{abstract}
We investigate single- and two-photon scattering dynamics in a system comprising two one-dimensional chiral waveguides coupled to a dissipative Rydberg atom pair. By employing the Laplace transform method, we analytically solve the scattering problem and characterize the transport properties. Our results demonstrate robust perfect single-photon diode effects in this system with or without dissipation, exhibiting unity transmittance in the forward direction and zero transmittance in the backward direction. Unlike previous studies where perfect nonreciprocity requires a specific  dissipation rate, our scheme achieves this effect without fine-tuning such parameters. Furthermore, we extend the nonreciprocal transport to multi-port configurations at the one- and two-photon levels. The system exhibits pronounced photon nonreciprocal behavior in lossy and lossless scenarios, without the need for precise parameter matching. This work provides a robust platform for on-chip nonreciprocal photonic devices.

\end{abstract}

\date{\today}

\maketitle

\section{Introduction}\label{I}

Quantum information science is a deeply interdisciplinary field encompassing cavity optomechanics \cite{1,2}, cold atom physics \cite{3,4}, and superconducting circuit quantum electrodynamics (circuit QED) \cite{5,6}. The construction of large-scale quantum networks based on integrated optoelectronic components stands as one of the most promising development directions for future information and communication technologies \cite{7}. To enable long-distance transmission and precise manipulation of photonic information, waveguide quantum electrodynamics (waveguide QED) \cite{8} has emerged as the core physical framework underpinning on-chip quantum networks. This system constructs efficient photon-atom interfaces by coupling nonlinear scatterers to one-dimensional waveguides, including two-level atoms \cite{9,10}, multi-level atoms \cite{11,12,13}, and multi--atom arrays \cite{14,15,16}. Leveraging strong coupling between traveling-wave fields and localized qubits, it provides a versatile platform for both theoretical and experimental investigations of photon correlation and transport properties, demonstrating broad application prospects in quantum communication.

As the integration density of on-chip quantum networks continues to rise, achieving unidirectional signal transmission and suppressing back-reflection noise have become core requirements for system stability, which has in turn driven research into functional devices such as optical diodes (optical isolators) \cite{17} that break Lorentz reciprocity \cite{18}. Over the past decade, researchers have conducted systematic studies on the nonreciprocal propagation of classical optical fields based on diverse physical mechanisms, including nonlinear optics \cite{19,20,21,22}, moving atomic gases \cite{23,24,25,26}, and non-Hermitian optics \cite{27,28,29,30}. In recent years, the research community has further extended these explorations to the quantum regime, successively proposing effects such as phase-controlled single-photon nonreciprocal transmission \cite{31}, nonreciprocal photon blockade \cite{32,33,34,35}, and nonreciprocal quantum entanglement \cite{36,37}, which open new avenues for the generation and manipulation of unidirectional nonclassical light fields.

Within the scope of quantum nonreciprocal devices, optical diodes operating in the few-photon regime are of irreplaceable fundamental importance. Single photons serve as ideal carriers for quantum information processing, and unidirectional transmission at the single-photon level forms the basis for building low-noise quantum links. Meanwhile, only by fully suppressing parasitic inter-device reflections at the single-photon level can performance degradation of interferometric quantum devices be avoided. Considerable progress has been made in the theoretical and experimental research on single-photon diodes in recent years \cite{38,39,40,41,42}; for example, one study realized two-photon optical diode functionality via the photon blockade effect \cite{38}. In chiral quantum optics systems \cite{43,44,45,46,47,48,49,50,51}, counter-propagating light fields exhibit intrinsically asymmetric coupling strengths with emitters, providing a native physical mechanism for unidirectional light transmission. Building on this property, researchers have demonstrated functional devices including few-photon routers \cite{52,53} and two-photon diodes \cite{54,55}. Among various implementation approaches, dissipation-engineered schemes can achieve a perfect single-photon diode effect while covering both the single-photon and two-photon regimes \cite{56,57,58}. However, such schemes generally suffer from notable limitations: on the one hand, their perfect nonreciprocity relies on strict parameter matching between the scatterer dissipation rate and the waveguide coupling strength (e.g., $\kappa
/\Gamma=1$). The nonreciprocal performance is highly sensitive to atomic dissipation fluctuations,  leading to poor parameter robustness. On the other hand, existing few-photon diode research is largely confined to basic two-port configurations, and systematic theoretical design and mechanism validation remain lacking for multi-port few-photon nonreciprocal schemes tailored to on-chip multi-channel networks.

Addressing the key challenges of insufficient parameter robustness and limited scalability of multi-port configurations in state-of-the-art chiral nonreciprocal systems, this work proposes a hybrid system consisting of two one-dimensional chiral waveguides coupled to a dissipative Rydberg atom pair, and systematically investigates the scattering dynamics and transport properties under single-photon and two-photon incidence. We analytically solve the scattering problem via the Laplace transform method, and our results demonstrate that the system can achieve a robust perfect single-photon optical diode effect between port 1 and 2: the forward transmittance  approaches unity, while backward transmittance is completely suppressed. In contrast to the schemes presented in Refs. \cite{56,57,58}, this nonreciprocal effect requires no fine-tuning of the $\kappa
/\Gamma=1$ parameter ratio, and its performance is immune to variations in atomic dissipation strength, yielding a remarkable improvement in parameter robustness. Building on this finding, we further extend the single-photon nonreciprocal transmission mechanism to multi-port configurations, and verify that the system exhibits high-contrast two-photon nonreciprocal behavior under both lossy and lossless conditions. Specifically, the two-photon diode performance likewise does not rely on precise parameter matching, and the multi-port nonreciprocal transmission effect also holds for photon pairs. The introduction of two-atom interaction can enhance the quantum correlation between photons. This study provides a novel implementation pathway for  highly robust chip-scale nonreciprocal photonic quantum devices, and bears important application potential in the fields of quantum communication and quantum information processing.

The remainder of this paper is organized as follows. Sec. \textcolor{blue}{II} establishes the theoretical model and presents the system Hamiltonian. In Sec. \textcolor{blue}{III}, we analytically derive the single-photon transmission and reflection amplitudes via the Laplace transform method. Then we analyze the nonreciprocal transmission characteristics under bidirectional port incidence, and extend the underlying mechanism to multi-port configurations. Sec. \textcolor{blue}{IV} addresses the two-photon scattering problem, comparing two-photon nonreciprocal behavior with and without loss, and verifies the parameter robustness of the two-photon diode as well as the multi-port transmission properties of photon pairs. Sec. \textcolor{blue}{V} discusses the experimental feasibility of the proposed scheme. Finally, Sec. \textcolor{blue}{VI}  concludes of this work.

\section{Model and Hamiltonian}\label{II}
The investigated system comprises two infinitely long 1D linear waveguides and a pair of two-level atoms coupled by Rydberg interaction (see Fig.\:\:\hyperref[Fig1]{1}). The system Hamiltonian reads $\left(\hbar=1\right)$:

\begin{figure}[t]
	\par
	\begin{center}
	 \includegraphics[scale=0.25]{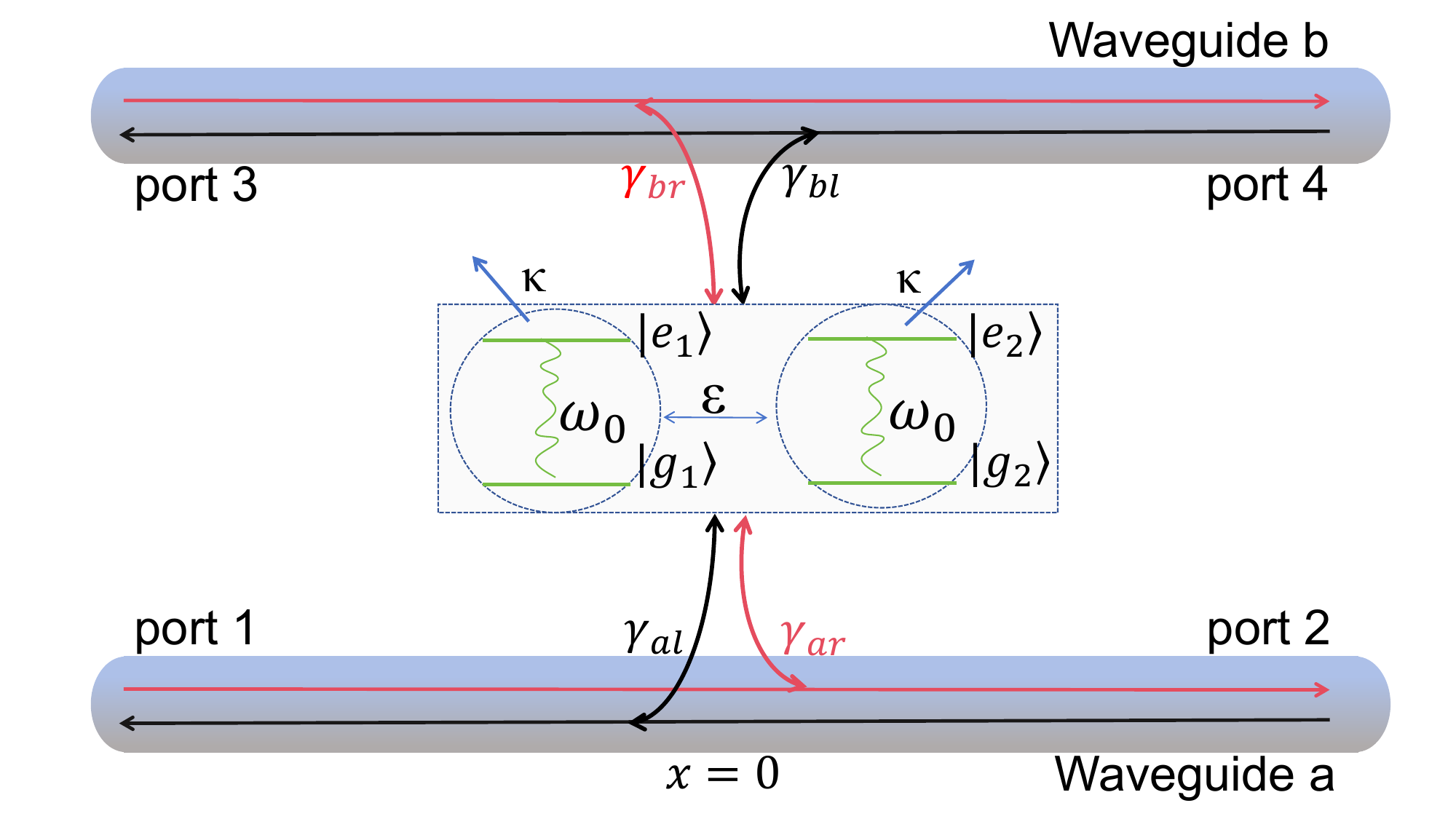}
	\end{center}
	\par
	\renewcommand\figurename{\textbf{FIG.}}
	\vspace{-0 mm}\caption{Schematic of two one-dimensional waveguides coupled to a Rydberg atom pair at the origin. Photons launched from the waveguide ports are scattered via interaction with the atom pair.}
	\label{Fig1}%
\end{figure}%

\begin{align}
	\hat{H} &  =\sum_{j=a,b}\int_{0}^{\infty}dk\omega_{jk}(\hat{r}_{jk}^{\dag
	}\hat{r}_{jk}+\hat{l}_{jk}^{\dag}\hat{l}_{jk})\nonumber\\
	&  +\sum_{i=1,2}\left(  \omega_{0}-i\kappa\right)  \hat{\sigma}_{i}^{+}%
	\hat{\sigma}_{i}^{-}\nonumber\\
	&  +\int_{0}^{\infty}dk\sum_{j=a,b}\sum_{i=1,2}[\sqrt{\frac{\gamma_{jr}}{2\pi
	}}\left(  \hat{\sigma}_{i}^{+}\hat{r}_{jk}+\hat{\sigma}_{i}^{-}\hat{r}%
	_{jk}^{\dag}\right)  \nonumber\\
	&  +\sqrt{\frac{\gamma_{jl}}{2\pi}}\left(  \hat{\sigma}_{i}^{+}\hat{l}%
	_{jk}+\hat{\sigma}_{i}^{-}\hat{l}_{jk}^{\dag}\right)  ]+\varepsilon\left\vert
	e\right\rangle _{1}\left\langle e\right\vert _{1}\otimes\left\vert
	e\right\rangle _{2}\left\langle e\right\vert _{2}. \label{1}
\end{align}
\qquad

The first term is the free-field Hamiltonian for the two waveguides. Here, $\hat
{r}_{jk}^{\dag}$($\hat{r}_{jk}$) and $\hat{l}_{jk}^{\dag}$ ($\hat{l}_{jk}%
$) denote the creation (annihilation) operators for right- and left-propagating photons with wavevector $k$ in the $j$-th waveguide ($j=a,b$), respectively, satisfying the dispersion relation $\omega
_{jk}=v_{p}k$. We set the group velocity $v_{p}=1$ for simplicity and assume identical dispersion for both waveguides, i.e., $\omega
_{ak}=\omega_{bk}=\omega_{k}$.   The second term is the free Hamiltonian for the two two-level atoms, each with level spacing $\omega
_{0}$ and dissipation rate $2\kappa$. The third term represents the chiral waveguide-atom couplings. $\sqrt
{\gamma_{jr}/\left(  2\pi\right)  }$ and $\sqrt{\gamma_{jl}/\left(
	2\pi\right)  }$ correspond to the coupling strengths of right- and left-propagating waves in the $j$-th waveguide to the two atoms, respectively; $\hat
	{\sigma}_{i}^{+}$ ($\hat{\sigma}_{i}^{-}$) are the raising (lowering) operators for the $i$-th atom ($i=1,2$). 
	The last term represents the Rydberg interaction between the Rydberg states $\left\vert
	e\right\rangle _{i}$ ($i=1,2$) of the two atoms, with strength $\varepsilon$. This coupling strength $\varepsilon$ is physically governed by the distance $r$ between the two atoms, scaling as $\varepsilon
	\sim1/r^{3}$ for a dipole-dipole interaction and as $\varepsilon
	\sim1/r^{6}$ for a van der Waals interaction \cite{16}.

We further introduce the following transformation
\begin{equation}
	\hat{e}_{jk}=\sqrt{\frac{\gamma_{jr}}{\Gamma_{j}}}\hat{r}_{jk}+\sqrt
	{\frac{\gamma_{jl}}{\Gamma_{j}}}\hat{l}_{jk}\text{, \ }\hat{o}_{jk}%
	=\sqrt{\frac{\gamma_{jl}}{\Gamma_{j}}}\hat{r}_{jk}-\sqrt{\frac{\gamma_{jr}%
		}{\Gamma_{j}}}\hat{l}_{jk},
\end{equation}
where $\Gamma_{j}=\gamma_{jr}+\gamma_{jl}$, then Eq. \eqref{1} can be rewritten as
\begin{align}
	\hat{H} &  =\sum_{j=a,b}\int_{0}^{\infty}dk\omega_{k}(\hat{e}_{jk}^{\dagger
	}\hat{e}_{jk}+\hat{o}_{jk}^{\dagger}\hat{o}_{jk})\nonumber\\
	&  +\sum_{i=1,2}\left(  \omega_{0}-i\kappa\right)  \hat{\sigma}_{i}^{+}%
	\hat{\sigma}_{i}^{-}+\varepsilon\left\vert e\right\rangle _{1}\left\langle
	e\right\vert _{1}\otimes\left\vert e\right\rangle _{2}\left\langle
	e\right\vert _{2}\nonumber\\
	&  +\sqrt{\frac{\Gamma_{j}}{2\pi}}\sum_{j=a,b}\sum_{i=1,2}\int_{0}^{\infty
	}\left(  \hat{\sigma}_{i}^{+}\hat{e}_{jk}+\hat{\sigma}_{i}^{-}\hat{e}%
	_{jk}^{\dagger}\right)  dk.\label{3}
\end{align}
Defining $\hat{c}_{k}$ and $\hat{d}_{k}$
as linear combinations of the field operators in the frequency space of the two waveguides, we obtain
\begin{equation}
	\hat{c}_{k}=\sqrt{\frac{\Gamma_{a}}{\Gamma}}\hat{e}_{ak}+\sqrt{\frac
		{\Gamma_{b}}{\Gamma}}\hat{e}_{bk},\text{ \ }\hat{d}_{k}=\sqrt{\frac{\Gamma
			_{b}}{\Gamma}}\hat{e}_{ak}-\sqrt{\frac{\Gamma_{a}}{\Gamma}}\hat{e}_{bk},
\end{equation}
where $\Gamma=\Gamma_{a}+\Gamma_{b}$ and is set to $1$ throughout the paper and all parameters are in units of $\Gamma$. We can rewrite Eq. \eqref{3} as 

\begin{equation}
	\hat{H}=\hat{H}_{c}+\hat{H}_{fd}%
\end{equation}
with
\begin{gather}\label{6}
	\begin{align} 
	\hat{H}_{c} &  =\int_{0}^{\infty}dk\omega_{k}\hat{c}_{k}^{\dagger}\hat{c}%
	_{k}+\sum_{i=1,2}\left(  \omega_{0}-i\kappa\right)  \hat{\sigma}_{i}^{+}%
	\hat{\sigma}_{i}^{-}\nonumber\\
	&  +\sqrt{\frac{\Gamma}{2\pi}}\int_{0}^{\infty}dk\left[  \left(  \hat{\sigma
	}_{1}^{+}+\hat{\sigma}_{2}^{+}\right)  \hat{c}_{k}+\left(  \hat{\sigma}%
	_{1}^{-}+\hat{\sigma}_{2}^{-}\right)  \hat{c}_{k}^{\dagger}\right]
	\nonumber\\
	&  +\varepsilon\left\vert e\right\rangle _{1}\left\langle e\right\vert
	_{1}\otimes\left\vert e\right\rangle _{2}\left\langle e\right\vert _{2}, 
\end{align}\\
	\hat{H}_{fd}=\int_{0}^{\infty}dk\omega_{k}\hat{d}_{k}^{\dagger}\hat{d}%
	_{k}+\sum_{j=a,b}\int_{0}^{\infty}dk\omega_{k}\hat{o}_{jk}^{\dagger}\hat
	{o}_{jk}.
\end{gather}
 It is evident that the odd-parity modes decouple from the atomic system and thus undergo free evolution. In subsequent analyses, we will focus on the dynamics of the even-parity modes.

In a rotating frame with respect to
\begin{equation}
	\hat{H}_{c}^{0}=\int_{0}^{\infty}dk\omega_{0}\hat{c}_{k}^{\dagger}\hat{c}%
	_{k}+\sum_{i=1,2}\omega_{0}\hat{\sigma}_{i}^{+}\hat{\sigma}_{i}^{-},
\end{equation}
Eq. \eqref{6} becomes
\begin{align}
	\hat{H}_{c}^{I} &  =\int_{0}^{\infty}dk\Delta_{k}\hat{c}_{k}^{\dagger}%
	\hat{c}_{k}-\sum_{i=1,2}i\kappa\hat{\sigma}_{i}^{+}\hat{\sigma}_{i}%
	^{-}\nonumber\\
	&  +\sqrt{\frac{\Gamma}{2\pi}}\int_{0}^{\infty}dk\left[  \left(  \hat{\sigma
	}_{1}^{+}+\hat{\sigma}_{2}^{+}\right)  \hat{c}_{k}+\left(  \hat{\sigma}%
	_{1}^{-}+\hat{\sigma}_{2}^{-}\right)  \hat{c}_{k}^{\dagger}\right]
	\nonumber\\
	&  +\varepsilon\left\vert e\right\rangle _{1}\left\langle e\right\vert
	_{1}\otimes\left\vert e\right\rangle _{2}\left\langle e\right\vert _{2}, \label{9}
\end{align}
where $\Delta_{k}=\omega_{k}-\omega_{0}$. Starting from Eq. \eqref{9}, we derive the analytical photon scattering solutions for the system. We first examine the scattering dynamics under single-photon incidence, and then investigate the transport properties in the two-photon scattering regime.

\section{Single-Photon Scattering}\label{III}

We first examine the single-photon scattering behavior under four-port incidence. Since photons in the $d$ and $o$ subspaces undergo only free propagation, scattering interactions are predominantly confined to the $c$ subspace. For the single-photon problem in this subspace, it suffices to restrict the analysis to the single-excitation subspace, where an arbitrary quantum state of the system is given by

\begin{align}
	\left\vert \phi\left(  t\right)  \right\rangle  &  =A_{1}\left(  t\right)
	\left\vert 0\right\rangle \left\vert e_{1}\right\rangle \left\vert
	g_{2}\right\rangle +A_{2}\left(  t\right)  \left\vert 0\right\rangle
	\left\vert g_{1}\right\rangle \left\vert e_{2}\right\rangle \nonumber\\
	&  +\int_{0}^{\infty}dkB_{k}\left(  t\right)  \left\vert 1_{k}\right\rangle
	\left\vert g_{1}\right\rangle \left\vert g_{2}\right\rangle .
\end{align}
Here, $\left\vert e\right\rangle$ and $\left\vert g\right\rangle$ represent the excited and ground states of the atoms, respectively. $\left\vert
1_{k}\right\rangle$ denotes a one-photon state in the $k$-th mode. $A_{1}
\left(  t\right)  $, $A_{2}\left(  t\right)  $, and $B_{k}\left(
t\right)  $ are the probability amplitudes of the respective states. By solving the Schr\"{o}dinger equation
\begin{equation}
	i\frac{\partial}{\partial t}\left\vert \phi\left(  t\right)  \right\rangle
	=\hat{H}_{c}^{I}\left\vert \phi\left(  t\right)  \right\rangle,
\end{equation}
we derive the equations of motion for the probability amplitudes
\begin{subequations} \label{10}
	\begin{align}
\dot{A}_{1}\left(  t\right)  =-i\sqrt{\frac{\Gamma}{2\pi}}\int_{0}^{\infty
}dkB_{k}\left(  t\right)  -\kappa A_{1}\left(  t\right),\\
\dot{A}_{2}\left(  t\right)  =-i\sqrt{\frac{\Gamma}{2\pi}}\int_{0}^{\infty
}dkB_{k}\left(  t\right)  -\kappa A_{2}\left(  t\right),\\
\dot{B}_{k}\left(  t\right)  =-i\Delta_{k}B_{k}\left(  t\right)
-i\sqrt{\frac{\Gamma}{2\pi}}\left[  A_{1}\left(  t\right)  +A_{2}\left(
t\right)  \right].
\end{align}
\end{subequations}
The Laplace transform is applied to Eq. \eqref{10} with the initial conditions $A_{1}\left(  0\right)  $, $A_{2}\left(  0\right)  $, and $B_{k}\left(  0\right)  $, which reduces to 
\begin{subequations} \label{11}
	\begin{align}
\left(  s+\kappa\right)  \tilde{A}_{1}\left(  s\right)  =A_{1}\left(
0\right)  -i\sqrt{\frac{\Gamma}{2\pi}}\int_{0}^{\infty}dk\tilde{B}_{k}\left(
s\right),\\
\left(  s+\kappa\right)  \tilde{A}_{2}\left(  s\right)  =A_{2}\left(
0\right)  -i\sqrt{\frac{\Gamma}{2\pi}}\int_{0}^{\infty}dk\tilde{B}_{k}\left(
s\right),\\
\left(  s+i\Delta_{k}\right)  \tilde{B}_{k}\left(  s\right)  =B_{k}\left(
0\right)  -i\sqrt{\frac{\Gamma}{2\pi}}\left[  \tilde{A}_{1}\left(  s\right)
+\tilde{A}_{2}\left(  s\right)  \right].
\end{align}
\end{subequations}

Considering only single-photon incidence via the four waveguide ports with a Lorentzian wave packet profile, we write the corresponding initial conditions as
\begin{subequations}
\begin{align}
A_{1}\left(  0\right)   &  =0,\text{ }A_{2}\left(  0\right)  =0,\\
B_{k}\left(  0\right)   &  =\frac{\sqrt{\epsilon/\pi}}{\Delta_{k}%
	-\delta+i\epsilon},
\end{align}
\end{subequations}
where $\delta$ and $\epsilon$ denote the central detuning and spectral width of the wave packet, respectively. Substituting the initial conditions into Eq. \eqref{11} yields
\begin{subequations} \label{14}
\begin{align}
\tilde{A}_{1}\left(  s\right)   &  =\frac{i\sqrt{2\epsilon\Gamma}}{\left(
	s+\kappa+\Gamma\right)  \left[  \delta-i(s+\epsilon)\right]  },\\
\tilde{A}_{2}\left(  s\right)   &  =\frac{i\sqrt{2\epsilon\Gamma}}{\left(
	s+\kappa+\Gamma\right)  \left[  \delta-i(s+\epsilon)\right]  },\\
\tilde{B}_{k}\left(  s\right)   &  =\frac{\sqrt{\epsilon/\pi}}{s+i\Delta
	_{k}}[\frac{1}{\triangle_{k}-\delta+i\epsilon}\nonumber\\
&  +\frac{2\Gamma}{\left(  s+\kappa+\Gamma\right)  \left[  \delta
	-i(s+\epsilon)\right]  }]. 
\end{align}
\end{subequations}
Taking the inverse Laplace transform of Eq. \eqref{14} and evaluating the long-time limit yields
\begin{subequations}\label{16}
\begin{align}
A_{1}\left(  t\rightarrow\infty\right)   &  =0,\text{ }A_{2}\left(
t\rightarrow\infty\right)  =0,\\
B_{k}\left(  t\rightarrow\infty\right)   &  =\bar{t}_{k}b_{k}\left(  0\right)
e^{-i\triangle_{k}t},
\end{align}
\end{subequations}
where
\begin{equation}
	\bar{t}_{k}=\frac{\Delta_{k}+i\left(  \kappa-\Gamma\right)  }{\Delta
		_{k}+i\left(  \kappa+\Gamma\right)  }.\label{17}
\end{equation}

In practice, single photons are launched in right-propagating ($\hat{r}
_{jk}$) or left-propagating ($\hat{l}_{jk}$) modes. For a single-photon wave packet incident from port 1, the initial quantum state reads
\begin{equation}
	\left\vert \psi\left(  0\right)  \right\rangle =\int_{0}^{\infty}%
	dkB_{k}\left(  0\right)  \hat{r}_{ak}^{\dag}\left\vert 0\right\rangle
	\left\vert g\right\rangle _{1}\left\vert g\right\rangle _{2}. \label{18}
\end{equation}
First, a photon incident from port 1 is absorbed by the Rydberg atom pair, and subsequently re-emitted into waveguides $a$ and $b$ in the form of reflected and transmitted components. By transforming Eq. \eqref{18} to the $c$, $d$, and $o$ basis, using Eqs. \eqref{16} and \eqref{17}, and transforming back to the $\hat{r}$ and $\hat{l}$, we obtain the scattering amplitudes
\begin{subequations}
\begin{gather}
t_{ak}^{0}=\frac{\Delta_{k}+i\left[  \kappa+\Gamma_{b}-\left(  \gamma
	_{ar}-\gamma_{al}\right)  \right]  }{\Delta_{k}+i\left(  \kappa+\Gamma\right)
}, \label{19a} \\
r_{ak}^{0}=\frac{-2i\sqrt{\gamma_{ar}\gamma_{al}}}{\Delta_{k}+i\left(
	\kappa+\Gamma\right)  },\\
t_{bk}^{0}=\frac{-2i\sqrt{\gamma_{ar}\gamma_{br}}}{\Delta_{k}+i\left(
		\kappa+\Gamma\right)  },r_{bk}^{0}=\frac{-2i\sqrt{\gamma_{ar}\gamma_{bl}}%
	}{\Delta_{k}+i\left(  \kappa+\Gamma\right)  }.%
\end{gather}
\end{subequations}
Here, $t_{ak}^{0}$ and $r_{ak}^{0}$ are the port-2 transmission and port-1 reflection amplitudes for a single photon incident at port 1, respectively, while $t_{bk}^{0}$ and $r_{bk}^{0}$ are the port-4 transmission and port-3 reflection amplitudes, respectively.

For a single-photon wave packet injected at port 2, its propagation state reads
\begin{equation}
\left\vert \psi\left(  0\right)  \right\rangle =\int_{0}^{\infty}%
dkB_{k}\left(  0\right)  \hat{l}_{ak}^{\dagger}\left\vert 0\right\rangle
\left\vert g\right\rangle _{1}\left\vert g\right\rangle _{2}. 
\end{equation}
The same derivation yields the expressions for the photon transmission and reflection amplitudes in waveguides $a$ and $b$
\begin{subequations}
	\begin{gather}
\tilde{t}_{ak}^{0}=\frac{\Delta_{k}+i\left[  \kappa+\Gamma_{b}-\left(
	\gamma_{al}-\gamma_{ar}\right)  \right]  }{\Delta_{k}+i\left(  \Gamma
	+\kappa\right)  }, \label{21a}\\
\tilde{r}_{ak}^{0}=\frac{-2i\sqrt{\gamma_{al}\gamma_{ar}}}{\Delta_{k}+i\left(
		\Gamma+\kappa\right)  },\\
\tilde{t}_{bk}^{0}=\frac{-2i\sqrt{\gamma_{al}\gamma_{bl}}}{\Delta_{k}+i\left(
			\Gamma+\kappa\right)  },\tilde{r}_{bk}^{0}=\frac{-2i\sqrt{\gamma_{al}%
				\gamma_{br}}}{\Delta_{k}+i\left(  \Gamma+\kappa\right)  }.
	 \label{21}
\end{gather}
\end{subequations}
Here, $\tilde{t}_{ak}^{0}$ and $\tilde{r}_{ak}^{0}$ are the port-1 transmission and port-2 reflection amplitudes for a single photon incident at port 2, respectively, while $\tilde{t}_{bk}^{0}
$ and $\tilde{r}_{bk}^{0}$ are the port-3 transmission and port-4 reflection amplitudes, respectively.

For a single-photon wave packet injected at port 3, its propagation state reads
\begin{equation}
	\left\vert \psi\left(  0\right)  \right\rangle =\int_{0}^{\infty}%
	dkB_{k}\left(  0\right)  \hat{r}_{bk}^{\dag}\left\vert 0\right\rangle
	\left\vert g\right\rangle _{1}\left\vert g\right\rangle _{2}. 
\end{equation}
The same derivation yields the expressions for the photon transmission and reflection amplitudes in waveguides $a$ and $b$
\begin{subequations}
	\begin{gather}
		t_{bk}^{1}=\frac{\Delta_{k}+i\left[  \kappa+\Gamma_{a}-\left(  \gamma
			_{br}-\gamma_{bl}\right)  \right]  }{\Delta_{k}+i\left(  \Gamma+\kappa\right)
		},\\
		r_{bk}^{1}=\frac{-2i\sqrt{\gamma_{br}\gamma_{bl}}}{\Delta_{k}+i\left(
			\Gamma+\kappa\right)  },\\
		t_{ak}^{1}=\frac{-2i\sqrt{\gamma_{br}\gamma_{ar}}}{\Delta_{k}+i\left(
			\Gamma+\kappa\right)  },r_{ak}^{1}=\frac{-2i\sqrt{\gamma_{br}\gamma_{al}}%
		}{\Delta_{k}+i\left(  \Gamma+\kappa\right)  }.
	\end{gather}
\end{subequations}
Here, $t_{bk}^{1}$ and $r_{bk}^{1}$ are the port-4 transmission and port-3 reflection amplitudes for a single photon incident at port 3, respectively, while $t_{ak}^{1}$ and $r_{ak}^{1}$ are the port-2 transmission and port-1 reflection amplitudes, respectively.

For a single-photon wave packet injected at port 4, its propagation state reads
\begin{equation}
	\left\vert \psi\left(  0\right)  \right\rangle =\int_{0}^{\infty}%
	dkB_{k}\left(  0\right)  \hat{l}_{bk}^{\dagger}\left\vert 0\right\rangle
	\left\vert g\right\rangle _{1}\left\vert g\right\rangle _{2}. 
\end{equation}
The same derivation yields the expressions for the photon transmission and reflection amplitudes in waveguides $a$ and $b$
\begin{subequations}
	\begin{gather}
		\tilde{t}_{bk}^{1}=\frac{\Delta_{k}+i\left[  \kappa+\Gamma_{a}-\left(
			\gamma_{bl}-\gamma_{br}\right)  \right]  }{\Delta_{k}+i\left(  \Gamma
			+\kappa\right)  },\\
		\tilde{r}_{bk}^{1}=\frac{-2i\sqrt{\gamma_{bl}\gamma_{br}}}{\Delta_{k}+i\left(
			\Gamma+\kappa\right)  },\\
		\tilde{t}_{ak}^{1}=\frac{-2i\sqrt{\gamma_{bl}\gamma_{al}}}{\Delta_{k}+i\left(
			\Gamma+\kappa\right)  },\tilde{r}_{ak}^{1}=\frac{-2i\sqrt{\gamma_{bl}%
				\gamma_{ar}}}{\Delta_{k}+i\left(  \Gamma+\kappa\right)  }.
	\end{gather}
\end{subequations}
Here, $\tilde{t}_{bk}^{1}$ and $\tilde{r}_{bk}^{1}$ are the port-3 transmission and port-4 reflection amplitudes for a single photon incident at port 4, respectively, while $\tilde{t}_{ak}^{1}$ and $\tilde{r}_{ak}^{1}$ are the port-1 transmission and port-2 reflection amplitudes, respectively.
\begin{figure}[t]
	\centering
	\par
	\begin{center}
		\includegraphics[scale=0.415]{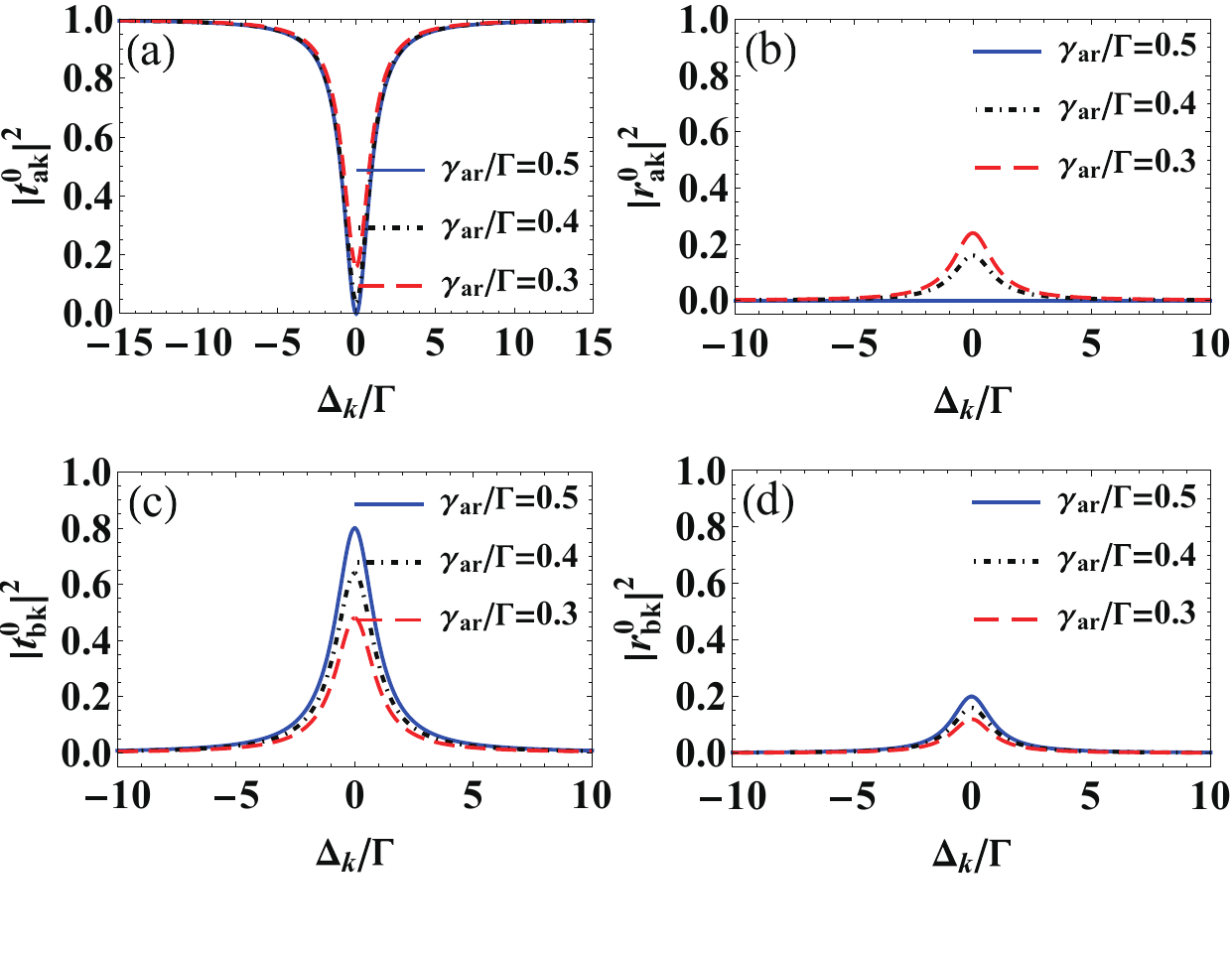}
	\end{center}
	\par
	\renewcommand\figurename{\textbf{FIG.}}
	\vspace{-10.0 mm}\caption{Transmittances $t_{ak}^{0}$, $t_{bk}^0$ (a, c) and reflectances $r_{ak}^{0}$, $r_{bk}^{0}$ (b, d) when a single photon is injected from port 1 are plotted against $\Delta_{k}$ for various $\gamma_{ar}$ values, at $\kappa=0$, $\Gamma_{b}=0.5$, $\gamma_{br}=0.4$, and $\gamma_{bl}=0.1$.}
	\label{Fig2}%
\end{figure}

Next, we systematically investigate the single-photon diode effect in a chiral coupling model of two waveguides and a Rydberg atom pair, with and without dissipation. For the latter case, we first analyze the transport properties of a single photon incident from port 1. As depicted in Fig.\:\:\hyperref[Fig2]{2}, the transmittances $\left\vert t_{ak}^{0}\right\vert ^{2}$ and $\left\vert t_{bk}^{0}\right\vert ^{2}$ at ports 2 and 4, together with the reflectances $\left\vert r_{ak}^{0}\right\vert
^{2}$ and $\left\vert r_{bk}^{0}\right\vert ^{2}$ at ports 1 and 3, are plotted as functions of the detuning $\Delta_{k}$ for different values of $\gamma_{ar}$. Since there are no dissipation channels, the system satisfies the probability conservation relation $P=(\left\vert t_{ak}^{0}\right\vert ^{2}+\left\vert r_{ak}^{0}\right\vert
^{2}+\left\vert t_{bk}^{0}\right\vert ^{2}+\left\vert r_{bk}^{0}\right\vert
^{2})=1$. When $\Delta_{k}=0$ and $\gamma_{ar}=0.5$, $\gamma_{al}=0$, we obtain $\left\vert
t_{ak}^{0}\right\vert ^{2}=\left\vert r_{ak}^{0}\right\vert ^{2}=0$, $\left\vert t_{bk}^{0}\right\vert ^{2}=0.8$ and $\left\vert r_{bk}%
^{0}\right\vert ^{2}=0.2$, meaning the single photon is fully transferred to waveguide $b$ and the propagation of right-propagating photons in waveguide $a$ is completely blocked. Furthermore, the transmittances of port 3 and port 4 are directly proportional to $\gamma_{bl}$ and $\gamma_{br}$, respectively, thereby enabling precise control over photon transmission. In contrast, when the incident photon is far off-resonant with the atomic frequency, the photon-atom interaction is negligible, allowing the photon to pass through waveguide $a$ losslessly.
\begin{figure}[t]
	\centering
	\par
	\begin{center}
		\includegraphics[scale=0.41]{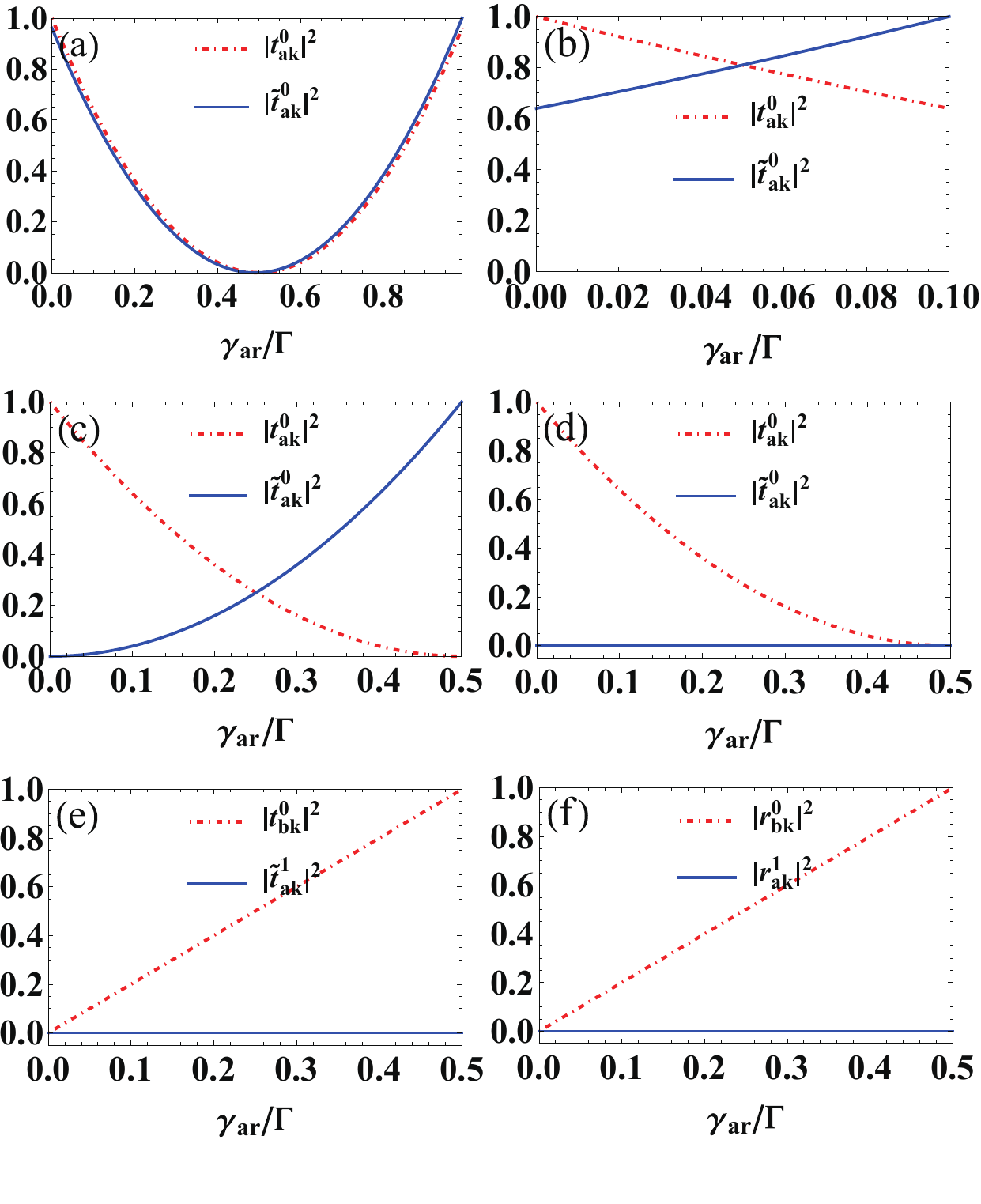}
	\end{center}
	\par
	\renewcommand\figurename{\textbf{FIG.}}
	\vspace{-10.0 mm}\caption{With $\Delta_{k}=0$ and $\kappa=0$, the transmittance and reflectance plotted as functions of
		$\gamma
		_{ar}$ for given $\Gamma_{b}=0.01$ in (a), $\Gamma_{b}=0.9$ in (b), $\Gamma_{b}=0.5$ in (c),
		$\Gamma_{b}=\gamma_{al}-\gamma_{ar}$ with $\gamma_{al}=0.5$ and $0\leq\gamma_{ar}%
		\leq0.5$ in (d),  $\gamma_{br}=0.5$ and $\gamma_{bl}=0$ in (e), and  $\gamma_{br}=0$, $\gamma_{bl}=0.5$ in (f), forming a diode between ports 1 and 3.}%
	\label{Fig3}%
\end{figure}

The chiral coupling leads to a nonreciprocal difference between the transmittances of left- and right-propagating single photons in waveguide $a$. Figs.\:\:\hyperref[Fig3]{3(a-d)} present the evolution of the bidirectional transmittances of waveguide $a$ versus $\gamma_{ar}$ for different $\Gamma
_{b}$ values at $\Delta_{k}=0$.  When $\Gamma_{b}$ is much smaller than $\Gamma_{a}$, the nonreciprocal effect is nearly absent, and the bidirectional transmittances are approximately equal, as indicated by Eqs. \eqref{19a} and \eqref{21a}. When $\Gamma_{b}$ is much larger than $\Gamma_{a}$, the coupling between waveguide $a$ and the atom is extremely weak, so single photons incident from both sides can be fully transmitted. The nonreciprocal effect is most pronounced when $\Gamma_{b}=\Gamma_{a}=0.5$. As shown in Fig.\:\:\hyperref[Fig3]{3(c)}, an ideal single-photon diode is achieved at $\gamma_{ar}=0$ or 0.5 under $\Gamma_{b}=0.5$, where photons propagating in one direction in waveguide $a$ are fully transmitted while those in the opposite direction are completely blocked. For the general case, the right-propagating transmittance drops to zero when $\Gamma_{b}=(\gamma_{ar}-\gamma_{al})$, and the left-propagating transmittance drops to zero when $\Gamma_{b}=(\gamma_{al}-\gamma_{ar})$, as can be seen from Eqs. \eqref{19a} and \eqref{21a}. This effect can be realized solely by chiral coupling ($\gamma
_{ar}\neq\gamma_{al}$), which is not reported in existing studies \cite{54,58}. The diode effects are displayed in Fig.\:\:\hyperref[Fig3]{3(d)}.
\begin{figure}[t]
	\centering
	\par
	\begin{center}
		\includegraphics[scale=0.41]{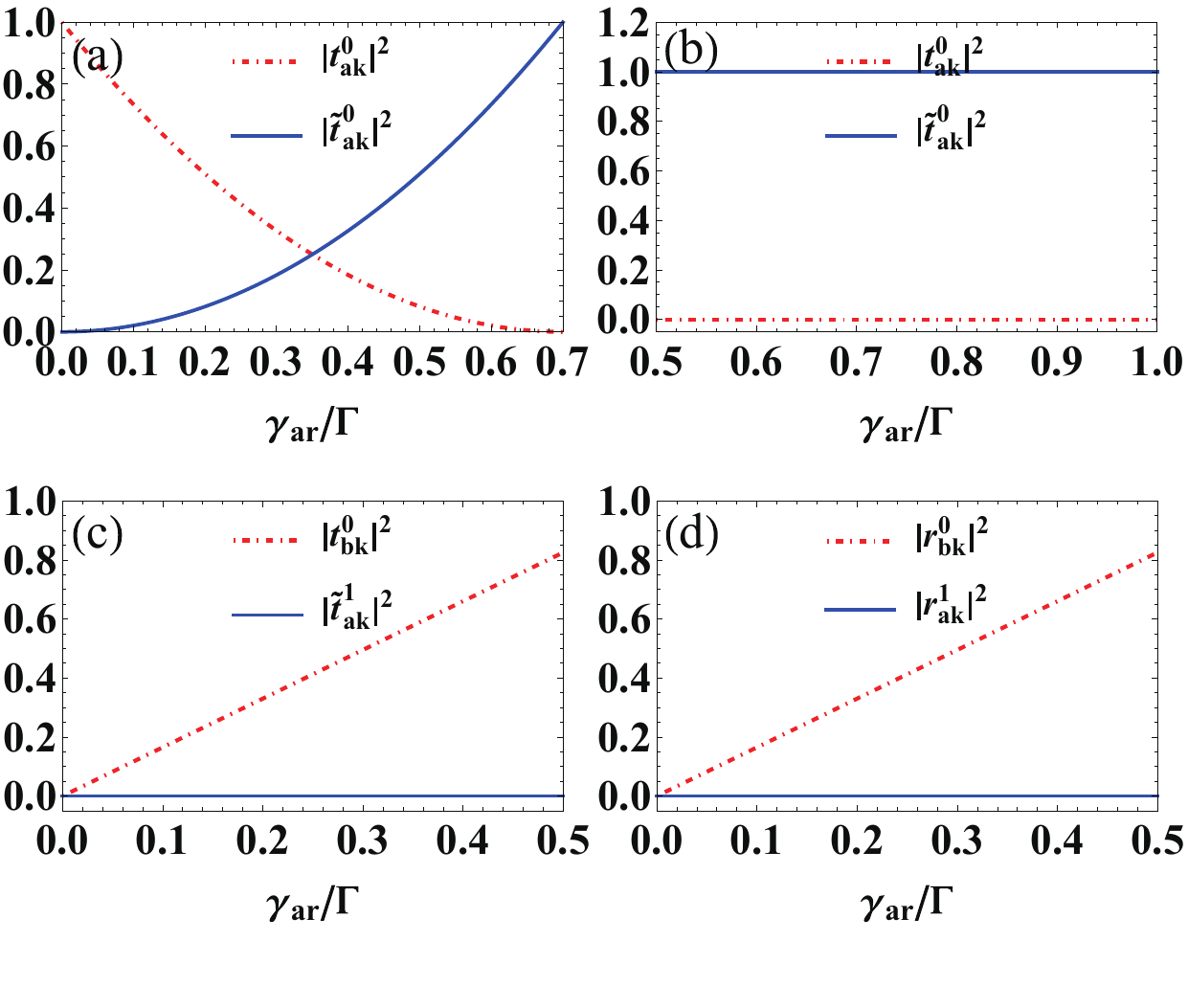}
	\end{center}
	\par
	\renewcommand\figurename{\textbf{FIG.}}
	\vspace{-10.0 mm}\caption{The transmittance and reflectance plotted as functions of
		$\gamma
		_{ar}$ for given $\Gamma_{b}=0.3$, $\kappa=0.4$ in (a), $\kappa=\gamma
		_{ar}-\gamma_{al}-\Gamma_{b}\geqslant0$, $\gamma_{al}=0$ in (b), $\kappa=0.1$, $\gamma_{br}=0.5$, $\gamma_{bl}=0$ in (c), and  $\kappa=0.1$, $\gamma_{br}=0$, $\gamma_{bl}=0.5$ in (d), with $\Delta_{k}=0$.}%
	\label{Fig4}%
\end{figure}

From a physical mechanism perspective, the modulation of transmittance by the atom is determined by its coupling to the waveguides. If $\gamma_{ar}>\gamma_{al}$, the right-propagating mode of waveguide $a$ interacts more strongly with the atom, making photons more prone to absorption and re-emission, while the left-propagating mode couples weakly to the atom, so photons in this mode undergo transmission more readily. When $\kappa=0$ and $\Gamma_{b}=0$, these two effects balance each other, resulting in equal bidirectional transmittances [see Eqs. \eqref{19a} and \eqref{21a}]. However, the atom-waveguide $b$ coupling $\Gamma_{b}$ breaks this balance, as the atom transports photons into waveguide $b$ and suppresses the transmittance in the corresponding propagation direction. As a limiting case, take $\gamma_{al}=0$: after absorbing a right-propagating photon, the atom re-emits it into waveguide $b$ and the right-propagating mode of waveguide $a$, eliminating the right-propagating transmittance. Meanwhile, left-propagating photons do not interact with the atom and maintain a transmittance of unity, thus giving rise to nonreciprocity. When $\Gamma_{b}=\Gamma_{a}=0.5$, the absorption rate from the waveguide to the atoms is equal to their emission rate to waveguide $b$; therefore, photons in waveguide $a$ are completely transferred to waveguide $b$, realizing a perfect single-photon diode between ports 1 and 2.

Notably, the proposed scheme can be extended to realize single-photon diode effects for the ports 1 to 4 and ports 1 to 3 configurations by setting $\gamma_{bl}=0$ or $\gamma_{br}=0$, respectively, as shown in  Figs.\:\:\hyperref[Fig3]{3(e)} and\:\:\hyperref[Fig3]{3(f)}. As $\gamma_{ar}$ increases, $\left\vert r_{ak}^{1}\right\vert ^{2}$ stays at zero while $\left\vert r_{bk}^{0}\right\vert ^{2}$ gradually increases to unity, forming a single-photon diode between ports 1 and 3 with perfect performance at $\gamma
_{ar}=0.5$. The physical mechanisms of these two multi-port diodes differ from that of the ports 1-2 diode. Taking the ports 1-4 configuration as an example, a photon incident from port 1 is absorbed by the atom, and constrained by the chiral coupling with $\gamma_{br}\neq0$ and $\gamma_{bl}=0$, can only couple to the right-propagating mode of waveguide $b$ and exit from port 4. In contrast, a photon incident from port 4 does not interact with the atom and transmits directly through port 3, resulting in nonreciprocal transmission. The working principle of the ports 1-3 diode is analogous.

After introducing atomic dissipation $\kappa$, we first examine the diode characteristics of the ports 1-2 configuration. Fig.\:\:\hyperref[Fig4]{4(a)} shows the evolution of the bidirectional transmittances of waveguide a with $\gamma_{ar}$ at $\Delta_{k}=0$, $\Gamma_{b}=0.3$ and $\kappa=0.4$. The system can still achieve a perfect single-photon diode effect: at $\gamma_{ar}=0.7$ and $\gamma_{al}=0$, the transmittance is unity in one direction and zero in the other, a parameter combination that cannot be realized in the dissipation-free case. For the general case, the right-propagating transmittance is zero and the left-propagating transmittance is unity when $\kappa=(\gamma_{ar}-\gamma_{al}-\Gamma_{b})$ with $\gamma_{al}=0$ and $0.5\leq\gamma_{ar}\leq1$. The opposite holds under the corresponding symmetric condition. This indicates that a perfect single-photon diode can also be realized by tuning the dissipation strength $\kappa$, a finding unreported in the existing literature \cite{54,58}. The corresponding results are presented in Fig.\:\:\hyperref[Fig4]{4(b)}. The underlying mechanism can be explained via the limiting case: when $\kappa=(\gamma_{ar}-\gamma_{al}-\Gamma_{b})$ and $\gamma_{al}=0$, after absorbing a photon, the atom transports part of the energy into waveguide $b$ and dissipates the rest into the environment, with no right-propagating photons re-emitted into waveguide $a$, hence the zero right-propagating transmittance. Left-propagating photons do not interact with the atom and maintain a transmittance of unity, thus forming perfect nonreciprocal transmission. Furthermore, atomic dissipation reduces the transmission amplitude of the ports 1-4 and ports 1-3 diodes [see Figs.\:\:\hyperref[Fig4]{4(c)} and\:\:\hyperref[Fig4]{4(d)}]. Taking the ports 1-4 diode as an example, after a photon incident from port 1 is absorbed by the atom, part of its energy is lost via dissipation besides the portion coupled into waveguide $b$, which lowers the probability of photons in waveguide $b$ exiting from port 4. Nevertheless, photons incident from port 4 do not interact with the atom and propagate directly through port 3, so the nonreciprocal effect persists with only a reduced transmission contrast. The working principle for the ports 1-3 diode is similar.

\section{Two-Photon Scattering}\label{IV}

In this section, we study scattering in the two-excitation subspace. Incident waveguide photons decompose into four modes $o_{a}$, $o_{b}$, $d$ and $c$. As photons in the $o_{a}$, $o_{b}$ and $d$ modes propagate freely, only the two-photon scattering states in the $c$ subspace need to be solved. This subspace includes four basis states: $\left\vert
0\right\rangle \left\vert e_{1}\right\rangle \left\vert e_{2}\right\rangle
$ is the field-vacuum, doubly-excited atomic state. $\left\vert
1_{k}\right\rangle \left\vert e_{1}\right\rangle \left\vert g_{2}\right\rangle
$ and $\left\vert 1_{k}\right\rangle \left\vert g_{1}\right\rangle
\left\vert e_{2}\right\rangle $ are single-photon (wave number $k$) states with one excited atom, respectively. $\left\vert
1_{p},1_{q}\right\rangle \left\vert g_{1}\right\rangle \left\vert
g_{2}\right\rangle $ is the two-photon (wave numbers $p$, $q$) state with both atoms in the ground state. Any state in this subspace reads
\begin{align}
	\left\vert \Phi\left(  t\right)  \right\rangle  &  =A\left(  t\right)
	\left\vert 0\right\rangle \left\vert e_{1}\right\rangle \left\vert
	e_{2}\right\rangle +\int_{0}^{\infty}dkB_{k}^{1}\left(  t\right)  \left\vert
	1_{k}\right\rangle \left\vert e_{1}\right\rangle \left\vert g_{2}\right\rangle
	\nonumber\\
	&  +\int_{0}^{\infty}dkB_{k}^{2}\left(  t\right)  \left\vert 1_{k}%
	\right\rangle \left\vert g_{1}\right\rangle \left\vert e_{2}\right\rangle
	\nonumber\\
	&  +\int_{0}^{\infty}dp\int_{0}^{p}dqC_{pq}\left(  t\right)  \left\vert
	1_{p},1_{q}\right\rangle \left\vert g_{1}\right\rangle \left\vert
	g_{2}\right\rangle,
\end{align}
where $A\left(  t\right)  $, $B_{k}^{1}\left(  t\right)
$, $B_{k}^{2}\left(  t\right)  $ and $C_{pq}\left(  t\right)
$ are the probability amplitudes. The time evolution of these amplitudes is governed by the Schr\"{o}dinger equation
\begin{subequations} \label{23}
\begin{align}
	\dot{A}\left(  t\right)   &  =-\left(  i\varepsilon+2\kappa\right)  A\left(
	t\right)  -i\sqrt{\frac{\Gamma}{2\pi}}\int_{0}^{\infty}dkB_{k}^{1}\left(
	t\right)  \nonumber\\
	&  -i\sqrt{\frac{\Gamma}{2\pi}}\int_{0}^{\infty}dkB_{k}^{2}\left(  t\right)  ,\\
	\dot{B}_{k}^{1}\left(  t\right)   &  =-\left(  i\Delta_{k}+\kappa\right)
	B_{k}^{1}\left(  t\right)  -i\sqrt{\frac{\Gamma}{2\pi}}A\left(  t\right)
	\nonumber\\
	& -i\sqrt{\frac{\Gamma}{2\pi}}\int_{0}^{\infty}dpC_{pk}\left(  t\right)  ,\\
	\dot{B}_{k}^{2}\left(  t\right)   &  =-\left(  i\Delta_{k}+\kappa\right)
	B_{k}^{2}\left(  t\right)  -i\sqrt{\frac{\Gamma}{2\pi}}A\left(  t\right)
	\nonumber\\
	& -i\sqrt{\frac{\Gamma}{2\pi}}\int_{0}^{\infty}dpC_{pk}\left(  t\right)  ,\\
	\dot{C}_{pq}\left(  t\right)   &  =-i\left(  \Delta_{p}+\Delta
	_{q}\right)  C_{pq}\left(  t\right)  -i\sqrt{\frac{\Gamma}{2\pi}}\left[
	B_{p}^{1}\left(  t\right)  +B_{q}^{1}\left(  t\right)  \right] \nonumber\\
	&  -i\sqrt{\frac{\Gamma}{2\pi}}\left[  B_{p}^{2}\left(  t\right)  +B_{q}%
	^{2}\left(  t\right)  \right]  .
\end{align}
\end{subequations}
Supposing two photons are injected, the corresponding initial conditions read
\begin{subequations} \label{24}
\begin{align}
	A\left(  0\right)   &  =0,B_{k}^{1}\left(  0\right)  =0,B_{k}^{2}\left(
	0\right)  =0,\\
C_{p,q}\left(  0\right)   &  =G(\frac{1}{\Delta_{p}-\delta_{1}+i\epsilon
}\frac{1}{\Delta_{q}-\delta_{2}+i\epsilon}\nonumber\\
&  +\frac{1}{\Delta_{q}-\delta_{1}+i\epsilon}\frac{1}{\Delta_{p}%
	-\delta_{2}+i\epsilon}), \label{24b}
\end{align}
\end{subequations}
where the normalization constant
\begin{equation}
	G=\frac{\epsilon}{\sqrt{2}\pi}\left[  1+\frac{4\epsilon^{2}}{\left(
		\delta_{1}-\delta_{2}\right)  ^{2}+4\epsilon^{2}}\right]  ^{-1/2}.
\end{equation}
With the above initial conditions, the long-time asymptotic solutions for these probability amplitudes read (see \hyperref[app:A]{Appendix A}):
\begin{subequations}
\label{27}
\begin{align}
	A\left(  t\rightarrow\infty\right)   &  =0,B_{k}^{1}\left(  t\rightarrow
	\infty\right)  =0,B_{k}^{2}\left(  t\rightarrow\infty\right)  =0,\\
	C_{p,q}\left(  t\rightarrow\infty\right)   &  =\left[  \bar{t}_{p}\bar{t}%
	_{q}C_{p,q}\left(  0\right)  +J_{p,q}\right]  e^{-i\left(  \triangle
		_{p}+\triangle_{q}\right)  t}. \label{27b}
\end{align}
\end{subequations}
Here, $\bar{t}_{p}$ and $\bar{t}_{q}$ are defined by Eq. (\textcolor{blue}{17}) and
\begin{align} \label{28}
	J_{p,q}  &  =\frac{4G\Gamma^{2}}{\left[  \Delta_{p}+i\left(  \kappa
		+\Gamma\right)  \right]  \left[  \Delta_{q}+i\left(  \kappa+\Gamma\right)
		\right]  }\nonumber\\
	&  \times\frac{\left(  \Delta_{p}+\Delta_{q}-2\varepsilon+2i\kappa
		\right)  }{\left[  \Delta_{p}+\Delta_{q}-\varepsilon+i\left(
		2\kappa+\Gamma\right)  \right]  }\nonumber\\
	&  \times\frac{1}{\left(  \Delta_{p}+\Delta_{q}-\delta_{1}-\delta
		_{2}+2i\epsilon\right)  }\nonumber\\
	&  \times\lbrack\frac{1}{\Delta_{p}+\Delta_{q}-\delta_{2}+i\epsilon
		+i\left(  \kappa+\Gamma\right)  }\nonumber\\
	&  +\frac{1}{\Delta_{p}+\Delta_{q}-\delta_{1}+i\epsilon+i\left(
		\kappa+\Gamma\right)  }].
\end{align}

The initial state of two right-moving photons incident from port 1 is
\begin{equation}
	\left\vert \psi\left(  0\right)  \right\rangle =\int_{0}^{\infty}dp\int%
	_{0}^{\infty}dqC_{pq}\left(  0\right)  \hat{r}_{ap}^{\dag}\hat{r}_{aq}^{\dag
	}\left\vert 0\right\rangle \left\vert g_{1}\right\rangle \left\vert
	g_{2}\right\rangle.
\end{equation}
In the basis of modes $c$, $d$, and $o$, we obtain the scattering state of two right-propagating photons according to Eq. \eqref{27b}. This state is described by the asymptotic solution in the long-time limit (see \hyperref[app:B]{Appendix B}). 

The initial state of two left-moving photons incident from port 2 is
\begin{equation}
	\left\vert \psi\left(  0\right)  \right\rangle =\int_{0}^{\infty}dp\int%
	_{0}^{\infty}dqC_{pq}\left(  0\right)  \hat{l}_{ap}^{\dagger}\hat{l}%
	_{aq}^{\dagger}\left\vert 0\right\rangle \left\vert g_{1}\right\rangle
	\left\vert g_{2}\right\rangle.
\end{equation}
Using the same approach, we can obtain the state after scattering in the long-time limit (see \hyperref[app:C]{Appendix C}).

The initial state of two right-moving photons incident from port 3 is
\begin{equation}
	\left\vert \psi\left(  0\right)  \right\rangle =\int_{0}^{\infty}dp\int%
	_{0}^{\infty}dqC_{pq}\left(  0\right)  \hat{r}_{bp}^{\dag}\hat{r}_{bq}^{\dag
	}\left\vert 0\right\rangle \left\vert g_{1}\right\rangle \left\vert
	g_{2}\right\rangle ,
\end{equation}
Using the same approach, we can obtain the state after scattering in the long-time limit (see \hyperref[app:D]{Appendix D}).

The initial state of two left-moving photons incident from port 4 is
\begin{equation}
	\left\vert \psi\left(  0\right)  \right\rangle =\int_{0}^{\infty}dp\int%
	_{0}^{\infty}dqC_{pq}\left(  0\right)  \hat{l}_{bp}^{\dagger}\hat{l}%
	_{bq}^{\dagger}\left\vert 0\right\rangle \left\vert g_{1}\right\rangle
	\left\vert g_{2}\right\rangle
\end{equation}
Using the same approach, we can obtain the state after scattering in the long-time limit (see \hyperref[app:E]{Appendix E}).

The initial state of two right-moving photons incident from ports 1 and 3 is
\begin{equation}
	\left\vert \psi\left(  0\right)  \right\rangle =\int_{0}^{\infty}dp\int%
	_{0}^{\infty}dqC_{pq}\left(  0\right)  \hat{r}_{ap}^{\dag}\hat{r}_{bq}^{\dag
	}\left\vert 0\right\rangle \left\vert g_{1}\right\rangle \left\vert
	g_{2}\right\rangle ,
\end{equation}
Using the same approach, we can obtain the state after scattering in the long-time limit (see \hyperref[app:F]{Appendix F}).

The initial state of two left-moving photons incident from ports 2 and 4 is
\begin{equation}
	\left\vert \psi\left(  0\right)  \right\rangle =\int_{0}^{\infty}dp\int%
	_{0}^{\infty}dqC_{pq}\left(  0\right)  \hat{l}_{ap}^{\dagger}\hat{l}%
	_{bq}^{\dagger}\left\vert 0\right\rangle \left\vert g_{1}\right\rangle
	\left\vert g_{2}\right\rangle
\end{equation}
Using the same approach, we can obtain the state after scattering in the long-time limit (see \hyperref[app:G]{Appendix G}).

For computational convenience, we assume that the two photons are initially independent of each other, i.e., we set $\delta
_{1}=\delta_{2}=\delta$. Consequently, the expressions in Eqs. \eqref{24b} and \eqref{28} can be rewritten respectively as
\begin{align}
	C_{p,q}\left(  0\right)   &  =\frac{\epsilon}{\pi}\frac{1}{\Delta
		_{p}-\delta+i\epsilon}\frac{1}{\Delta_{q}-\delta+i\epsilon},\label{38}\\
	J_{p,q}  &  =\frac{\epsilon}{\pi}\frac{4\Gamma^{2}}{\left[  \Delta
		_{p}+i\left(  \kappa+\Gamma\right)  \right]  \left[  \Delta_{q}+i\left(
		\kappa+\Gamma\right)  \right]  }\nonumber\\
	&  \times\frac{\left(  \Delta_{p}+\Delta_{q}-2\varepsilon+2i\kappa
		\right)  }{\left[  \Delta_{p}+\Delta_{q}-\varepsilon+i\left(
		2\kappa+\Gamma\right)  \right]  }\nonumber\\
	&  \times\frac{1}{\left(  \Delta_{p}+\Delta_{q}-2\delta+2i\epsilon
		\right)  }\nonumber\\
	&  \times\frac{1}{\left[  \Delta_{p}+\Delta_{q}-\delta+i\epsilon
		+i\left(  \kappa+\Gamma\right)  \right]  }.\label{39}
\end{align}
The first term in Eq. \eqref{27b} corresponds to the independent scattering process of two photons, while the second term describes the photon correlations induced by scattering. Notably, even when the parameter $\varepsilon
=0$, we have $J_{p,q}\neq0$. This indicates that photon correlation effects can still be observed in the absence of Rydberg interactions. The contrast of the two-photon diode effect is defined as follows \cite{58}
\begin{equation}
	\chi=\frac{\left\vert \left\vert D_{p,q}^{r_{a}r_{a}}\right\vert
		^{2}-\left\vert \tilde{D}_{p,q}^{l_{a}l_{a}}\right\vert ^{2}\right\vert
	}{\left\vert D_{p,q}^{r_{a}r_{a}}\right\vert ^{2}+\left\vert \tilde{D}%
		_{p,q}^{l_{a}l_{a}}\right\vert ^{2}}.%
\end{equation}

\subsection{Diode effect at ports 1-2 without dissipation}\label{A}

\begin{figure}[t]
	\centering
	\par
	\begin{center}
		\includegraphics[scale=0.41]{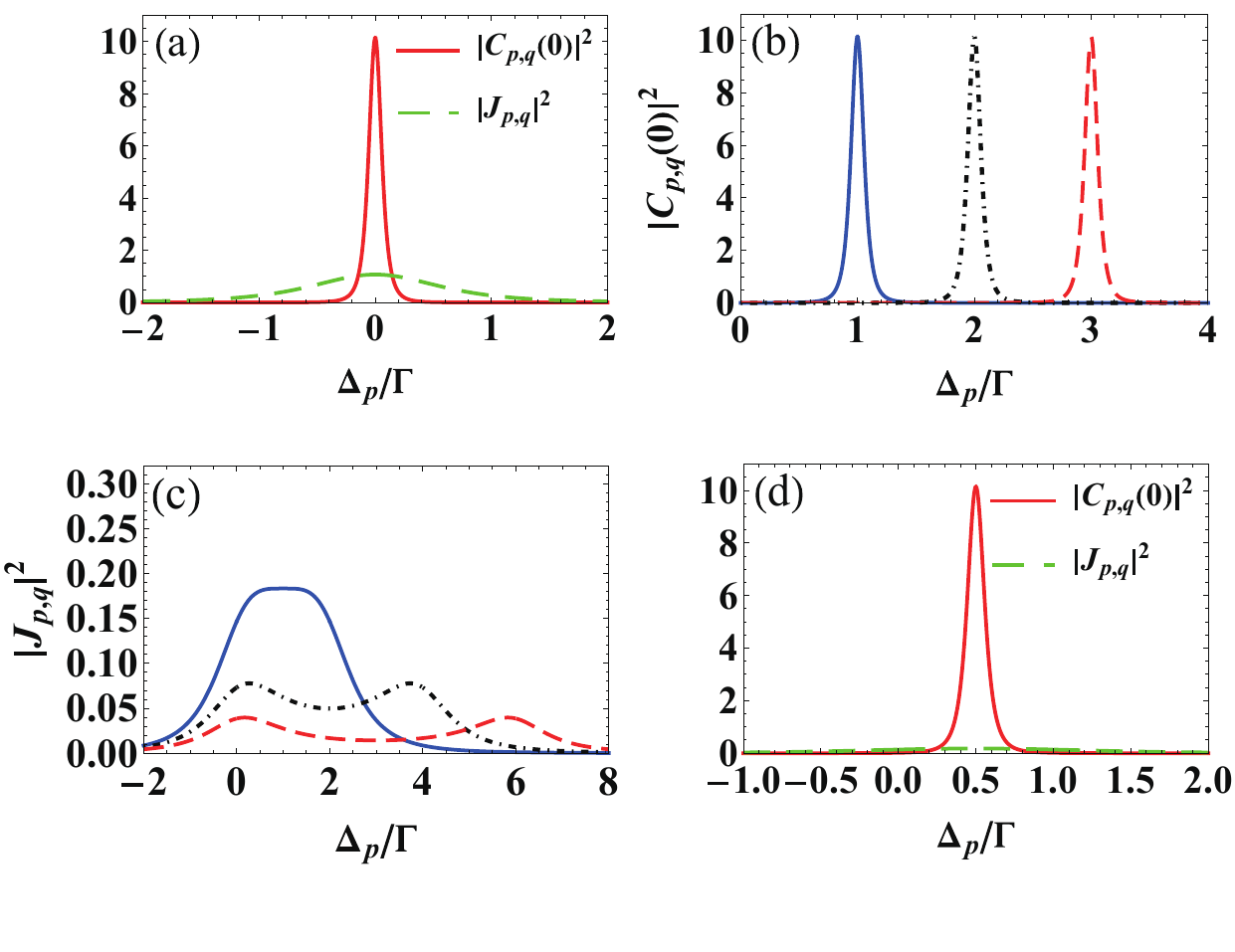}
	\end{center}
	\par
	\renewcommand\figurename{\textbf{FIG.}}
	\vspace{-10.0 mm}\caption{$\left\vert C_{p,q}\left(  0\right)  \right\vert ^{2}$ and
		$\left\vert
		J_{p,q}\right\vert ^{2}$ plotted as functions
		of $\Delta_{p}$. Other parameters $\epsilon=0.1$,  $\kappa=0$. (a) ($\Delta$, $\delta$, $\varepsilon$) = (0, 0, 2), (d) ($\Delta$, $\delta$, $\varepsilon$) = (1, 0.5, 1). In (b) and (c), the blue line is for ($\Delta$, $\delta$, $\varepsilon$) = (2, 1, 2), the black dot-dashed line is for ($\Delta$, $\delta$, $\varepsilon$) = (4, 2, 4), the red dashed line is for ($\Delta$, $\delta$, $\varepsilon$) = (6, 3, 6).}
	\label{Fig5}
\end{figure}

\begin{figure}[t]
	\centering
	\par
	\begin{center}
		\includegraphics[scale=0.41]{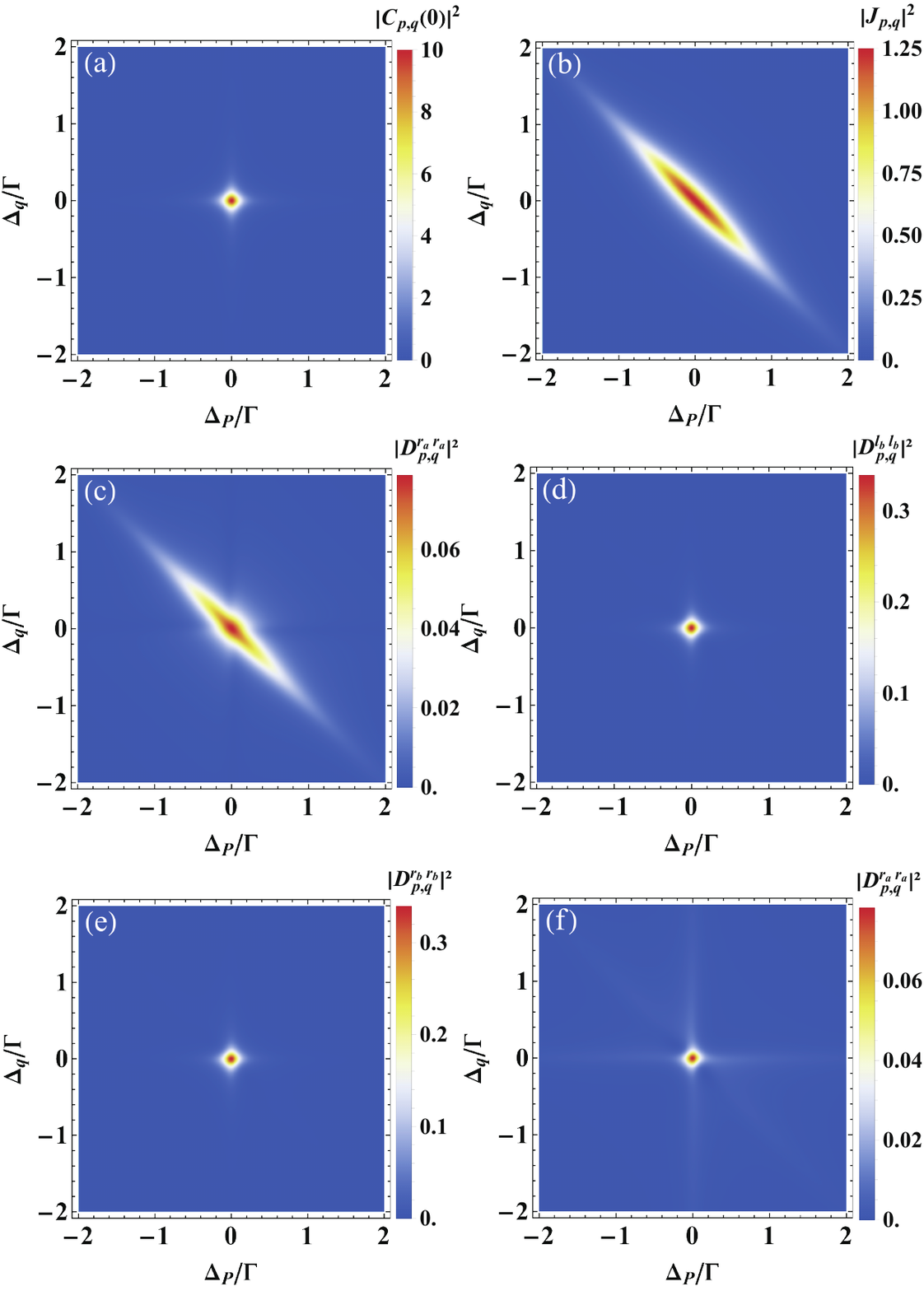}
	\end{center}
	\par
	\renewcommand\figurename{\textbf{FIG.}}
	\vspace{-5.0 mm}\caption{Scattering process of two photons injected from port 1.  Plots of $\left\vert
		C_{p,q}\left(  0\right)  \right\vert ^{2}$ (a), $\left\vert J_{p,q}%
		\right\vert ^{2}$ (b), $\left\vert D_{p,q}^{r_{a}r_{a}}\right\vert
		^{2}$ (c), $\left\vert D_{p,q}^{l_{b}l_{b}}\right\vert ^{2}$ (d), $\left\vert D_{p,q}^{r_{b}r_{b}}\right\vert ^{2}$ (e), and $\left\vert
		D_{p,q}^{r_{a}r_{a}}\right\vert ^{2}$ (f) versus $\Delta_{p}$ and $\Delta_{q}$. (a), (b), (c), (d), (e) $\kappa=0$ and (f) $\kappa=0.5$ . Other parameters are set as $\delta
		=0$, $\epsilon=0.1$, $\varepsilon=4$, $\Gamma_{b}=0.2$, $\gamma_{ar}=0.5$, $\gamma_{al}=0.3$, $\gamma_{br}=0.1$ and $\gamma_{bl}=0.1$. All
		parameters are in units of $\Gamma$.}
	\label{Fig6}
\end{figure}

In this subsection, we investigate the two-photon diode effect at ports 1-2 in the absence of atomic dissipation. The two-photon transmission probability density $\left\vert
D_{p,q}^{r_{a}r_{a}}\right\vert ^{2}$ shown in Appendix B arises from the coherent superposition of two components: the independent propagation term of non-interacting photons proportional to $\left\vert
C_{p,q}\left(  0\right)  \right\vert ^{2}$ and the two-photon correlation term proportional to $\left\vert J_{p,q}\right\vert ^{2}$. We therefore first analyze them separately, and then discuss the full transmitted wave function comprehensively. To explore the modulation effect of the Rydberg interaction on two-photon correlations, we fix the total energy $\Delta=\Delta
_{p}+\Delta_{q}$, width $\epsilon
$, dissipation $\kappa$ and detuning $\delta$. Fig.\:\:\hyperref[Fig5]{5} shows the evolution of the two probability components as a function of the single-photon detuning $\Delta
_{p}$. Under the single-photon resonance condition ($\Delta=0$, $\delta=0$), both components peak at $\Delta
_{p}=\Delta_{q}=0$, which corresponds to the quantum interference region of the two components [see Fig.\:\:\hyperref[Fig5]{5(a)}]. In the two-photon resonance case ($2\delta
=\varepsilon$), the peak of $\left\vert
C_{p,q}\left(  0\right)  \right\vert ^{2}$ shifts linearly with $\varepsilon$: for $\varepsilon=2,4,6$, the peaks occur at $\Delta_{p}=\Delta_{q}=1,2,3$, respectively [see Fig.\:\:\hyperref[Fig5]{5(b)}]. Analytical analysis of $J_{p,q}$ reveals that, under $\Delta=\varepsilon
=2\delta$, the peak of $\left\vert J_{p,q}\right\vert ^{2}$ lies at $\Delta_{p}=\Delta
_{q}=\varepsilon/2$ for $0\leq\varepsilon\leq2$. When $\varepsilon\geq2$, the peak splits into two symmetric branches at $\Delta
_{p_{\max\pm}}=\left(  \varepsilon\pm\sqrt{\varepsilon^{2}-4}\right)
/2$; as $\varepsilon$ increases, the two peaks approach 0 and $\varepsilon$ respectively, and they merge as $\varepsilon$ decreases [see Figs.\:\:\hyperref[Fig5]{5(c)} and\:\:\hyperref[Fig5]{5(d)}]. This double-peak structure, corresponding to two-photon energy redistribution, is directly associated with the interatomic Rydberg interaction.

To characterize the quantum correlations of transmitted photons at each port, Fig.\:\:\hyperref[Fig6]{6} presents the two-dimensional distributions of $\left\vert C_{p,q}\left(  0\right)  \right\vert ^{2}$, $\left\vert J_{p,q}\right\vert ^{2}$ and the port-resolved two-photon transmission probability densities as functions of $\Delta
_{p}$ and $\Delta_{q}$. At $\delta=0$, $\left\vert C_{p,q}\left(  0\right)  \right\vert ^{2}$ shows uncorrelated behavior, while $\left\vert
J_{p,q}\right\vert ^{2}$ is concentrated along the line $\Delta_{p}+\Delta
_{q}=0$, exhibiting strong two-photon correlations [see Figs.\:\:\hyperref[Fig6]{6(a)} and\:\:\hyperref[Fig6]{6(b)}]. In the dissipation-free regime, with chiral coupling parameters $\gamma
_{ar}=0.5$, $\gamma_{al}=0.3$, $\gamma_{br}=0.1$ and $\gamma_{bl}=0.1$, two photons incident from port 1 exhibit correlated transmission at port 2, but show no appreciable correlation at ports 3 and 4 [see Figs.\:\:\hyperref[Fig6]{6(c)},\:\:\hyperref[Fig6]{6(d)}, and\:\:\hyperref[Fig6]{6(e)}]. Upon introducing atomic dissipation $\kappa
=0.5$, the correlation of photons transmitted at port 2 vanishes, indicating that atomic dissipation can effectively tailor the photon correlation properties at output ports [see Fig.\:\:\hyperref[Fig6]{6(f)}]. 

\begin{figure}[t]
	\centering
	\par
	\begin{center}
		\includegraphics[scale=0.41]{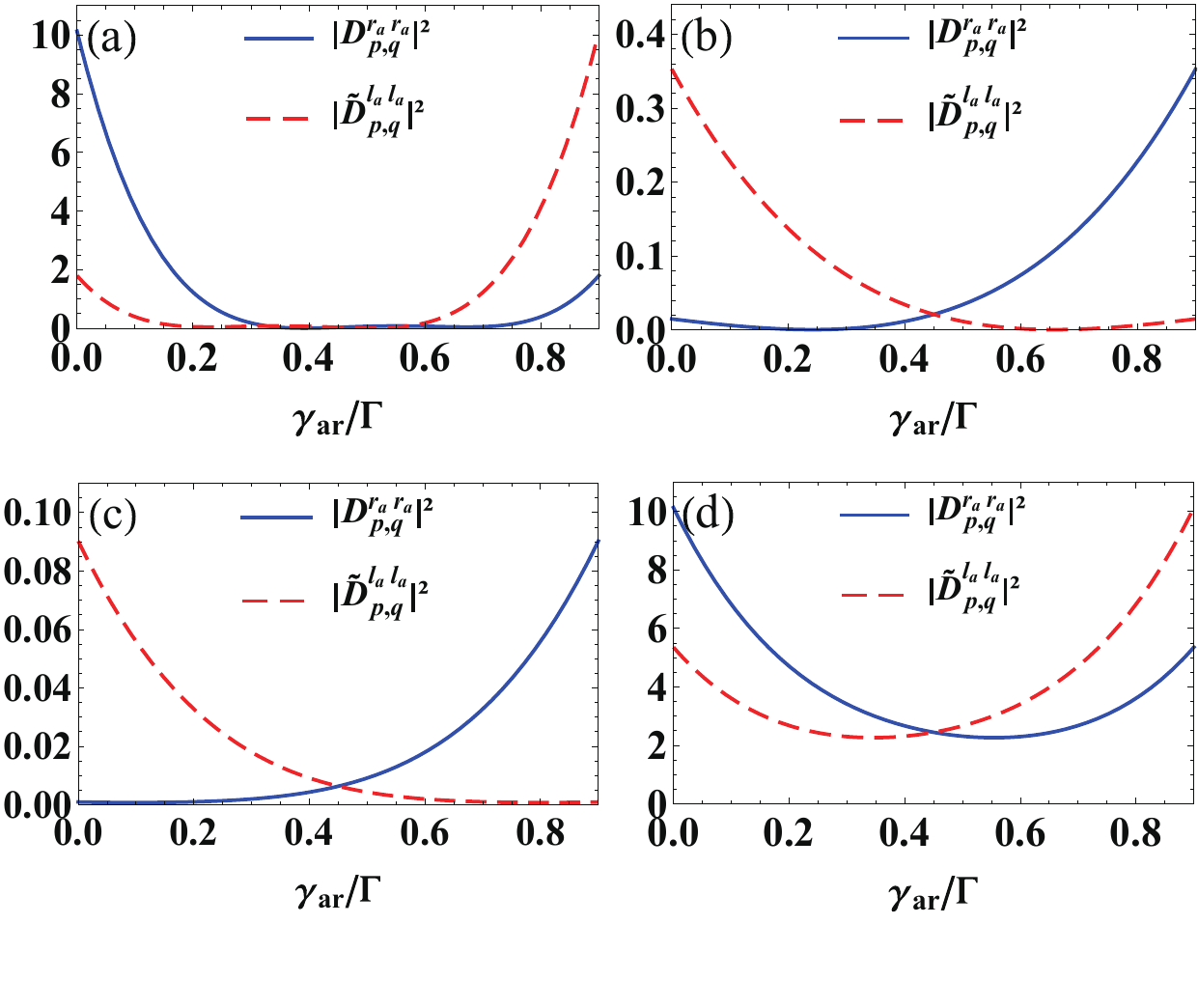}
	\end{center}
	\par
	\renewcommand\figurename{\textbf{FIG.}}
	\vspace{-10.0 mm}\caption{$\left\vert D_{p,q}^{r_{a}r_{a}}\right\vert ^{2}$ and
		$\left\vert \tilde
		{D}_{p,q}^{l_{a}l_{a}}\right\vert ^{2}$ plotted as functions
		of $\gamma_{ar}$ with $\epsilon=0.1$, $\Gamma_{b}=0.1$ and $\kappa=0$. The upper panel deals with
		the single-photon resonance case $\delta=0$. (a) ($\Delta_{p}$, $\Delta_{q}$, $\varepsilon$) = (0, 0, 2) and (b) ($\Delta_{p}$, $\Delta_{q}$, $\varepsilon$) = (0.5, -0.5, 2). The lower panel deals with the two-photon resonance case $2\delta=\varepsilon$. (c) ($\Delta_{p}$, $\Delta_{q}$, $\delta$, $\varepsilon$) = (2, 0, 1, 2) and (d) ($\Delta_{p}$, $\Delta_{q}$, $\delta$, $\varepsilon$) = (1, 1, 1, 2). All
		parameters are in units of $\Gamma$.}%
	\label{Fig7}
\end{figure}
\begin{figure}[b]
	\centering
	\par
	\begin{center}
		\includegraphics[scale=0.41]{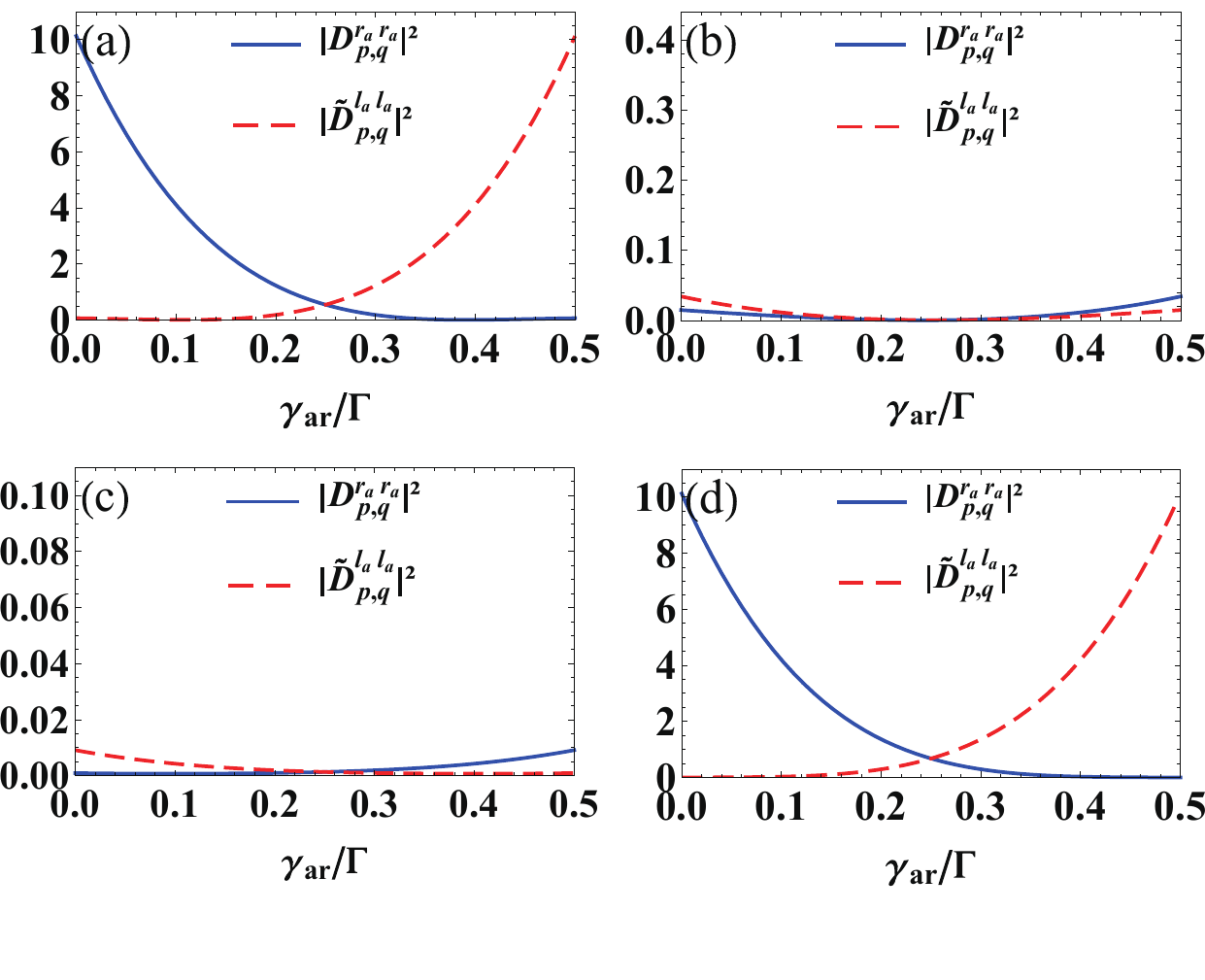}
	\end{center}
	\par
	\renewcommand\figurename{\textbf{FIG.}}
	\vspace{-10.0 mm}\caption{$\left\vert D_{p,q}^{r_{a}r_{a}}\right\vert ^{2}$ and
		$\left\vert \tilde
		{D}_{p,q}^{l_{a}l_{a}}\right\vert ^{2}$ plotted as functions
		of $\gamma_{ar}$ with $\epsilon=0.1$, $\Gamma_{b}=0.5$ and $\kappa=0$. The upper panel deals with
		the single-photon resonance case $\delta=0$. (a) ($\Delta_{p}$, $\Delta_{q}$, $\varepsilon$) = (0, 0, 2) and (b) ($\Delta_{p}$, $\Delta_{q}$, $\varepsilon$) = (0.5, -0.5, 2). The lower panel deals with the two-photon resonance case $2\delta=\varepsilon$. (c) ($\Delta_{p}$, $\Delta_{q}$, $\delta$, $\varepsilon$) = (2, 0, 1, 2) and (d) ($\Delta_{p}$, $\Delta_{q}$, $\delta$, $\varepsilon$) = (0.1, 0.1, 0.1, 0.2). All
		parameters are in units of $\Gamma$.}%
	\label{Fig8}
\end{figure}

For the two-photon diode effect at ports 1-2, Fig.\:\:\hyperref[Fig7]{7} plots the bidirectional transmission probability densities versus $\gamma
_{ar}$ at $\Gamma_{b}=0.1$. Under single-photon resonance, the bidirectional transmission probabilities differ markedly at $\gamma
_{ar}=0.2$, enabling nonreciprocal transmission [see Fig.\:\:\hyperref[Fig7]{7(a)}]. A profound diode effect with perfect unidirectional transmission is achieved at a specific $\gamma
_{ar}$. For instance, clear diode behavior is observed at input detuning $(\Delta
_{p},\Delta_{q})=(0.5,-0.5)$ [see Fig.\:\:\hyperref[Fig7]{7(b)}]. Under two-photon resonance, diode effects also emerge at input detunings \((\Delta_p,\Delta_q)=(0,2)\) and \((2,0)\) [see Fig.\:\:\hyperref[Fig7]{7(c)}]. However, when the input detuning is set to $\Delta_{p}=\Delta_{q}=1$ where the two components share the same peak and the uncorrelated part is prominent, as shown in Figs.\:\:\hyperref[Fig5]{5(b)} and\:\:\hyperref[Fig5]{5(c)}, destructive interference between them cancels out the diode effect [see Fig.\:\:\hyperref[Fig7]{7(d)}]. Under this parameter set, the transmission is dominated by the two-photon uncorrelated term, leading to a relatively high transmission probability density. To enhance the transmission probability density, Fig.\:\:\hyperref[Fig8]{8} shows the corresponding results at $\Gamma_{b}=0.5$. With identical parameters, the diode effect remains prominent in the single-photon resonance case [see Fig.\:\:\hyperref[Fig8]{8(a)}], but the transmission probability density of the correlation-dominated diode is substantially lower than that at $\Gamma
_{b}=0.1$ [see Figs.\:\:\hyperref[Fig8]{8(b)} and\:\:\hyperref[Fig8]{8(c)}]. Here the diode effect is instead governed by the independent photon propagation term. By tuning the Rydberg interaction strength $\varepsilon
$, such an independent-term-dominated two-photon diode effect can also be realized under two-photon resonance [see Fig.\:\:\hyperref[Fig8]{8(d)}].

\begin{figure}[t]
	\centering
	\par
	\begin{center}
		\includegraphics[scale=0.41]{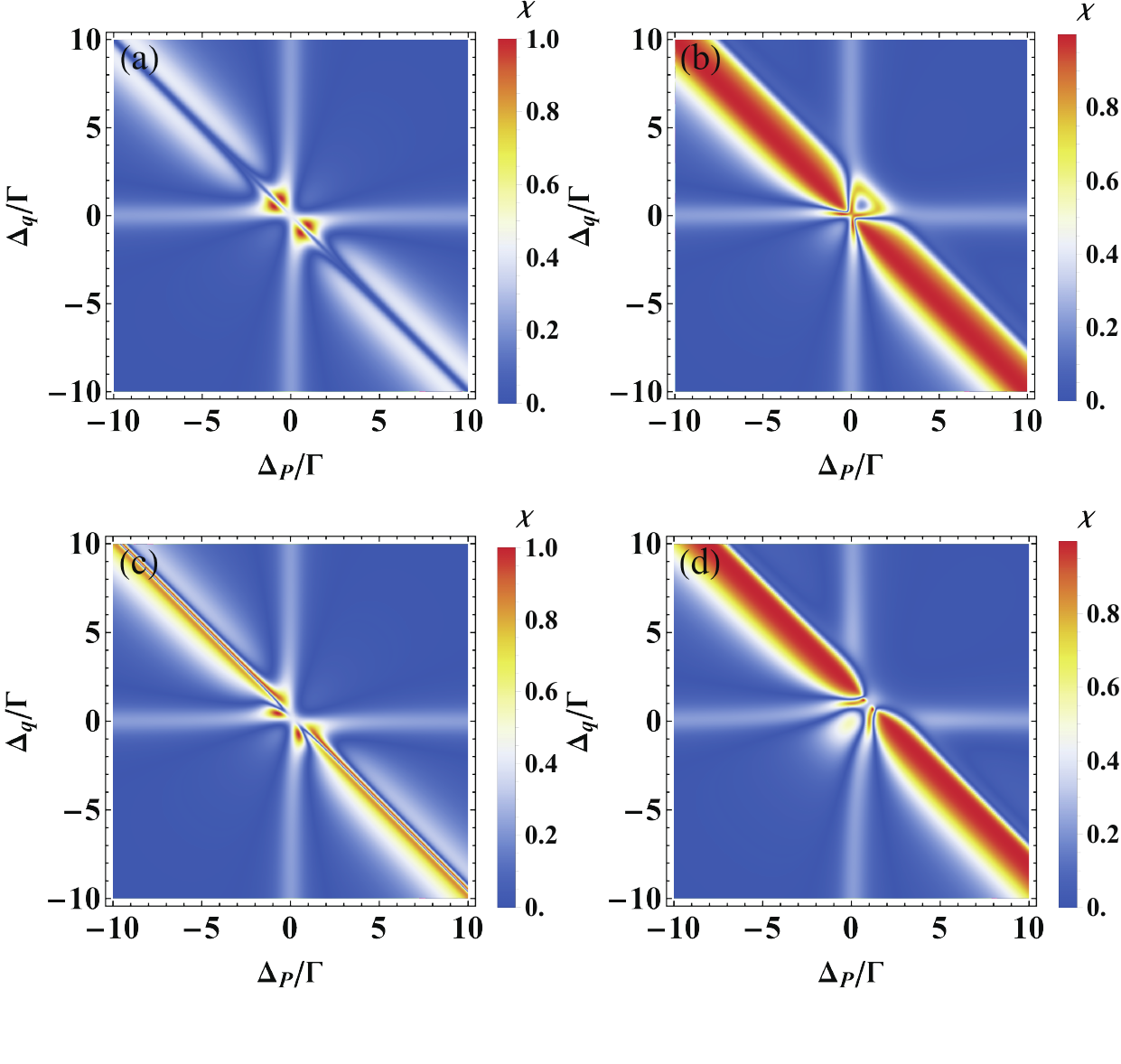}
	\end{center}
	\par
	\renewcommand\figurename{\textbf{FIG.}}
	\vspace{-10.0 mm}\caption{Contrast $\chi
		$ of the two-photon diode effect versus $\Delta_{p}$ and $\Delta_{q}$. (a) ($\varepsilon
		$, $\delta$) = (0, 0) and  (b) ($\varepsilon
		$, $\delta$) = (2, 0). (c) ($\varepsilon
		$, $\delta$) = (0.2, 0.1) and  (d) ($\varepsilon
		$, $\delta$) = (2, 1). Other parameters are set as  $\epsilon=0.1$,  $\Gamma_{b}=0.1$, $\gamma_{ar}=0.9$, $\gamma_{al}=0$ and $\kappa=0$. All
		parameters are in units of $\Gamma$.}%
	\label{Fig9}
\end{figure}

Here we show that the diode effect at $\Gamma_{b}=0.1$ originates from the correlation component, and that the Rydberg interaction effectively adjusts it. Fig.\:\:\hyperref[Fig9]{9} maps the two-photon diode contrast $\chi
$ as a function of $\Delta_{p}$ and $\Delta_{q}$. At $\delta=0$ and $\varepsilon
=0$, the diode operating region distributes along $\Delta
_{p}+\Delta_{q}=0$. Compared with Fig.\:\:\hyperref[Fig6]{6}, this shows that the correlation component dominates the diode effects. With vanishing Rydberg interaction, the correlation stems from quantum interference between the two atomic transition channels, yielding a narrow operating region but a unity contrast [see Fig.\:\:\hyperref[Fig9]{9(a)}]. At $\varepsilon=2$, the operating region is significantly broadened [see Fig.\:\:\hyperref[Fig9]{9(b)}]. The same trend holds under two-photon resonance $2\delta
=\varepsilon$, the operating region is narrow with unity contrast at $\varepsilon
=0.2$ [see Fig.\:\:\hyperref[Fig9]{9(c)}], and expands noticeably at $\varepsilon=2$ [see Fig.\:\:\hyperref[Fig9]{9(d)}]. Moreover, the diode operating region distributes along
$\Delta_{p}+\Delta_{q}=\varepsilon$, as can be drawn from Eqs. \eqref{38} and \eqref{39}. These results confirm that the two-photon diode effect at $\Gamma
_{b}=0.1$ is dominated by the photon correlation component, and tuning the Rydberg interaction can effectively widen its operating bandwidth. Notably, the proposed scheme can realize diode effects dominated by either the photon correlation mechanism or the independent photon propagation mechanism, without requiring atomic dissipation.

\subsection{Diode effect at ports 1-2 with dissipation}\label{B}

\begin{figure}[tbp]
	\centering
	\par
	\begin{center}
		\includegraphics[scale=0.41]{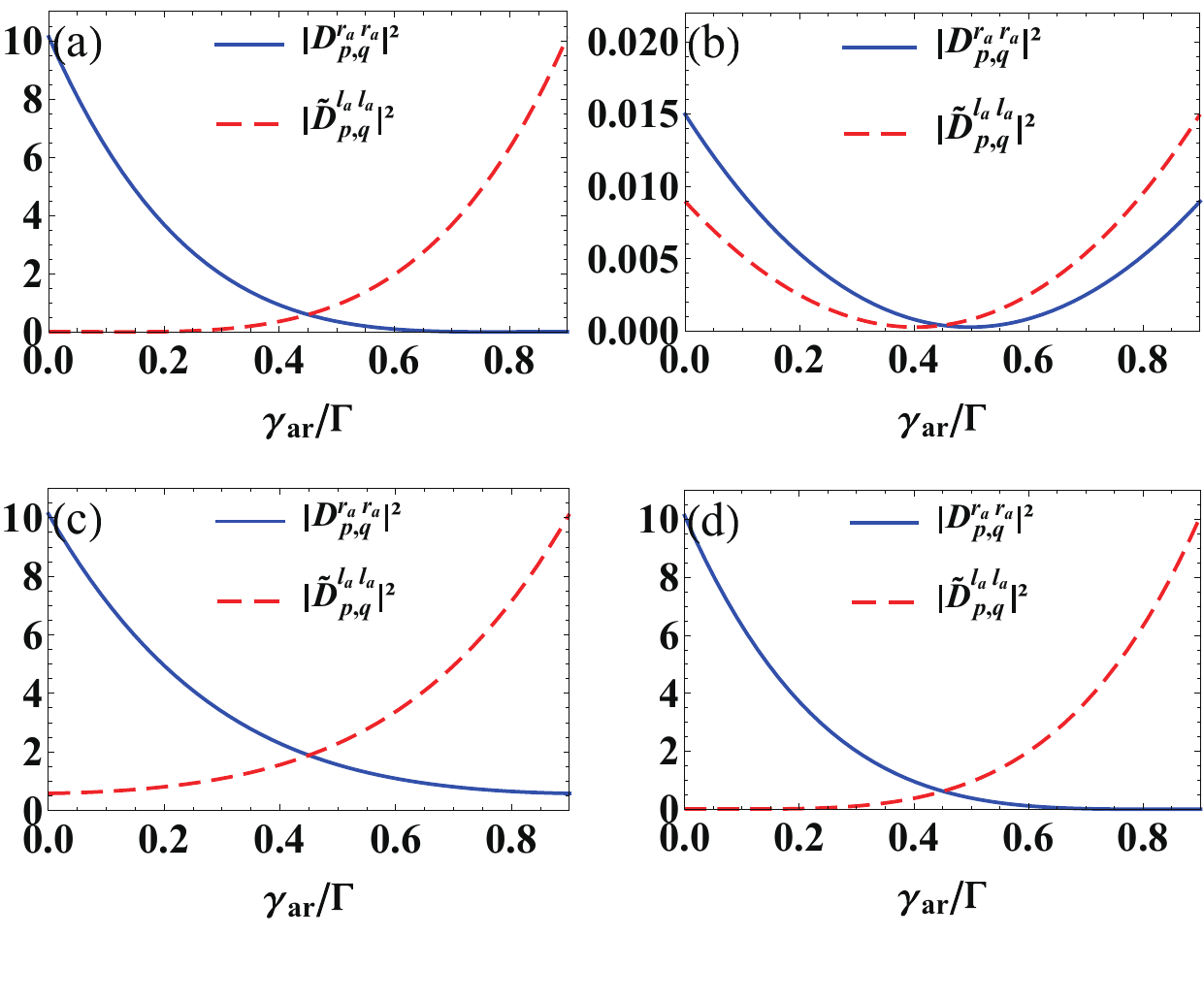}
	\end{center}
	\par
	\renewcommand\figurename{\textbf{FIG.}}
	\vspace{-10.0 mm}\caption{$\left\vert D_{p,q}^{r_{a}r_{a}}\right\vert ^{2}$ and
		$\left\vert \tilde
		{D}_{p,q}^{l_{a}l_{a}}\right\vert ^{2}$ plotted as functions
		of $\gamma_{ar}$ with $\epsilon=0.1$, $\Gamma_{b}=0.1$ and $\kappa=0.8$. The upper panel deals with
		the single-photon resonance case $\delta=0$. (a) ($\Delta_{p}$, $\Delta_{q}$, $\varepsilon$) = (0, 0, 2) and (b) ($\Delta_{p}$, $\Delta_{q}$, $\varepsilon$) = (0.5, -0.5, 2). The lower panel deals with the two-photon resonance case $2\delta=\varepsilon$. (c) ($\Delta_{p}$, $\Delta_{q}$, $\delta$, $\varepsilon$) = (1, 1, 1, 2) and (d) ($\Delta_{p}$, $\Delta_{q}$, $\delta$, $\varepsilon$) = (0.1, 0.1, 0.1, 0.2). All
		parameters are in units of $\Gamma$.}%
	\label{Fig10}
\end{figure}

In this subsection, we investigate the two-photon diode effect at ports 1-2 in the presence of atomic dissipation, which can enhance the diode effects in some cases. With parameters fixed at $\Gamma
_{b}=0.1$ and $\kappa=0.8$, the evolution of the bidirectional transmission probability densities $\left\vert
D_{p,q}^{r_{a}r_{a}}\right\vert ^{2}$ and $\left\vert \tilde{D}_{p,q}^{l_{a}l_{a}}\right\vert ^{2}$ as a function of $\gamma_{ar}$ is presented in Fig.\:\:\hyperref[Fig10]{10}. By comparison with the dissipation-free results in Figs.\:\:\hyperref[Fig7]{7(a)} and\:\:\hyperref[Fig7]{7(b)}, a clear diode effect persists after introducing dissipation under the single-photon resonance condition [see Fig.\:\:\hyperref[Fig10]{10(a)}]. However, for parameter regimes dominated by the photon correlation component, dissipation not only reduces the transmission probability density but also deteriorates the original diode effect [see Fig.\:\:\hyperref[Fig10]{10(b)}]. To show the benefit of dissipation on the diode effect, we perform analysis under two-photon resonance at detuning $\Delta
_{p}=\Delta_{q}=1$, with parameters identical to those in Fig.\:\:\hyperref[Fig7]{7(d)}, where the scattering is dominated by the free propagation part. Compared with the dissipation-free results, the system still exhibits pronounced nonreciprocal transmission [see Fig.\:\:\hyperref[Fig10]{10(c)}], but the diode performance is less pronounced than in the single-photon resonance case [see Fig.\:\:\hyperref[Fig10]{10(a)}]. By further tuning the Rydberg interaction strength $\varepsilon
$, a clear diode effect can also be realized [see Fig.\:\:\hyperref[Fig10]{10(d)}]. From a physical mechanism perspective, under the combined action of dissipation $\kappa
$ and coupling strength $\Gamma_{b}$, the independent-propagation-dominated diode effect shares the same physical origin as the single-photon diode. It can therefore be realized by flexibly tuning $\kappa
$ and $\Gamma_{b}$, which reduces the difficulty of experimental implementation.

\begin{figure}[tbp]
	\centering
	\par
	\begin{center}
		\includegraphics[scale=0.41]{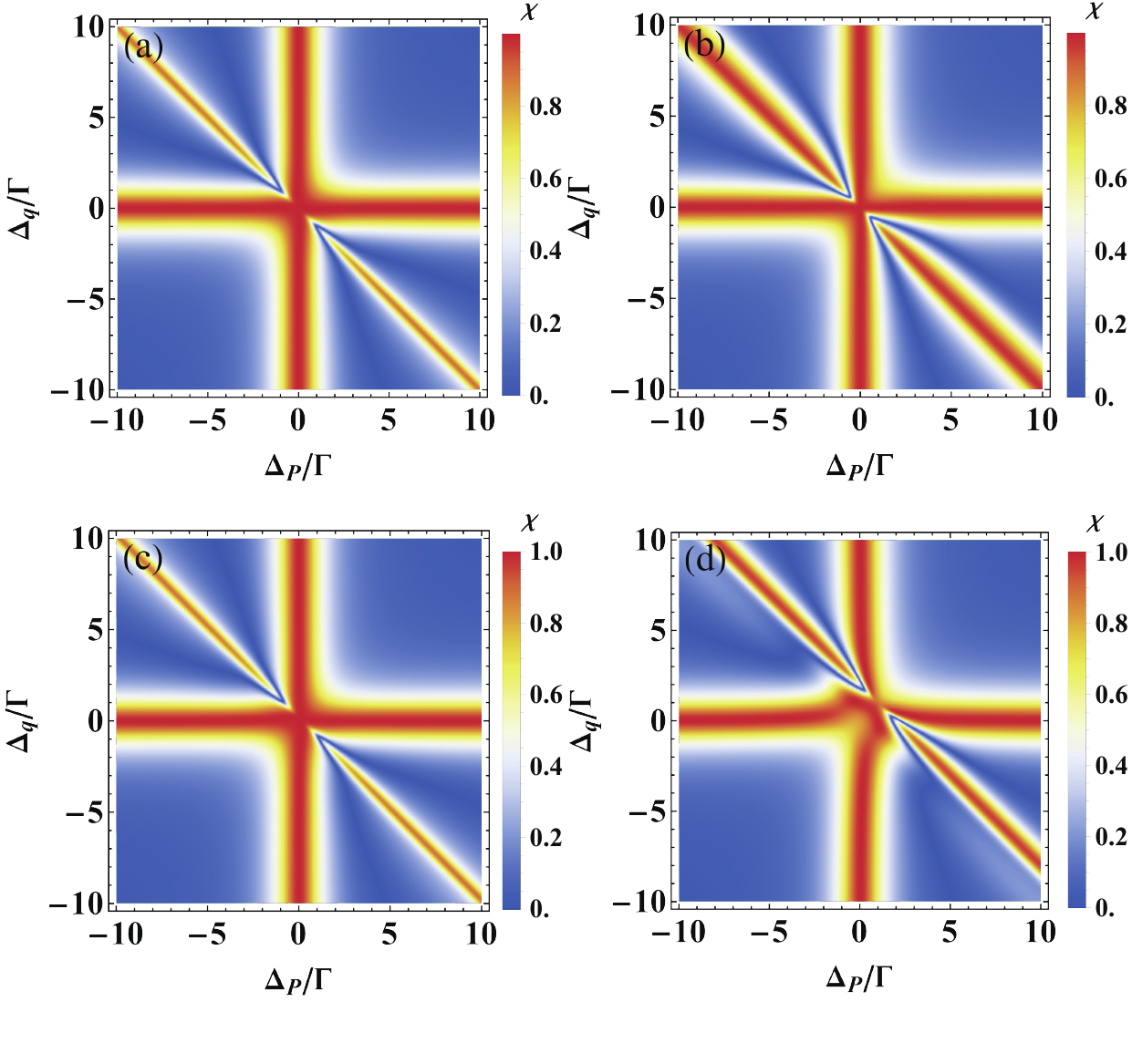}
	\end{center}
	\par
	\renewcommand\figurename{\textbf{FIG.}}
	\vspace{-10.0 mm}\caption{Contrast $\chi
		$ of the two-photon diode effect versus $\Delta_{p}$ and $\Delta_{q}$. (a) ($\varepsilon
		$, $\delta$) = (0, 0) and  (b) ($\varepsilon
		$, $\delta$) = (2, 0). (c) ($\varepsilon
		$, $\delta$) = (0.2, 0.1) and  (d) ($\varepsilon
		$, $\delta$) = (2, 1). Other parameters are set as  $\epsilon=0.1$,  $\Gamma_{b}=0.1$, $\gamma_{ar}=0.9$, $\gamma_{al}=0$ and $\kappa=0.8$. All
		parameters are in units of $\Gamma$.}%
	\label{Fig11}
\end{figure}

To clarify the physical origin of the optimal diode operating region at $\Gamma_{b}=0.1$ and $\kappa=0.8$, Fig.\:\:\hyperref[Fig11]{11} shows the two-dimensional distribution of the two-photon diode contrast $\chi
$ as a function of $\Delta_{p}$ and $\Delta_{q}$. At $\delta=0$ and $\varepsilon=0$, the diode operating regions appear both at $\Delta
_{p}=0$, $\Delta_{q}=0$ (independent-part dominated) and along the line $\Delta_{p}+\Delta
_{q}=0$ (correlated-part dominated). The former exhibits higher contrast and serves as the optimal operating region [see Fig.\:\:\hyperref[Fig11]{11(a)}]. When $\varepsilon=2$, the line-shaped region corresponding to the correlation component broadens with enhanced contrast, but the transmission probability density is much less than that of the independent-propagating part, so the optimal operating region remains at $\Delta
_{p}=0$ or $\Delta_{q}=0$ [see Fig.\:\:\hyperref[Fig11]{11(b)}]. A consistent trend is observed under two-photon resonance $2\delta
=\varepsilon$: at $\varepsilon
=0.2$, the optimal operating region lies at $\Delta
_{p}=0$ or $\Delta_{q}=0$ with a unity contrast [see Fig.\:\:\hyperref[Fig11]{11(c)}]. At $\varepsilon=2$, a diode effect with unity contrast can also be achieved by selecting appropriate photon detunings [see Fig.\:\:\hyperref[Fig11]{11(d)}]. However, now the diode operating region for the correlated-part dominated scattering distributes along
$\Delta_{p}+\Delta
_{q}=\varepsilon$.
These results demonstrate that the two-photon diode effect can be realized via parameter optimization under both single-photon and two-photon resonance conditions, and the optimal diode effect under this dissipative condition is dominated by the independent photon propagation component.

\subsection{Diode effects at other ports}\label{c}

\begin{figure}[t]
	\centering
	\par
	\begin{center}
		\includegraphics[scale=0.41]{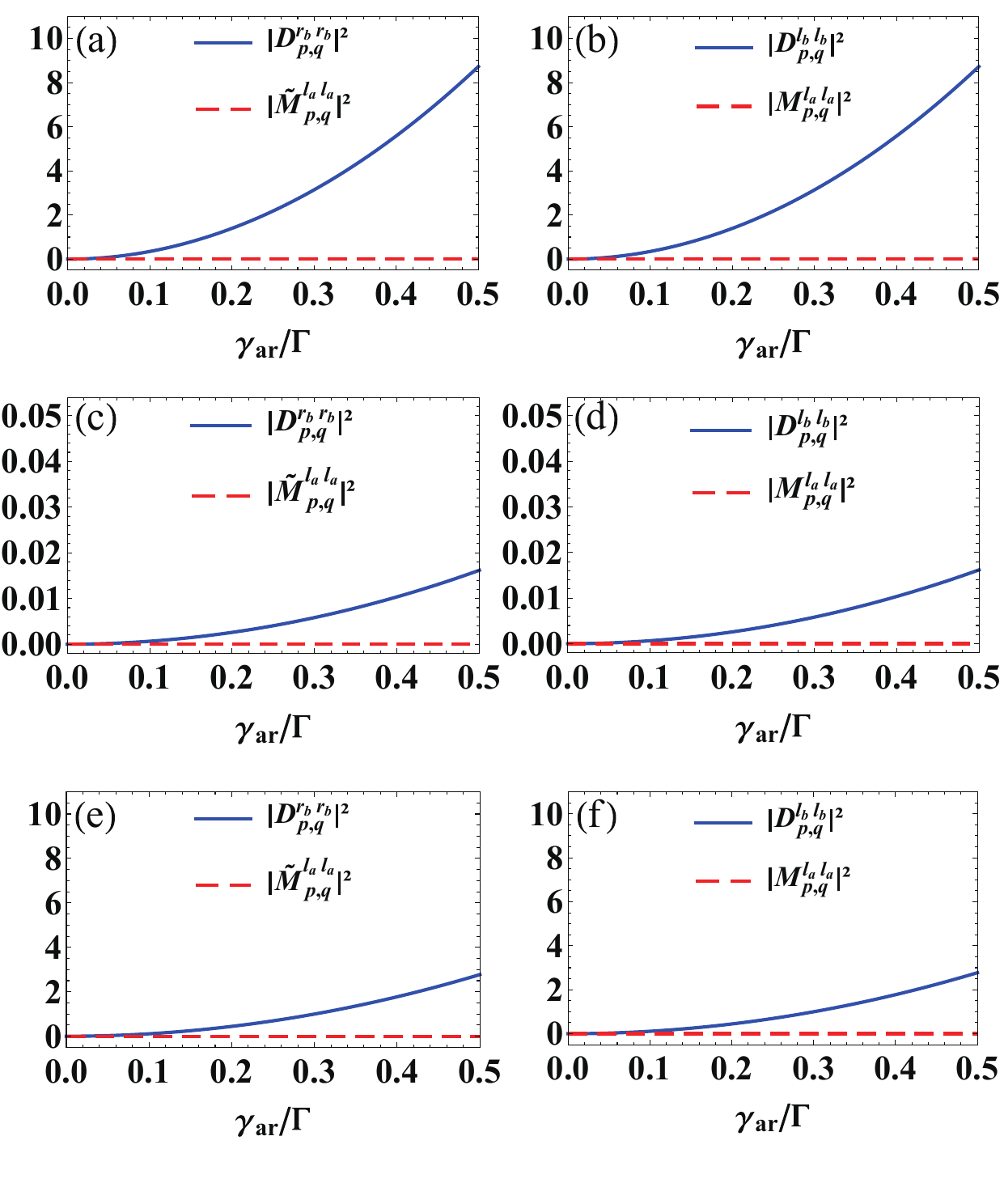}
	\end{center}
	\par
	\renewcommand\figurename{\textbf{FIG.}}
	\vspace{-10.0 mm}\caption{(a, c, e) $\left\vert D_{p,q}^{r_{b}r_{b}}\right\vert
		^{2}$ and $\left\vert \tilde{M}_{p,q}^{l_{a}l_{a}}\right\vert ^{2}
		$ versus $\gamma_{ar}$; (b, d, f) $\left\vert
		D_{p,q}^{l_{b}l_{b}}\right\vert ^{2}$ and $\left\vert M_{p,q}^{l_{a}l_{a}
		}\right\vert ^{2}$ versus $\gamma_{ar}$. Panels (a)--(d): single-photon resonance at $\delta=0$. (a) ($\Delta_{p}$, $\Delta_{q}$, 
		$\gamma_{br}$, $\gamma_{bl}$) = (0, 0,  0.5, 0) and (b) ($\Delta_{p}$, $\Delta_{q}$,
		$\gamma_{br}$, $\gamma_{bl}$) = (0, 0, 0, 0.5). (c) ($\Delta_{p}$, $\Delta_{q}$, $\gamma_{br}$, $\gamma_{bl}$) = (0.5, -0.5, 0.5,
		0) and (d) ($\Delta_{p}$, $\Delta_{q}$, $\gamma_{br}$, $\gamma_{bl}$) = (0.5, -0.5, 0, 0.5). Panels (e)--(f): two-photon resonance at $2\delta=\varepsilon$. (e) ($\Delta_{p}$, $\Delta_{q}$, 
		$\delta$, $\gamma_{br}$, $\gamma_{bl}$) = (1, 1, 1, 0.5,
		0) and (f) ($\Delta_{p}$, $\Delta_{q}$,
		$\delta$, $\gamma_{br}$, $\gamma_{bl}$) = (1, 1, 1, 0,
		0.5). Other parameters are set as  $\epsilon=0.1$, $\varepsilon
		=2$ and $\kappa=0$. All
		parameters are in units of $\Gamma$. $\left\vert D_{p,q}^{r_{b}r_{b}}\right\vert
		^{2}$ represents the transmission probability density from port 1 to 4. $\left\vert \tilde{M}_{p,q}^{l_{a}l_{a}}\right\vert ^{2}
		$ represents the transmission probability density from port 4 to 1. $\left\vert
		D_{p,q}^{l_{b}l_{b}}\right\vert ^{2}$ represents the transmission probability density from port 1 to 3. $\left\vert M_{p,q}^{l_{a}l_{a}
		}\right\vert ^{2}$ represents the transmission probability density from port 3 to 1.}
	\label{Fig12}
\end{figure}

In this subsection, we investigate the two-photon diode effects at other port configurations both in the absence and presence of atomic dissipation. First, the transmission properties between port 3 and 4 are similar to 1 and 2. Then we study the two symmetric structures of ports 1-4 and ports 1-3. The numerical results under dissipation-free conditions are presented in Fig.\:\:\hyperref[Fig12]{12}. Figs.\:\:\hyperref[Fig12]{12(a)},\:\:\hyperref[Fig12]{12(c)} and\:\:\hyperref[Fig12]{12(e)} show the bidirectional transmission probability densities as functions of the coupling strength $\gamma
_{ar}$ for the port 1-4 configuration, while Figs.\:\:\hyperref[Fig12]{12(b)},\:\:\hyperref[Fig12]{12(d)} and\:\:\hyperref[Fig12]{12(f)} display the corresponding results for the ports 1-3 configuration. Diode effects dominated by different mechanisms can be realized at both port configurations by matching the chiral coupling parameters with the input photon detunings. For the ports 1-4 configuration with chiral couplings $\gamma
_{br}=0.5$ and $\gamma_{bl}=0$, under single-photon resonance and $\Delta
_{p}=\Delta_{q}=0$ where the independent propagation dominates, the backward transmission probability density remains zero throughout, while the forward transmission probability density increases monotonically with $\gamma
_{ar}$ and reaches optimal performance at $\gamma
_{ar}=0.5$ [see Fig.\:\:\hyperref[Fig12]{12(a)}]. When the correlated-part dominated transmission at detunings $\Delta
_{p}=0.5$ and $\Delta_{q}=-0.5$ is chosen, the diode effect is still observable but with a drastically reduced transmission probability density [see Fig.\:\:\hyperref[Fig12]{12(c)}]. Under two-photon resonance and $\Delta
_{p}=\Delta_{q}=1$, an independent-propagation-dominated diode effect emerges [see Fig.\:\:\hyperref[Fig12]{12(e)}], with a lower maximum transmission probability density than in the single-photon resonance case [see Fig.\:\:\hyperref[Fig12]{12(a)}]. The ports 1-3 configuration with chiral couplings $\gamma
_{br}=0$ and $\gamma_{bl}=0.5$ exhibits a fully symmetric diode behavior: under single-photon resonance, it supports both an independent-propagation-dominated [see Fig.\:\:\hyperref[Fig12]{12(b)}] and a correlation-dominated [see Fig.\:\:\hyperref[Fig12]{12(d)}] diode effect, and under two-photon resonance it shows an independent-propagation-dominated diode effect [see Fig.\:\:\hyperref[Fig12]{12(f)}]. Upon introducing atomic dissipation, the transmission probability densities of both port configurations 1-3 and 1-4 decrease, following the same physical mechanism as the dissipation effect in single-photon diodes.
\begin{figure}[t]
	\centering
	\par
	\begin{center}
		\includegraphics[scale=0.41]{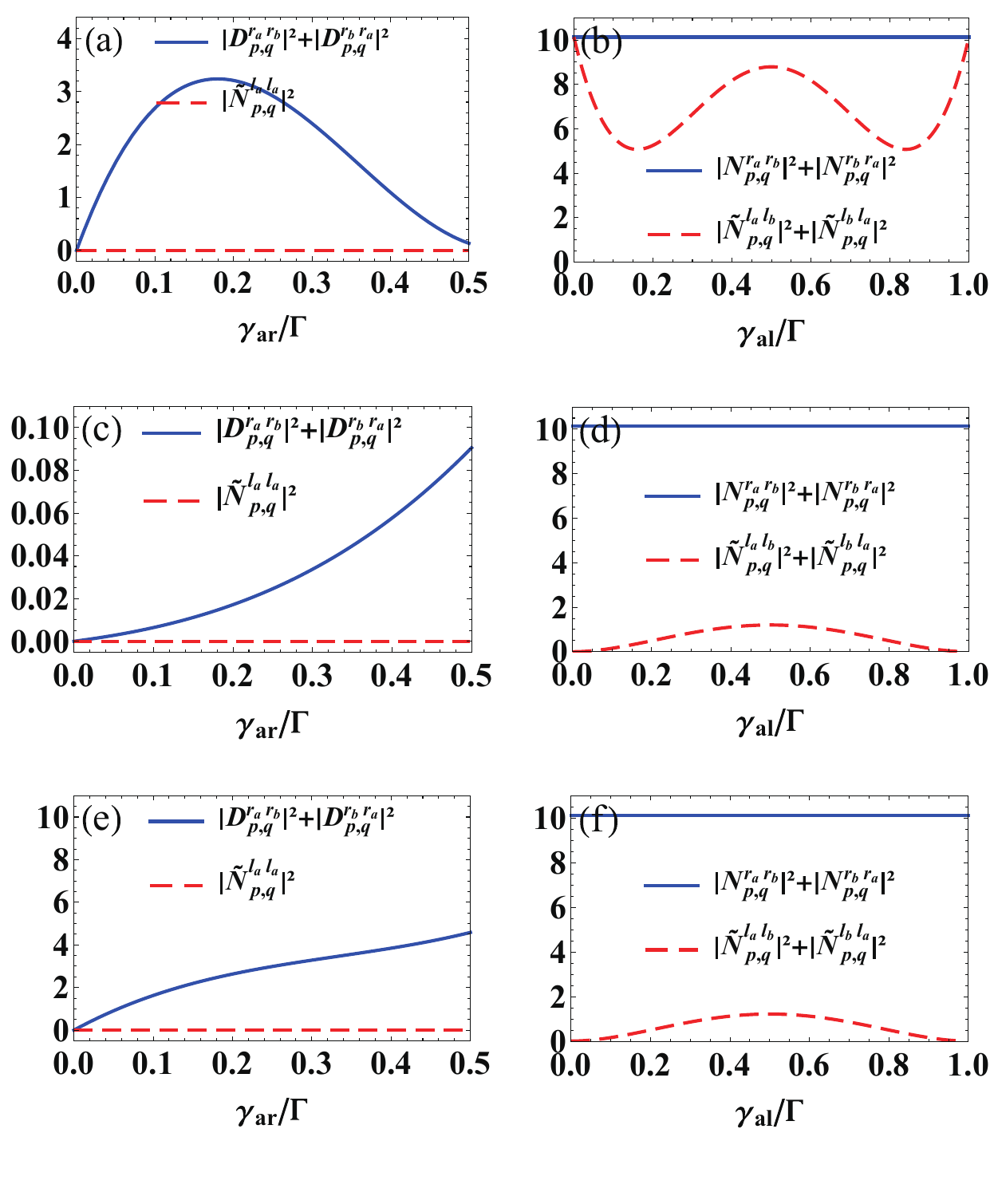}
	\end{center}
	\par
	\renewcommand\figurename{\textbf{FIG.}}
	\vspace{-10.0 mm}\caption{(a, c, e) $\left\vert D_{p,q}^{r_{a}r_{b}}\right\vert
		^{2}+\left\vert D_{p,q}^{r_{b}r_{a}}\right\vert ^{2}$ and $\left\vert
		\tilde{N}_{p,q}^{l_{a}l_{a}}\right\vert ^{2}$ versus $\gamma_{ar}$; (b, d, f) $\left\vert
		N_{p,q}^{r_{a}r_{b}}\right\vert ^{2}+\left\vert N_{p,q}^{r_{b}r_{a}
		}\right\vert ^{2}$ and $\left\vert \tilde{N}_{p,q}^{l_{a}l_{b}}\right\vert
		^{2}+\left\vert \tilde{N}_{p,q}^{l_{b}l_{a}}\right\vert ^{2}$ versus $\gamma_{al}$. $\left\vert D_{p,q}^{r_{a}r_{b}}\right\vert
		^{2}+\left\vert D_{p,q}^{r_{b}r_{a}}\right\vert ^{2}$ represents the transmission probability density from port 1 to 2, 4. $\left\vert
		\tilde{N}_{p,q}^{l_{a}l_{a}}\right\vert ^{2}$ represents the transmission probability density from ports 2, 4 to 1.
		$\left\vert
		N_{p,q}^{r_{a}r_{b}}\right\vert ^{2}+\left\vert N_{p,q}^{r_{b}r_{a}
		}\right\vert ^{2}$ represents the transmission probability density from ports 1, 3 to 2, 4. $\left\vert \tilde{N}_{p,q}^{l_{a}l_{b}}\right\vert
		^{2}+\left\vert \tilde{N}_{p,q}^{l_{b}l_{a}}\right\vert ^{2}$ represents the transmission probability density from ports 2, 4 to 1, 3. Panels (a)--(d): single-photon resonance at $\delta=0$. (a) ($\Delta_{p}$, $\Delta_{q}$, 
		$\varepsilon$, $\gamma_{br}$, $\gamma_{bl}$,
		$\kappa$) = (0, 0, 2, 0.5, 0, 0) and (b) ($\Delta_{p}$, $\Delta_{q}$, 
		$\varepsilon$, $\gamma_{ar}$, $\gamma_{br}$,
		$\kappa$) = (0, 0, 2, 0, 0, 0). (c) ($\Delta_{p}$, $\Delta_{q}$, 
		$\varepsilon$, $\gamma_{br}$, $\gamma_{bl}$,
		$\kappa$) = (0.5, -0.5, 2, 0.5, 0, 0) and (d) ($\Delta_{p}$, $\Delta_{q}$, 
		$\varepsilon$, $\gamma_{ar}$, $\gamma_{br}$,
		$\kappa$) = (0, 0, 2, 0, 0, 1). Panels (e)--(f): two-photon resonance at $2\delta=\varepsilon$. (e) ($\Delta_{p}$, $\Delta_{q}$, 
		$\varepsilon$,
		$\delta$, 
		$\gamma_{br}$, $\gamma_{bl}$,
		$\kappa$) = (1, 1, 2, 1, 0.5, 0, 0) and (f) ($\Delta_{p}$, $\Delta_{q}$, 
		$\varepsilon$,
		$\delta$, 
		$\gamma_{ar}$, $\gamma_{br}$,
		$\kappa$) = (0.1, 0.1, 0.2, 0.1, 0, 0, 1). Other parameters $\epsilon=0.1$. All
		parameters are in units of $\Gamma$.}%
	\label{Fig13}
\end{figure}

In the ports 1-2, 4 configuration, two photons incident from port 1 in the forward direction exit separately from ports 2 and 4, whereas two photons incident separately from ports 2 and 4 in the backward direction exit jointly from port 1 [see Figs.\:\:\hyperref[Fig13]{13(a)},\:\:\hyperref[Fig13]{13(c)}, and\:\:\hyperref[Fig13]{13(e)}]. With chiral couplings $\gamma
_{br}=0.5$ and $\gamma_{bl}=0$, the backward two-photon transmission probability density is identically zero for all parameter sets. Under single-photon resonance at $\Delta
_{p}=\Delta_{q}=0$, the forward transmission probability
density first increases and then decreases with $\gamma
_{ar}$ [see Fig.\:\:\hyperref[Fig13]{13(a)}]. At detunings $\Delta_{p}=0.5$ and $\Delta_{q}=-0.5$, it increases monotonically with $\gamma
_{ar}$ [see Fig.\:\:\hyperref[Fig13]{13(c)}]. Under two-photon resonance at $\Delta
_{p}=\Delta_{q}=1$, it also grows monotonically with $\gamma
_{ar}$ [see Fig.\:\:\hyperref[Fig13]{13(e)}]. However, this behavior does not constitute a strict two-photon diode effect. From a physical perspective, because of the constraint imposed by chiral coupling, photons incident from port 4 do not interact with the atoms and are instead transmitted directly to port 3. Only photons incident from port 2 can be output from port 1, which means that photons are always output from port 1. Therefore, this does not conform to the physical definition of a two-photon diode in the sense of optical isolation. To resolve this issue, atomic dissipation is introduced to suppress the transmission from port 2 to port 1. With $\kappa=0.2$, $\gamma_{br}=0.3$ and $\gamma_{bl}=0$, a valid diode effect is achieved only at $\gamma
_{ar}=0.1$ under single-photon resonance. This can be understood from Eq. \eqref{21a}. A photon incident from port 2 is absorbed by the atom. Part of its energy is dissipated into the environment, and the remaining energy is re-emitted into waveguide b and finally leaves through port 4. In this case, photons input from ports 2 and 4 are not transmitted through port 1, and the diode effect is therefore achieved. No parameters satisfying the diode requirement exist under two-photon resonance.

For the ports 1, 3-2, 4 configuration, where photons incident separately from ports 1 and 3 in the forward direction exit from ports 2 and 4 separately, while photons incident from ports 2 and 4 in the backward direction exit from ports 1 and 3 separately, the dissipation-free result is presented in Fig.\:\:\hyperref[Fig13]{13(b)}. Under single-photon resonance with $\gamma
_{ar}=0$, $\gamma_{br}=0$ and input detunings $\Delta_{p}=\Delta_{q}=0$, the forward transmission probability density remains at its maximum value, while the backward transmission probability density oscillates with the coupling strength $\gamma
_{al}$, showing no two-photon diode effect [see Fig.\:\:\hyperref[Fig13]{13(b)}]. After introducing dissipation $\kappa
=1$, as shown in Fig.\:\:\hyperref[Fig13]{13(d)}, the backward transmission probability density drops to zero at $\gamma
_{al}=0$ and $\gamma_{al}=1$. Although this numerical behavior superficially resembles a diode effect, it is not a genuine nonreciprocal isolation effect. As an example, consider $\gamma
_{al}=1$: photons incident from port 4 pass directly to port 3 without interacting with the atom, while photons incident from port 2 are fully dissipated into the environment after absorption, leaving no photons to exit from port 1 and thus yielding zero total two-photon backward transmission. However, the photon propagation is not blocked in this direction. Under two-photon resonance at input detunings $\Delta
_{p}=\Delta_{q}=0.1$, the result with $\kappa
=1$ shown in Fig.\:\:\hyperref[Fig13]{13(f)} follows the same trend as in the single-photon resonance case.

\section{Feasibility of an Experiment}\label{V}

In this section we examine whether the chiral coupling scheme proposed above can be implemented with current technology. Fermi's golden rule tells us that the coupling rates between the Rydberg atom pair and the two waveguides, $\gamma_{jr}$ and $\gamma_{jl}$ ($j=a,b$), vary inversely with the group velocity of the mode propagating to the right and to the left, respectively. An ordinary waveguide offers the same group velocity in either direction, a consequence of time-reversal symmetry (TRS) \cite{59}. Maxwell's equations are invariant under time reversal, so the reversed image of any valid electromagnetic field configuration is again a valid solution; light consequently propagates reciprocally. For a local system that is isolated, this constraint can be evaded once its TRS is lifted, for instance by an external drive or by a magnetic field applied to the sample \cite{60}. One then writes $H_{sys}=H_{0}-\vec{\mu}\cdot\vec{B}_{ext}$, with $H_{0}$ the free Hamiltonian that respects the symmetry and $\vec{\mu}$ the magnetic moment, which flips sign under time reversal according to $T\vec{\mu}T^{-1}=-\vec{\mu}$, while $\vec{B}_{ext}$ denotes a static external field acting on the isolated local system. It follows that $TH_{sys}T^{-1}=H_{0}+\vec{\mu}\cdot\vec{B}_{ext}$, so the local system no longer possesses TRS \cite{56}. A global time-reversal operation inverts the sign of the magnetic field yet leaves Maxwell's equations unchanged \cite{60}. Once TRS is lifted, one cannot assume that reversing a solution of Maxwell's equations yields another solution with an identical frequency. Put differently, the wave that would travel backwards may either be missing or appear at a different frequency. Counter-propagating photons then become separately addressable, which is exactly what makes a system chiral. Such a situation arises when the photonic medium responds magneto-optically to an applied field, or when it is coupled to an emitter whose levels are shifted by a magnetic field through the Zeeman effect \cite{60}.

The strongly TRS-broken chiral waveguides invoked in this work have already been realized on gyromagnetic photonic-crystal platforms \cite{61,62} and employed for quantum information tasks \cite{49}. Chiral light-matter interaction is likewise available in nanobeam ridge waveguides \cite{63}, in conventional line-defect photonic crystal waveguides \cite{64}, and in topological spin-Hall \cite{65,66} and valley-Hall \cite{67,68} photonic crystal waveguides. Chiral couplings have also been demonstrated with multi-atom arrays \cite{14,15,16}, with superconducting circuit QED architectures \cite{5,6}, and with optomechanical systems \cite{1,2}, exploiting spin-orbit interaction or synthetic gauge fields. Our model is therefore within reach of several mature experimental platforms.

\section{Conclusion}\label{VI}

In summary, this work systematically investigates few-photon diode effects across multi-port configurations, spanning both single-photon and two-photon transport and presenting a thorough comparison between dissipation-free and dissipative atomic scenarios, on the basis of a chiral coupling architecture composed of dual waveguides and a Rydberg atom pair.

At the single-photon level, an ideal unidirectional single-photon diode can be realized at ports 1-2, 1-3 and 1-4 purely via chiral couplings, with no requirement for atomic dissipation which is not found in existing references. Its transmission performance can be flexibly tuned through the coupling ratios between waveguides. With atomic dissipation introduced, a perfect diode effect can still be realized by matching the dissipation strength to the coupling parameters, offering an additional degree of freedom for device design. This scheme can be further extended to multi-port setups such as ports 1-4 and ports 1-3, all with clear physical mechanisms and favorable experimental feasibility.

At the two-photon level, robust diode effects are governed by two distinct mechanisms, the photon correlation term and the independent photon propagation term, which can be realized between various port configurations (e.g., ports 1-2, 1-3, 1-4) with and without dissipation, in a wide range of parameters, which Refs. \cite{56,57} fail
to achieve. For ports 1-2, 4, a limited yet functional diode effect can only be achieved under specific dissipation parameters. The interatomic Rydberg interaction effectively broadens the operating bandwidth of the correlation-dominated diode. 
This work provides a systematic study of few-photon scattering among multiple ports in chiral coupling systems and offers theoretical guidance for the design and experimental implementation of multi-port quantum nonreciprocal devices.

\begin{widetext}
	\section*{Appendix A: Solution of Eq. (27)}
	\phantomsection\label{app:A}
	
Starting from the
equations of motion in
Eq. \eqref{23}, together with
the initial conditions
in Eq. \eqref{24}, we apply
the Laplace transform to
obtain the long-time
two-photon scattering
solution given in Eq. \eqref{27}.
	\begin{gather}
		s\tilde{A}\left(  s\right)  =-\left(  i\varepsilon+2\kappa\right)  \tilde
		{A}\left(  s\right)  -i\sqrt{\frac{\Gamma}{2\pi}}\int_{0}^{\infty}dk\left[
		\tilde{B}_{k}^{1}\left(  s\right)  +\tilde{B}_{k}^{2}\left(  s\right)
		\right]
		,\tag{A1}\\
		s\tilde{B}_{k}^{1}\left(  s\right)  =-\left(  i\Delta_{k}+\kappa\right)
		\tilde{B}_{k}^{1}\left(  s\right)  -i\sqrt{\frac{\Gamma}{2\pi}}\tilde
		{A}\left(  s\right)  -i\sqrt{\frac{\Gamma}{2\pi}}\int_{0}^{\infty}dp\tilde
		{C}_{pk}\left(  s\right)  ,\tag{A2}\\
		s\tilde{B}_{k}^{2}\left(  s\right)  =-\left(  i\Delta_{k}+\kappa\right)
		\tilde{B}_{k}^{2}\left(  s\right)  -i\sqrt{\frac{\Gamma}{2\pi}}\tilde
		{A}\left(  s\right)  -i\sqrt{\frac{\Gamma}{2\pi}}\int_{0}^{\infty}dp\tilde
		{C}_{pk}\left(  s\right)  ,\tag{A3}\\
		s\tilde{C}_{pq}\left(  s\right)  =C_{pq}\left(  0\right)  -i\left(
		\Delta_{p}+\Delta_{q}\right)  \tilde{C}_{pq}\left(  s\right)
		-i\sqrt{\frac{\Gamma}{2\pi}}\left[  \tilde{B}_{p}^{1}\left(  s\right)
		+\tilde{B}_{q}^{1}\left(  s\right)  \right]  -i\sqrt{\frac{\Gamma}{2\pi}%
		}\left[  \tilde{B}_{p}^{2}\left(  s\right)  +\tilde{B}_{q}^{2}\left(
		s\right)  \right]
		.\tag{A4}
\end{gather}
Thus, the equation for the variable $\tilde{B}%
_{k}^{1}\left(  s\right)  $ can be derived:
\begin{align}
	\left(  s+\kappa+\Gamma+i\Delta_{k}\right)  \tilde{B}_{k}^{1}\left(
	s\right)   &  =-\frac{\Gamma}{\pi}\int_{-\infty}^{\infty}d\Delta_{k}\left[
	\frac{1}{\left(  s+2\kappa+i\varepsilon\right)  }+\frac{1}{s+i\left(
		\Delta_{p}+\Delta_{k}\right)  }\right]  \tilde{B}_{p}^{1}\left(
	s\right) \nonumber\\
	&  -i\sqrt{\frac{\Gamma}{2\pi}}\int_{-\infty}^{\infty}d\Delta_{p}%
	\frac{C_{pk}\left(  0\right)  }{\left[  s+i\left(  \Delta_{p}+\Delta
		_{k}\right)  \right]  }.\tag{A5}%
\end{align}
Finally, we solve for $\tilde{B}_{k}%
^{1}\left(  s\right)  $ in the form:
\begin{equation}
	\tilde{B}_{k}^{1}\left(  s\right)  =\frac{\sqrt{2\pi\Gamma}G}{\Delta
		_{k}-i\left(  s+\kappa+\Gamma\right)  }\left[  \tilde{F}_{1}+2\Gamma\left(
	\tilde{F}_{2}+\tilde{F}_{3}\right)  \right]  .\tag{A6}%
\end{equation}
The parameters are as follows
\begin{gather}
	\tilde{F}_{1}=\frac{1}{\left[  \Delta_{k}+\delta_{1}-i\left(
		\epsilon+s\right)  \right]  \left(  \Delta_{k}-\delta_{2}+i\epsilon\right)
	}+\frac{1}{\left[  \Delta_{k}+\delta_{2}-i\left(  \epsilon+s\right)
		\right]  \left(  \Delta_{k}-\delta_{1}+i\epsilon\right)  },\tag{A7}\\
	\begin{align}
		\tilde{F}_{2}  &  =\frac{1}{\left[  \Delta_{k}+i\left(  \kappa
			+\Gamma\right)  \right]  \left(  \Delta_{k}-\delta_{2}+i\epsilon\right)
			\left[  s+i\left(  \Delta_{k}+\delta_{1}\right)  +\epsilon\right]
		}\nonumber\\
		&  +\frac{1}{\left(  \kappa+\Gamma-\epsilon-i\delta_{2}\right)  \left(
			s+\kappa+\Gamma+\epsilon+i\delta_{1}\right)  }\left(  \frac{1}{i\Delta
			_{k}-\kappa-\Gamma}+\frac{1}{s+2\kappa+\Gamma+i\varepsilon}\right)  ,\tag{A8}\\
	\tilde{F}_{3}  &  =\frac{1}{s+i\left(  \delta_{1}+\delta_{2}\right)
		+2\epsilon}[\frac{1}{\left(  \delta_{2}+i\kappa+i\Gamma-i\epsilon\right)
		\left(  \delta_{2}-i\epsilon-\Delta_{k}\right)  }-\frac{1}{\left(
		i\delta_{2}+\epsilon+s+\kappa+\Gamma\right)  \left(  s+2\kappa+\Gamma
		+i\varepsilon\right)  }\nonumber\\
	&  -\frac{1}{\left(  \kappa+\Gamma-i\delta_{2}-\epsilon\right)  \left(
		s+2\kappa+\Gamma+i\varepsilon\right)  }-\frac{1}{\left(  i\delta_{2}%
		+\epsilon+s+\kappa+\Gamma\right)  \left(  i\delta_{2}+\epsilon+i\Delta
		_{k}+s\right)  }].\tag{A9}%
\end{align}
\end{gather}
Then we have
\begin{equation}
	\tilde{C}_{pq}\left(  s\right)  =\frac{C_{pq}\left(  0\right)  -i\sqrt
		{\frac{2\Gamma}{\pi}}\left[  \tilde{B}_{p}^{1}\left(  s\right)  +\tilde{B}%
		_{q}^{1}\left(  s\right)  \right]  }{s+i\left(  \Delta_{p}+\Delta
		_{q}\right)  }.\tag{A10}%
\end{equation}
The transient solution $C_{pq}\left(  t\right)  $ can be obtained via the inverse Laplace transform, while the study of photon scattering only requires examining the long-time asymptotic solution given by Eq. \eqref{27} under the conditions $t\gg\Gamma^{-1}$ and $\epsilon^{-1}$.

\end{widetext}

\begin{widetext}
\section*{Appendix B: the two-photon scattering state for two-photon incident from port 1}
\phantomsection\label{app:B}
\end{widetext}
\[
\psi(t\rightarrow\infty)=\int_{0}^{\infty}\int_{0}^{\infty}dpdqD_{p,q}%
^{r_{a}r_{a}}e^{-i\left(  \Delta_{p}+\Delta_{q}\right)  t}\hat{r}%
_{ap}^{\dagger}\hat{r}_{aq}^{\dagger}\left\vert 0\right\rangle
\]
\[
+\int_{0}^{\infty}\int_{0}^{\infty}dpdqD_{p,q}^{r_{a}l_{a}}e^{-i\left(
	\Delta_{p}+\Delta_{q}\right)  t}\hat{r}_{ap}^{\dagger}\hat{l}_{aq}^{\dagger
}\left\vert 0\right\rangle
\]
\[
+\int_{0}^{\infty}\int_{0}^{\infty}dpdqD_{p,q}^{l_{a}r_{a}}e^{-i\left(
	\Delta_{p}+\Delta_{q}\right)  t}\hat{l}_{ap}^{\dagger}\hat{r}_{aq}^{\dagger
}\left\vert 0\right\rangle
\]
\[
+\int_{0}^{\infty}\int_{0}^{\infty}dpdqD_{p,q}^{l_{a}l_{a}}e^{-i\left(
	\Delta_{p}+\Delta_{q}\right)  t}\hat{l}_{ap}^{\dagger}\hat{l}_{aq}^{\dagger
}\left\vert 0\right\rangle
\]
\[
+\int_{0}^{\infty}\int_{0}^{\infty}dpdqD_{p,q}^{r_{a}r_{b}}e^{-i\left(
	\Delta_{p}+\Delta_{q}\right)  t}\hat{r}_{ap}^{\dagger}\hat{r}_{bq}^{\dagger
}\left\vert 0\right\rangle
\]
\[
+\int_{0}^{\infty}\int_{0}^{\infty}dpdqD_{p,q}^{r_{a}l_{b}}e^{-i\left(
	\Delta_{p}+\Delta_{q}\right)  t}\hat{r}_{ap}^{\dagger}\hat{l}_{bq}^{\dagger
}\left\vert 0\right\rangle
\]
\[
+\int_{0}^{\infty}\int_{0}^{\infty}dpdqD_{p,q}^{l_{a}r_{b}}e^{-i\left(
	\Delta_{p}+\Delta_{q}\right)  t}\hat{l}_{ap}^{\dagger}\hat{r}_{bq}^{\dagger
}\left\vert 0\right\rangle
\]
\[
+\int_{0}^{\infty}\int_{0}^{\infty}dpdqD_{p,q}^{l_{a}l_{b}}e^{-i\left(
	\Delta_{p}+\Delta_{q}\right)  t}\hat{l}_{ap}^{\dagger}\hat{l}_{bq}^{\dagger
}\left\vert 0\right\rangle
\]
\[
+\int_{0}^{\infty}\int_{0}^{\infty}dpdqD_{p,q}^{r_{b}r_{a}}e^{-i\left(
	\Delta_{p}+\Delta_{q}\right)  t}\hat{r}_{bp}^{\dagger}\hat{r}_{aq}^{\dagger
}\left\vert 0\right\rangle
\]
\[
+\int_{0}^{\infty}\int_{0}^{\infty}dpdqD_{p,q}^{r_{b}l_{a}}e^{-i\left(
	\Delta_{p}+\Delta_{q}\right)  t}\hat{r}_{bp}^{\dagger}\hat{l}_{aq}^{\dagger
}\left\vert 0\right\rangle
\]
\[
+\int_{0}^{\infty}\int_{0}^{\infty}dpdqD_{p,q}^{l_{b}r_{a}}e^{-i\left(
	\Delta_{p}+\Delta_{q}\right)  t}\hat{l}_{bp}^{\dagger}\hat{r}_{aq}^{\dagger
}\left\vert 0\right\rangle
\]
\[
+\int_{0}^{\infty}\int_{0}^{\infty}dpdqD_{p,q}^{l_{b}l_{a}}e^{-i\left(
	\Delta_{p}+\Delta_{q}\right)  t}\hat{l}_{bp}^{\dagger}\hat{l}_{aq}^{\dagger
}\left\vert 0\right\rangle
\]
\[
+\int_{0}^{\infty}\int_{0}^{\infty}dpdqD_{p,q}^{r_{b}r_{b}}e^{-i\left(
	\Delta_{p}+\Delta_{q}\right)  t}\hat{r}_{bp}^{\dagger}\hat{r}_{bq}^{\dagger
}\left\vert 0\right\rangle
\]
\[
+\int_{0}^{\infty}\int_{0}^{\infty}dpdqD_{p,q}^{r_{b}l_{b}}e^{-i\left(
	\Delta_{p}+\Delta_{q}\right)  t}\hat{r}_{bp}^{\dagger}\hat{l}_{bq}^{\dagger
}\left\vert 0\right\rangle
\]
\[
+\int_{0}^{\infty}\int_{0}^{\infty}dpdqD_{p,q}^{l_{b}r_{b}}e^{-i\left(
	\Delta_{p}+\Delta_{q}\right)  t}\hat{l}_{bp}^{\dagger}\hat{r}_{bq}^{\dagger
}\left\vert 0\right\rangle
\]
\[
+\int_{0}^{\infty}\int_{0}^{\infty}dpdqD_{p,q}^{l_{b}l_{b}}e^{-i\left(
	\Delta_{p}+\Delta_{q}\right)  t}\hat{l}_{bp}^{\dagger}\hat{l}_{bq}^{\dagger
}\left\vert 0\right\rangle
\]
where
\begin{subequations}
		\begin{equation}
			D_{p,q}^{r_{a}r_{a}}=t_{ap}^{0}t_{aq}^{0}C_{p,q}(0)+\frac{\gamma_{ar}^{2}%
			}{\Gamma^{2}}J_{p,q},\tag{B1}
		\end{equation}
		\begin{equation}
			D_{p,q}^{r_{a}l_{a}}=t_{ap}^{0}r_{aq}^{0}C_{p,q}(0)+\frac{\gamma_{ar}%
				\sqrt{\gamma_{ar}\gamma_{al}}}{\Gamma^{2}}J_{p,q},\tag{B2}
		\end{equation}
		\begin{equation}
			D_{p,q}^{l_{a}r_{a}}=r_{ap}^{0}t_{aq}^{0}C_{p,q}(0)+\frac{\gamma_{ar}%
				\sqrt{\gamma_{ar}\gamma_{al}}}{\Gamma^{2}}J_{p,q},\tag{B3}
		\end{equation}
		\begin{equation}
			D_{p,q}^{l_{a}l_{a}}=r_{ap}^{0}r_{aq}^{0}C_{p,q}(0)+\frac{\gamma_{ar}%
				\gamma_{al}}{\Gamma^{2}}J_{p,q},\tag{B4}
		\end{equation}

		\begin{equation}
			D_{p,q}^{r_{a}r_{b}}=t_{ap}^{0}t_{bq}^{0}C_{p,q}(0)+\frac{\gamma_{ar}%
				\sqrt{\gamma_{ar}\gamma_{br}}}{\Gamma^{2}}J_{p,q},\tag{B5}
		\end{equation}%
		\begin{equation}
			D_{p,q}^{r_{a}l_{b}}=t_{ap}^{0}r_{bq}^{0}C_{p,q}(0)+\frac{\gamma_{ar}%
				\sqrt{\gamma_{ar}\gamma_{bl}}}{\Gamma^{2}}J_{p,q},\tag{B6}
		\end{equation}

		\begin{equation}
			D_{p,q}^{l_{a}r_{b}}=r_{ap}^{0}t_{bq}^{0}C_{p,q}(0)+\frac{\gamma_{ar}%
				\sqrt{\gamma_{al}\gamma_{br}}}{\Gamma^{2}}J_{p,q},\tag{B7}
		\end{equation}

		\begin{equation}
			D_{p,q}^{l_{a}l_{b}}=r_{ap}^{0}r_{bq}^{0}C_{p,q}(0)+\frac{\gamma_{ar}%
				\sqrt{\gamma_{al}\gamma_{bl}}}{\Gamma^{2}}J_{p,q},\tag{B8}
		\end{equation}%
		\begin{equation}
			D_{p,q}^{r_{b}r_{a}}=t_{bp}^{0}t_{aq}^{0}C_{p,q}(0)+\frac{\gamma_{ar}%
				\sqrt{\gamma_{ar}\gamma_{br}}}{\Gamma^{2}}J_{p,q},\tag{B9}
		\end{equation}%
		\begin{equation}
			D_{p,q}^{r_{b}l_{a}}=t_{bp}^{0}r_{aq}^{0}C_{p,q}(0)+\frac{\gamma_{ar}%
				\sqrt{\gamma_{al}\gamma_{br}}}{\Gamma^{2}}J_{p,q},\tag{B10}
		\end{equation}%
		\begin{equation}
			D_{p,q}^{l_{b}r_{a}}=r_{bp}^{0}t_{aq}^{0}C_{p,q}(0)+\frac{\gamma_{ar}%
				\sqrt{\gamma_{ar}\gamma_{bl}}}{\Gamma^{2}}J_{p,q},\tag{B11}
		\end{equation}%
		\begin{equation}
			D_{p,q}^{l_{b}l_{a}}=r_{bp}^{0}r_{aq}^{0}C_{p,q}(0)+\frac{\gamma_{ar}%
				\sqrt{\gamma_{al}\gamma_{bl}}}{\Gamma^{2}}J_{p,q},\tag{B12}
		\end{equation}%
		\begin{equation}
			D_{p,q}^{r_{b}r_{b}}=t_{bp}^{0}t_{bq}^{0}C_{p,q}(0)+\frac{\gamma_{ar}%
				\gamma_{br}}{\Gamma^{2}}J_{p,q},\tag{B13}
		\end{equation}%
		\begin{equation}
			D_{p,q}^{r_{b}l_{b}}=t_{bp}^{0}r_{bq}^{0}C_{p,q}(0)+\frac{\gamma_{ar}%
				\sqrt{\gamma_{br}\gamma_{bl}}}{\Gamma^{2}}J_{p,q},\tag{B14}
		\end{equation}%
		\begin{equation}
			D_{p,q}^{l_{b}r_{b}}=r_{bp}^{0}t_{bq}^{0}C_{p,q}(0)+\frac{\gamma_{ar}%
				\sqrt{\gamma_{br}\gamma_{bl}}}{\Gamma^{2}}J_{p,q},\tag{B15}
		\end{equation}%
		\begin{equation}
			D_{p,q}^{l_{b}l_{b}}=r_{bp}^{0}r_{bq}^{0}C_{p,q}(0)+\frac{\gamma_{ar}%
				\gamma_{bl}}{\Gamma^{2}}J_{p,q}.\tag{B16}
		\end{equation}	
\end{subequations}
Here, $D_{p,q}
^{r_{a}r_{a}}$ and $D_{p,q}^{l_{a}l_{a}}$ (and correspondingly $D_{p,q}^{r_{b}r_{b}}
$ and $D_{p,q}^{l_{b}l_{b}}$) denote the transmission and reflection amplitudes of two photons in waveguide $a$ (or $b$), respectively. $D_{p,q}%
^{r_{a}l_{a}}$ (or $D_{p,q}^{l_{a}r_{a}}$ ) describe the amplitude for a photon with wavenumber $p$ (or $q$) being transmitted while a photon with wavenumber $q$ (or $p$) is reflected in waveguide $a$. $D_{p,q}^{r_{b}l_{b}}%
$(or $D_{p,q}^{l_{b}r_{b}}$) can be similarly defined in waveguide $b$. 
The terms $D_{p,q}%
^{r_{a}r_{b}}$
(or $D_{p,q}^{r_{b}r_{a}}$ ) give the amplitude for a photon with wavenumber $p$ (or $q$) being transmitted in waveguide $a$ and a photon with wavenumber $q$ (or $p$) being transmitted in waveguide $b$.
The terms $D_{p,q}^{l_{a}l_{b}}$
(or $D_{p,q}
^{l_{b}l_{a}}$) give the amplitude for a photon with wavenumber $p$ (or $q$) being reflected in waveguide $a$ and a photon with wavenumber $q$ (or $p$) being reflected in waveguide $b$. 
$D_{p,q}%
^{r_{a}l_{b}}$ (or $D_{p,q}^{l_{b}r_{a}}$ )  describe the amplitude for a photon with wavenumber $p$ (or $q$) being transmitted in waveguide $a$ and a photon with wavenumber $q$ (or $p$) being reflected in waveguide $b$. 
$D_{p,q}^{r_{b}l_{a}}%
$(or $D_{p,q}^{l_{a}r_{b}}$) describe the amplitude for a photon with wavenumber $p$ (or $q$) being transmitted in waveguide $b$ and a photon with wavenumber $q$ (or $p$) being reflected in waveguide $a$.

\begin{widetext}
\section*{Appendix C: the two-photon scattering state for photons incident from port 2}
\phantomsection\label{app:C}
\end{widetext}
	\[\psi(t   \rightarrow\infty)=\int_{0}^{\infty}\int_{0}^{\infty}dpdq\tilde
	{D}_{p,q}^{r_{a}r_{a}}e^{-i\left(  \Delta_{p}+\Delta_{q}\right)  t}\hat
	{r}_{ap}^{\dagger}\hat{r}_{aq}^{\dagger}\left\vert 0\right\rangle \]
	
	 \[+\int_{0}^{\infty}\int_{0}^{\infty}dpdq\tilde{D}_{p,q}^{r_{a}l_{a}%
	}e^{-i\left(  \Delta_{p}+\Delta_{q}\right)  t}\hat{r}_{ap}^{\dagger}\hat
	{l}_{aq}^{\dagger}\left\vert 0\right\rangle \]
	
	 \[+\int_{0}^{\infty}\int_{0}^{\infty}dpdq\tilde{D}_{p,q}^{l_{a}r_{a}%
	}e^{-i\left(  \Delta_{p}+\Delta_{q}\right)  t}\hat{l}_{ap}^{\dagger}\hat
	{r}_{aq}^{\dagger}\left\vert 0\right\rangle \]
	
	 \[+\int_{0}^{\infty}\int_{0}^{\infty}dpdq\tilde{D}_{p,q}^{l_{a}l_{a}%
	}e^{-i\left(  \Delta_{p}+\Delta_{q}\right)  t}\hat{l}_{ap}^{\dagger}\hat
	{l}_{aq}^{\dagger}\left\vert 0\right\rangle \]
   
   \[+\int_{0}^{\infty}\int_{0}^{\infty}dpdq\tilde{D}_{p,q}^{r_{a}r_{b}%
	}e^{-i\left(  \Delta_{p}+\Delta_{q}\right)  t}\hat{r}_{ap}^{\dagger}\hat
	{r}_{bq}^{\dagger}\left\vert 0\right\rangle \]
	
	  \[+\int_{0}^{\infty}\int_{0}^{\infty}dpdq\tilde{D}_{p,q}^{r_{a}l_{b}%
	}e^{-i\left(  \Delta_{p}+\Delta_{q}\right)  t}\hat{r}_{ap}^{\dagger}\hat
	{l}_{bq}^{\dagger}\left\vert 0\right\rangle \]
	
	  \[+\int_{0}^{\infty}\int_{0}^{\infty}dpdq\tilde{D}_{p,q}^{l_{a}r_{b}%
	}e^{-i\left(  \Delta_{p}+\Delta_{q}\right)  t}\hat{l}_{ap}^{\dagger}\hat
	{r}_{bq}^{\dagger}\left\vert 0\right\rangle \]
	
	 \[+\int_{0}^{\infty}\int_{0}^{\infty}dpdq\tilde{D}_{p,q}^{l_{a}l_{b}%
	}e^{-i\left(  \Delta_{p}+\Delta_{q}\right)  t}\hat{l}_{ap}^{\dagger}\hat
	{l}_{bq}^{\dagger}\left\vert 0\right\rangle \]
	
	 \[+\int_{0}^{\infty}\int_{0}^{\infty}dpdq\tilde{D}_{p,q}^{r_{b}r_{a}%
	}e^{-i\left(  \Delta_{p}+\Delta_{q}\right)  t}\hat{r}_{bp}^{\dagger}\hat
	{r}_{aq}^{\dagger}\left\vert 0\right\rangle \]
	
	  \[+\int_{0}^{\infty}\int_{0}^{\infty}dpdq\tilde{D}_{p,q}^{r_{b}l_{a}%
	}e^{-i\left(  \Delta_{p}+\Delta_{q}\right)  t}\hat{r}_{bp}^{\dagger}\hat
	{l}_{aq}^{\dagger}\left\vert 0\right\rangle \]
	
	  \[+\int_{0}^{\infty}\int_{0}^{\infty}dpdq\tilde{D}_{p,q}^{l_{b}r_{a}%
	}e^{-i\left(  \Delta_{p}+\Delta_{q}\right)  t}\hat{l}_{bp}^{\dagger}\hat
	{r}_{aq}^{\dagger}\left\vert 0\right\rangle \]
	
	 \[+\int_{0}^{\infty}\int_{0}^{\infty}dpdq\tilde{D}_{p,q}^{l_{b}l_{a}%
	}e^{-i\left(  \Delta_{p}+\Delta_{q}\right)  t}\hat{l}_{bp}^{\dagger}\hat
	{l}_{aq}^{\dagger}\left\vert 0\right\rangle \]
	
	 \[+\int_{0}^{\infty}\int_{0}^{\infty}dpdq\tilde{D}_{p,q}^{r_{b}r_{b}%
	}e^{-i\left(  \Delta_{p}+\Delta_{q}\right)  t}\hat{r}_{bp}^{\dagger}\hat
	{r}_{bq}^{\dagger}\left\vert 0\right\rangle \]
   
   \[+\int_{0}^{\infty}\int_{0}^{\infty}dpdq\tilde{D}_{p,q}^{r_{b}l_{b}%
	}e^{-i\left(  \Delta_{p}+\Delta_{q}\right)  t}\hat{r}_{bp}^{\dagger}\hat
	{l}_{bq}^{\dagger}\left\vert 0\right\rangle \]
	
	 \[+\int_{0}^{\infty}\int_{0}^{\infty}dpdq\tilde{D}_{p,q}^{l_{b}r_{b}%
	}e^{-i\left(  \Delta_{p}+\Delta_{q}\right)  t}\hat{l}_{bp}^{\dagger}\hat
	{r}_{bq}^{\dagger}\left\vert 0\right\rangle \]
	
	 \[+\int_{0}^{\infty}\int_{0}^{\infty}dpdq\tilde{D}_{p,q}^{l_{b}l_{b}%
	}e^{-i\left(  \Delta_{p}+\Delta_{q}\right)  t}\hat{l}_{bp}^{\dagger}\hat
	{l}_{bq}^{\dagger}\left\vert 0\right\rangle\]
where
\begin{subequations}
	\begin{equation}
		\tilde{D}_{p,q}^{r_{a}r_{a}}   =\tilde{r}_{ap}^{0}\tilde{r}_{aq}^{0}%
		C_{p,q}(0)+\frac{\gamma_{ar}\gamma_{al}}{\Gamma^{2}}J_{p,q},\tag{C1}
	\end{equation}
		\begin{equation}
	    \tilde{D}_{p,q}^{r_{a}l_{a}}  =\tilde{r}_{ap}^{0}\tilde{t}_{aq}^{0}%
		C_{p,q}(0)+\frac{\gamma_{al}\sqrt{\gamma_{ar}\gamma_{al}}}{\Gamma^{2}}%
		J_{p,q},\tag{C2}
		\end{equation}
			\begin{equation}	
		\tilde{D}_{p,q}^{l_{a}r_{a}}  =\tilde{t}_{ap}^{0}\tilde{r}_{aq}^{0}%
		C_{p,q}(0)+\frac{\gamma_{al}\sqrt{\gamma_{ar}\gamma_{al}}}{\Gamma^{2}}%
		J_{p,q},\tag{C3}
		\end{equation}
			\begin{equation}
		\tilde{D}_{p,q}^{l_{a}l_{a}}  =\tilde{t}_{ap}^{0}\tilde{t}_{aq}^{0}%
		C_{p,q}(0)+\frac{\gamma_{al}^{2}}{\Gamma^{2}}J_{p,q},\tag{C4}
		\end{equation}
			\begin{equation}
        \tilde{D}_{p,q}^{r_{a}r_{b}}   =\tilde{r}_{ap}^{0}\tilde{r}_{bq}^{0}%
		C_{p,q}(0)+\frac{\gamma_{al}\sqrt{\gamma_{ar}\gamma_{br}}}{\Gamma^{2}}%
		J_{p,q},\tag{C5}
		\end{equation}
			\begin{equation}
		\tilde{D}_{p,q}^{r_{a}l_{b}}  =\tilde{r}_{ap}^{0}\tilde{t}_{bq}^{0}%
		C_{p,q}(0)+\frac{\gamma_{al}\sqrt{\gamma_{ar}\gamma_{bl}}}{\Gamma^{2}}%
		J_{p,q},\tag{C6}
		\end{equation}
			\begin{equation}
		\tilde{D}_{p,q}^{l_{a}r_{b}}  =\tilde{t}_{ap}^{0}\tilde{r}_{bq}^{0}%
		C_{p,q}(0)+\frac{\gamma_{al}\sqrt{\gamma_{al}\gamma_{br}}}{\Gamma^{2}}%
		J_{p,q},\tag{C7}
		\end{equation}
			\begin{equation}
		\tilde{D}_{p,q}^{l_{a}l_{b}}  =\tilde{t}_{ap}^{0}\tilde{t}_{bq}^{0}%
		C_{p,q}(0)+\frac{\gamma_{al}\sqrt{\gamma_{al}\gamma_{bl}}}{\Gamma^{2}}%
		J_{p,q},\tag{C8}
		\end{equation}
			\begin{equation}
		\tilde{D}_{p,q}^{r_{b}r_{a}}   =\tilde{r}_{bp}^{0}\tilde{r}_{aq}^{0}%
		C_{p,q}(0)+\frac{\gamma_{al}\sqrt{\gamma_{ar}\gamma_{br}}}{\Gamma^{2}}%
		J_{p,q},\tag{C9}
		\end{equation}
		\begin{equation}
		\tilde{D}_{p,q}^{r_{b}l_{a}}   =\tilde{r}_{bp}^{0}\tilde{t}_{aq}^{0}%
		C_{p,q}(0)+\frac{\gamma_{al}\sqrt{\gamma_{al}\gamma_{br}}}{\Gamma^{2}}%
		J_{p,q},\tag{C10}
		\end{equation}
			\begin{equation}
		\tilde{D}_{p,q}^{l_{b}r_{a}}   =\tilde{t}_{bp}^{0}\tilde{r}_{aq}^{0}%
		C_{p,q}(0)+\frac{\gamma_{al}\sqrt{\gamma_{ar}\gamma_{bl}}}{\Gamma^{2}}%
		J_{p,q},\tag{C11}
		\end{equation}
			\begin{equation}
		\tilde{D}_{p,q}^{l_{b}l_{a}}   =\tilde{t}_{bp}^{0}\tilde{t}_{aq}^{0}%
		C_{p,q}(0)+\frac{\gamma_{al}\sqrt{\gamma_{al}\gamma_{bl}}}{\Gamma^{2}}%
		J_{p,q},\tag{C12}
		\end{equation}
			\begin{equation}
		\tilde{D}_{p,q}^{r_{b}r_{b}}   =\tilde{r}_{bp}^{0}\tilde{r}_{bq}^{0}%
		C_{p,q}(0)+\frac{\gamma_{al}\gamma_{br}}{\Gamma^{2}}J_{p,q},\tag{C13}
		\end{equation}
			\begin{equation}
		\tilde{D}_{p,q}^{r_{b}l_{b}}  =\tilde{r}_{bp}^{0}\tilde{t}_{bq}^{0}%
		C_{p,q}(0)+\frac{\gamma_{al}\sqrt{\gamma_{br}\gamma_{bl}}}{\Gamma^{2}}%
		J_{p,q},\tag{C14}
		\end{equation}
			\begin{equation}
		\tilde{D}_{p,q}^{l_{b}r_{b}}   =\tilde{t}_{bp}^{0}\tilde{r}_{bq}^{0}%
		C_{p,q}(0)+\frac{\gamma_{al}\sqrt{\gamma_{br}\gamma_{bl}}}{\Gamma^{2}}%
		J_{p,q},\tag{C15}
		\end{equation}
			\begin{equation}
		\tilde{D}_{p,q}^{l_{b}l_{b}}   =\tilde{t}_{bp}^{0}\tilde{t}_{bq}^{0}%
		C_{p,q}(0)+\frac{\gamma_{al}\gamma_{bl}}{\Gamma^{2}}J_{p,q}.\tag{C16}
		\end{equation}
\end{subequations}
Here, $\tilde{D}_{p,q}^{r_{a}r_{a}}$ and $\tilde{D}%
_{p,q}^{l_{a}l_{a}}$
(and correspondingly $\tilde{D}_{p,q}^{r_{b}r_{b}}$
and $\tilde
{D}_{p,q}^{l_{b}l_{b}}$
) denote the reflection and transmission amplitudes of two photons in waveguide $a$ (or $b$), respectively.
$\tilde
{D}_{p,q}^{r_{a}l_{a}}$
(or $\tilde{D}_{p,q}^{l_{a}r_{a}}$
) describe the amplitude for a photon with wavenumber $p$ (or $q$) being reflected while a photon with wavenumber $q$ (or $p$) is transmitted in waveguide $a$. $\tilde
{D}_{p,q}^{r_{b}l_{b}}$
(or$\tilde{D}_{p,q}^{l_{b}r_{b}}$) can be similarly defined in waveguide $b$.
The terms $\tilde
{D}_{p,q}^{r_{a}r_{b}}$
(or $\tilde{D}_{p,q}^{r_{b}r_{a}}$
) give the amplitude for a photon with wavenumber $p$ (or $q$) being reflected in waveguide $a$ and a photon with wavenumber $q$ (or $p$) being reflected in waveguide $b$. 
The terms $\tilde{D}
_{p,q}^{l_{a}l_{b}}$
(or $\tilde{D}_{p,q}^{l_{b}l_{a}}$
) give the amplitude for a photon with wavenumber $p$ (or $q$) being transmitted in waveguide $a$ and a photon with wavenumber $q$ (or $p$) being transmitted in waveguide $b$. 
$\tilde
{D}_{p,q}^{r_{a}l_{b}}$
(or $\tilde{D}_{p,q}^{l_{b}r_{a}}$) describe the amplitude for a photon with wavenumber $p$ (or $q$) being reflected in waveguide $a$ and a photon with wavenumber $q$ (or $p$) being transmitted in waveguide $b$.
$\tilde
{D}_{p,q}^{r_{b}l_{a}}$
(or $\tilde{D}_{p,q}^{l_{a}r_{b}}$
) describe the amplitude for a photon with wavenumber $p$ (or $q$) being reflected in waveguide $b$ and a photon with wavenumber $q$ (or $p$) being transmitted in waveguide $a$.

\begin{widetext}
\section*{Appendix D: the two-photon scattering state for photons incident from port 3}
\phantomsection\label{app:D}
\end{widetext}

	\[\psi(t   \rightarrow\infty)=\int_{0}^{\infty}\int_{0}^{\infty}dpdqM_{p,q}%
	^{r_{a}r_{a}}e^{-i\left(  \Delta_{p}+\Delta_{q}\right)  t}\hat{r}%
	_{ap}^{\dagger}\hat{r}_{aq}^{\dagger}\left\vert 0\right\rangle \]
	
	 \[+\int_{0}^{\infty}\int_{0}^{\infty}dpdqM_{p,q}^{r_{a}l_{a}}e^{-i\left(
		\Delta_{p}+\Delta_{q}\right)  t}\hat{r}_{ap}^{\dagger}\hat{l}_{aq}^{\dagger
	}\left\vert 0\right\rangle \]
	
	 \[+\int_{0}^{\infty}\int_{0}^{\infty}dpdqM_{p,q}^{l_{a}r_{a}}e^{-i\left(
		\Delta_{p}+\Delta_{q}\right)  t}\hat{l}_{ap}^{\dagger}\hat{r}_{aq}^{\dagger
	}\left\vert 0\right\rangle \]
	
	 \[+\int_{0}^{\infty}\int_{0}^{\infty}dpdqM_{p,q}^{l_{a}l_{a}}e^{-i\left(
		\Delta_{p}+\Delta_{q}\right)  t}\hat{l}_{ap}^{\dagger}\hat{l}_{aq}^{\dagger
	}\left\vert 0\right\rangle \]
	
	 \[+\int_{0}^{\infty}\int_{0}^{\infty}dpdqM_{p,q}^{r_{a}r_{b}}e^{-i\left(
		\Delta_{p}+\Delta_{q}\right)  t}\hat{r}_{ap}^{\dagger}\hat{r}_{bq}^{\dagger
	}\left\vert 0\right\rangle \]
	
	\[+\int_{0}^{\infty}\int_{0}^{\infty}dpdqM_{p,q}^{r_{a}l_{b}}e^{-i\left(
		\Delta_{p}+\Delta_{q}\right)  t}\hat{r}_{ap}^{\dagger}\hat{l}_{bq}^{\dagger
	}\left\vert 0\right\rangle \]
	
	 \[+\int_{0}^{\infty}\int_{0}^{\infty}dpdqM_{p,q}^{l_{a}r_{b}}e^{-i\left(
		\Delta_{p}+\Delta_{q}\right)  t}\hat{l}_{ap}^{\dagger}\hat{r}_{bq}^{\dagger
	}\left\vert 0\right\rangle \]
	
	 \[+\int_{0}^{\infty}\int_{0}^{\infty}dpdqM_{p,q}^{l_{a}l_{b}}e^{-i\left(
		\Delta_{p}+\Delta_{q}\right)  t}\hat{l}_{ap}^{\dagger}\hat{l}_{bq}^{\dagger
	}\left\vert 0\right\rangle \]
	
	 \[+\int_{0}^{\infty}\int_{0}^{\infty}dpdqM_{p,q}^{r_{b}r_{a}}e^{-i\left(
		\Delta_{p}+\Delta_{q}\right)  t}\hat{r}_{bp}^{\dagger}\hat{r}_{aq}^{\dagger
	}\left\vert 0\right\rangle \]
	
	 \[+\int_{0}^{\infty}\int_{0}^{\infty}dpdqM_{p,q}^{r_{b}l_{a}}e^{-i\left(
		\Delta_{p}+\Delta_{q}\right)  t}\hat{r}_{bp}^{\dagger}\hat{l}_{aq}^{\dagger
	}\left\vert 0\right\rangle \]
	
	 \[+\int_{0}^{\infty}\int_{0}^{\infty}dpdqM_{p,q}^{l_{b}r_{a}}e^{-i\left(
		\Delta_{p}+\Delta_{q}\right)  t}\hat{l}_{bp}^{\dagger}\hat{r}_{aq}^{\dagger
	}\left\vert 0\right\rangle \]
	
	 \[+\int_{0}^{\infty}\int_{0}^{\infty}dpdqM_{p,q}^{l_{b}l_{a}}e^{-i\left(
		\Delta_{p}+\Delta_{q}\right)  t}\hat{l}_{bp}^{\dagger}\hat{l}_{aq}^{\dagger
	}\left\vert 0\right\rangle \]
	
	 \[+\int_{0}^{\infty}\int_{0}^{\infty}dpdqM_{p,q}^{r_{b}r_{b}}e^{-i\left(
		\Delta_{p}+\Delta_{q}\right)  t}\hat{r}_{bp}^{\dagger}\hat{r}_{bq}^{\dagger
	}\left\vert 0\right\rangle \]
	
	 \[+\int_{0}^{\infty}\int_{0}^{\infty}dpdqM_{p,q}^{r_{b}l_{b}}e^{-i\left(
		\Delta_{p}+\Delta_{q}\right)  t}\hat{r}_{bp}^{\dagger}\hat{l}_{bq}^{\dagger
	}\left\vert 0\right\rangle \]
	
	  \[+\int_{0}^{\infty}\int_{0}^{\infty}dpdqM_{p,q}^{l_{b}r_{b}}e^{-i\left(
		\Delta_{p}+\Delta_{q}\right)  t}\hat{l}_{bp}^{\dagger}\hat{r}_{bq}^{\dagger
	}\left\vert 0\right\rangle \]
	
	 \[+\int_{0}^{\infty}\int_{0}^{\infty}dpdqM_{p,q}^{l_{b}l_{b}}e^{-i\left(
		\Delta_{p}+\Delta_{q}\right)  t}\hat{l}_{bp}^{\dagger}\hat{l}_{bq}^{\dagger
	}\left\vert 0\right\rangle\]

where
\begin{subequations}
	\begin{equation}
		M_{p,q}^{r_{a}r_{a}}   =t_{ap}^{1}t_{aq}^{1}C_{p,q}(0)+\frac{\gamma
			_{br}\gamma_{ar}}{\Gamma^{2}}J_{p,q},\tag{D1}
			\end{equation}
			\begin{equation}
		M_{p,q}^{r_{a}l_{a}}   =t_{ap}^{1}r_{aq}^{1}C_{p,q}(0)+\frac{\gamma_{br}%
			\sqrt{\gamma_{ar}\gamma_{al}}}{\Gamma^{2}}J_{p,q},\tag{D2}
			\end{equation}
				\begin{equation}
		M_{p,q}^{l_{a}r_{a}}   =r_{ap}^{1}t_{aq}^{1}C_{p,q}(0)+\frac{\gamma_{br}%
			\sqrt{\gamma_{ar}\gamma_{al}}}{\Gamma^{2}}J_{p,q},\tag{D3}
			\end{equation}
				\begin{equation}
		M_{p,q}^{l_{a}l_{a}}   =r_{ap}^{1}r_{aq}^{1}C_{p,q}(0)+\frac{\gamma
			_{br}\gamma_{al}}{\Gamma^{2}}J_{p,q},\tag{D4}
			\end{equation}
				\begin{equation}
		M_{p,q}^{r_{a}r_{b}}   =t_{ap}^{1}t_{bq}^{1}C_{p,q}(0)+\frac{\gamma_{br}%
			\sqrt{\gamma_{ar}\gamma_{br}}}{\Gamma^{2}}J_{p,q},\tag{D5}
			\end{equation}
			\begin{equation}
		M_{p,q}^{r_{a}l_{b}}   =t_{ap}^{1}r_{bq}^{1}C_{p,q}(0)+\frac{\gamma_{br}%
			\sqrt{\gamma_{ar}\gamma_{bl}}}{\Gamma^{2}}J_{p,q},\tag{D6}
			\end{equation}
				\begin{equation}
		M_{p,q}^{l_{a}r_{b}}   =r_{ap}^{1}t_{bq}^{1}C_{p,q}(0)+\frac{\gamma_{br}%
			\sqrt{\gamma_{al}\gamma_{br}}}{\Gamma^{2}}J_{p,q},\tag{D7}
			\end{equation}
				\begin{equation}
		M_{p,q}^{l_{a}l_{b}}   =r_{ap}^{1}r_{bq}^{1}C_{p,q}(0)+\frac{\gamma_{br}%
			\sqrt{\gamma_{al}\gamma_{bl}}}{\Gamma^{2}}J_{p,q},\tag{D8}
			\end{equation}
				\begin{equation}
		M_{p,q}^{r_{b}r_{a}}   =t_{bp}^{1}t_{aq}^{1}C_{p,q}(0)+\frac{\gamma_{br}%
			\sqrt{\gamma_{ar}\gamma_{br}}}{\Gamma^{2}}J_{p,q},\tag{D9}
			\end{equation}
				\begin{equation}
		M_{p,q}^{r_{b}l_{a}}   =t_{bp}^{1}r_{aq}^{1}C_{p,q}(0)+\frac{\gamma_{br}%
			\sqrt{\gamma_{al}\gamma_{br}}}{\Gamma^{2}}J_{p,q},\tag{D10}
			\end{equation}
				\begin{equation}
		M_{p,q}^{l_{b}r_{a}}   =r_{bp}^{1}t_{aq}^{1}C_{p,q}(0)+\frac{\gamma_{br}%
			\sqrt{\gamma_{ar}\gamma_{bl}}}{\Gamma^{2}}J_{p,q},\tag{D11}
			\end{equation}
				\begin{equation}
		M_{p,q}^{l_{b}l_{a}}   =r_{bp}^{1}r_{aq}^{1}C_{p,q}(0)+\frac{\gamma_{br}%
			\sqrt{\gamma_{al}\gamma_{bl}}}{\Gamma^{2}}J_{p,q},\tag{D12}
			\end{equation}
			\begin{equation}
		M_{p,q}^{r_{b}r_{b}}   =t_{bp}^{1}t_{bq}^{1}C_{p,q}(0)+\frac{\gamma_{br}^{2}%
		}{\Gamma^{2}}J_{p,q},\tag{D13}
		\end{equation}
			\begin{equation}
		M_{p,q}^{r_{b}l_{b}}   =t_{bp}^{1}r_{bq}^{1}C_{p,q}(0)+\frac{\gamma_{br}%
			\sqrt{\gamma_{br}\gamma_{bl}}}{\Gamma^{2}}J_{p,q},\tag{D14}
			\end{equation}
			\begin{equation}
		M_{p,q}^{l_{b}r_{b}}   =r_{bp}^{1}t_{bq}^{1}C_{p,q}(0)+\frac{\gamma_{br}%
			\sqrt{\gamma_{br}\gamma_{bl}}}{\Gamma^{2}}J_{p,q},\tag{D15}
			\end{equation}
			\begin{equation}
		M_{p,q}^{l_{b}l_{b}}   =r_{bp}^{1}r_{bq}^{1}C_{p,q}(0)+\frac{\gamma
			_{br}\gamma_{bl}}{\Gamma^{2}}J_{p,q}.\tag{D16}
			\end{equation}
\end{subequations}
Here, $M_{p,q}^{r_{a}r_{a}}$
and $M_{p,q}^{l_{a}l_{a}}
$
(and correspondingly $M_{p,q}^{r_{b}r_{b}}$
and $M_{p,q}^{l_{b}l_{b}}$
) denote the transmission and reflection amplitudes of two photons in waveguide $a$ (or $b$), respectively. 
$M_{p,q}^{r_{a}l_{a}}$
(or $M_{p,q}^{l_{a}r_{a}}$) describe the amplitude for a photon with wavenumber $p$ (or $q$) being transmitted while a photon with wavenumber $q$ (or $p$) is reflected in waveguide $a$. $M_{p,q}^{r_{b}l_{b}}$
(or $M_{p,q}^{l_{b}r_{b}}$
) can be similarly defined in waveguide $b$.
$M_{p,q}^{r_{a}r_{b}}$
(or $M_{p,q}^{r_{b}r_{a}}$
) give the amplitude for a photon with wavenumber $p$ (or $q$) being transmitted in waveguide $a$ and a photon with wavenumber $q$ (or $p$) being transmitted in waveguide $b$. 
$M_{p,q}^{l_{a}l_{b}}$
(or $M_{p,q}^{l_{b}l_{a}}$) give the amplitude for a photon with wavenumber $p$ (or $q$) being reflected in waveguide $a$ and a photon with wavenumber $q$ (or $p$) being reflected in waveguide $b$. 
$M_{p,q}^{r_{a}l_{b}}$
(or $M_{p,q}^{l_{b}r_{a}}$
) describe the amplitude for a photon with wavenumber $p$ (or $q$) being transmitted in waveguide $a$ and a photon with wavenumber $q$ (or $p$) being reflected in waveguide $b$.
$M_{p,q}^{r_{b}l_{a}}$
(or $M_{p,q}^{l_{a}r_{b}}$
) describe the amplitude for a photon with wavenumber $p$ (or $q$) being transmitted in waveguide $b$ and a photon with wavenumber $q$ (or $p$) being reflected in waveguide $a$.

\begin{widetext}
\section*{Appendix E: the two-photon scattering state for photons incident from port 4}
\phantomsection\label{app:E}
\end{widetext}

	\[\psi(t   \rightarrow\infty)=\int_{0}^{\infty}\int_{0}^{\infty}dpdq\tilde
	{M}_{p,q}^{r_{a}r_{a}}e^{-i\left(  \Delta_{p}+\Delta_{q}\right)  t}\hat
	{r}_{ap}^{\dagger}\hat{r}_{aq}^{\dagger}\left\vert 0\right\rangle \]
	
	\[ +\int_{0}^{\infty}\int_{0}^{\infty}dpdq\tilde{M}_{p,q}^{r_{a}l_{a}%
	}e^{-i\left(  \Delta_{p}+\Delta_{q}\right)  t}\hat{r}_{ap}^{\dagger}\hat
	{l}_{aq}^{\dagger}\left\vert 0\right\rangle \]
\[ +\int_{0}^{\infty}\int_{0}^{\infty}dpdq\tilde{M}_{p,q}^{l_{a}r_{a}%
	}e^{-i\left(  \Delta_{p}+\Delta_{q}\right)  t}\hat{l}_{ap}^{\dagger}\hat
	{r}_{aq}^{\dagger}\left\vert 0\right\rangle \]
\[  +\int_{0}^{\infty}\int_{0}^{\infty}dpdq\tilde{M}_{p,q}^{l_{a}l_{a}%
	}e^{-i\left(  \Delta_{p}+\Delta_{q}\right)  t}\hat{l}_{ap}^{\dagger}\hat
	{l}_{aq}^{\dagger}\left\vert 0\right\rangle \]
\[	  +\int_{0}^{\infty}\int_{0}^{\infty}dpdq\tilde{M}_{p,q}^{r_{a}r_{b}%
	}e^{-i\left(  \Delta_{p}+\Delta_{q}\right)  t}\hat{r}_{ap}^{\dagger}\hat
	{r}_{bq}^{\dagger}\left\vert 0\right\rangle \]
\[	 +\int_{0}^{\infty}\int_{0}^{\infty}dpdq\tilde{M}_{p,q}^{r_{a}l_{b}%
	}e^{-i\left(  \Delta_{p}+\Delta_{q}\right)  t}\hat{r}_{ap}^{\dagger}\hat
	{l}_{bq}^{\dagger}\left\vert 0\right\rangle \]
\[	 +\int_{0}^{\infty}\int_{0}^{\infty}dpdq\tilde{M}_{p,q}^{l_{a}r_{b}%
	}e^{-i\left(  \Delta_{p}+\Delta_{q}\right)  t}\hat{l}_{ap}^{\dagger}\hat
	{r}_{bq}^{\dagger}\left\vert 0\right\rangle \]
\[	 +\int_{0}^{\infty}\int_{0}^{\infty}dpdq\tilde{M}_{p,q}^{l_{a}l_{b}%
	}e^{-i\left(  \Delta_{p}+\Delta_{q}\right)  t}\hat{l}_{ap}^{\dagger}\hat
	{l}_{bq}^{\dagger}\left\vert 0\right\rangle \]
\[	 +\int_{0}^{\infty}\int_{0}^{\infty}dpdq\tilde{M}_{p,q}^{r_{b}r_{a}%
	}e^{-i\left(  \Delta_{p}+\Delta_{q}\right)  t}\hat{r}_{bp}^{\dagger}\hat
	{r}_{aq}^{\dagger}\left\vert 0\right\rangle \]
\[	 +\int_{0}^{\infty}\int_{0}^{\infty}dpdq\tilde{M}_{p,q}^{r_{b}l_{a}%
	}e^{-i\left(  \Delta_{p}+\Delta_{q}\right)  t}\hat{r}_{bp}^{\dagger}\hat
	{l}_{aq}^{\dagger}\left\vert 0\right\rangle \]
	\[  +\int_{0}^{\infty}\int_{0}^{\infty}dpdq\tilde{M}_{p,q}^{l_{b}r_{a}%
	}e^{-i\left(  \Delta_{p}+\Delta_{q}\right)  t}\hat{l}_{bp}^{\dagger}\hat
	{r}_{aq}^{\dagger}\left\vert 0\right\rangle \]
\[	  +\int_{0}^{\infty}\int_{0}^{\infty}dpdq\tilde{M}_{p,q}^{l_{b}l_{a}%
	}e^{-i\left(  \Delta_{p}+\Delta_{q}\right)  t}\hat{l}_{bp}^{\dagger}\hat
	{l}_{aq}^{\dagger}\left\vert 0\right\rangle \]
\[	 +\int_{0}^{\infty}\int_{0}^{\infty}dpdq\tilde{M}_{p,q}^{r_{b}r_{b}%
	}e^{-i\left(  \Delta_{p}+\Delta_{q}\right)  t}\hat{r}_{bp}^{\dagger}\hat
	{r}_{bq}^{\dagger}\left\vert 0\right\rangle \]
\[	 +\int_{0}^{\infty}\int_{0}^{\infty}dpdq\tilde{M}_{p,q}^{r_{b}l_{b}%
	}e^{-i\left(  \Delta_{p}+\Delta_{q}\right)  t}\hat{r}_{bp}^{\dagger}\hat
	{l}_{bq}^{\dagger}\left\vert 0\right\rangle \]
	\[ +\int_{0}^{\infty}\int_{0}^{\infty}dpdq\tilde{M}_{p,q}^{l_{b}r_{b}%
	}e^{-i\left(  \Delta_{p}+\Delta_{q}\right)  t}\hat{l}_{bp}^{\dagger}\hat
	{r}_{bq}^{\dagger}\left\vert 0\right\rangle \]
\[  +\int_{0}^{\infty}\int_{0}^{\infty}dpdq\tilde{M}_{p,q}^{l_{b}l_{b}%
	}e^{-i\left(  \Delta_{p}+\Delta_{q}\right)  t}\hat{l}_{bp}^{\dagger}\hat
	{l}_{bq}^{\dagger}\left\vert 0\right\rangle\]

where
\begin{subequations}
\begin{equation}
	\tilde{M}_{p,q}^{r_{a}r_{a}}   =\tilde{r}_{ap}^{1}\tilde{r}_{aq}^{1}%
	C_{p,q}(0)+\frac{\gamma_{bl}\gamma_{ar}}{\Gamma^{2}}J_{p,q},\tag{E1}
	\end{equation}
		\begin{equation}
	\tilde{M}_{p,q}^{r_{a}l_{a}}   =\tilde{r}_{ap}^{1}\tilde{t}_{aq}^{1}%
	C_{p,q}(0)+\frac{\gamma_{bl}\sqrt{\gamma_{ar}\gamma_{al}}}{\Gamma^{2}}%
	J_{p,q},\tag{E2}
	\end{equation}
		\begin{equation}
	\tilde{M}_{p,q}^{l_{a}r_{a}}   =\tilde{t}_{ap}^{1}\tilde{r}_{aq}^{1}%
	C_{p,q}(0)+\frac{\gamma_{bl}\sqrt{\gamma_{ar}\gamma_{al}}}{\Gamma^{2}}%
	J_{p,q},\tag{E3}
	\end{equation}
		\begin{equation}
	\tilde{M}_{p,q}^{l_{a}l_{a}}   =\tilde{t}_{ap}^{1}\tilde{t}_{aq}^{1}%
	C_{p,q}(0)+\frac{\gamma_{bl}\gamma_{al}}{\Gamma^{2}}J_{p,q},\tag{E4}
	\end{equation}
		\begin{equation}
	\tilde{M}_{p,q}^{r_{a}r_{b}}   =\tilde{r}_{ap}^{1}\tilde{r}_{bq}^{1}%
	C_{p,q}(0)+\frac{\gamma_{bl}\sqrt{\gamma_{ar}\gamma_{br}}}{\Gamma^{2}}%
	J_{p,q},\tag{E5}
	\end{equation}
		\begin{equation}
	\tilde{M}_{p,q}^{r_{a}l_{b}}   =\tilde{r}_{ap}^{1}\tilde{t}_{bq}^{1}%
	C_{p,q}(0)+\frac{\gamma_{bl}\sqrt{\gamma_{ar}\gamma_{bl}}}{\Gamma^{2}}%
	J_{p,q},\tag{E6}
	\end{equation}
		\begin{equation}
	\tilde{M}_{p,q}^{l_{a}r_{b}}   =\tilde{t}_{ap}^{1}\tilde{r}_{bq}^{1}%
	C_{p,q}(0)+\frac{\gamma_{bl}\sqrt{\gamma_{al}\gamma_{br}}}{\Gamma^{2}}%
	J_{p,q},\tag{E7}
	\end{equation}
		\begin{equation}
	\tilde{M}_{p,q}^{l_{a}l_{b}}   =\tilde{t}_{ap}^{1}\tilde{t}_{bq}^{1}%
	C_{p,q}(0)+\frac{\gamma_{bl}\sqrt{\gamma_{al}\gamma_{bl}}}{\Gamma^{2}}%
	J_{p,q},\tag{E8}
	\end{equation}
		\begin{equation}
	\tilde{M}_{p,q}^{r_{b}r_{a}}   =\tilde{r}_{bp}^{1}\tilde{r}_{aq}^{1}%
	C_{p,q}(0)+\frac{\gamma_{bl}\sqrt{\gamma_{ar}\gamma_{br}}}{\Gamma^{2}}%
	J_{p,q},\tag{E9}
	\end{equation}
		\begin{equation}
	\tilde{M}_{p,q}^{r_{b}l_{a}}   =\tilde{r}_{bp}^{1}\tilde{t}_{aq}^{1}%
	C_{p,q}(0)+\frac{\gamma_{bl}\sqrt{\gamma_{al}\gamma_{br}}}{\Gamma^{2}}%
	J_{p,q},\tag{E10}
	\end{equation}
		\begin{equation}
	\tilde{M}_{p,q}^{l_{b}r_{a}}   =\tilde{t}_{bp}^{1}\tilde{r}_{aq}^{1}%
	C_{p,q}(0)+\frac{\gamma_{bl}\sqrt{\gamma_{ar}\gamma_{bl}}}{\Gamma^{2}}%
	J_{p,q},\tag{E11}
	\end{equation}
		\begin{equation}
	\tilde{M}_{p,q}^{l_{b}l_{a}}   =\tilde{t}_{bp}^{1}\tilde{t}_{aq}^{1}%
	C_{p,q}(0)+\frac{\gamma_{bl}\sqrt{\gamma_{al}\gamma_{bl}}}{\Gamma^{2}}%
	J_{p,q},\tag{E12}
	\end{equation}
		\begin{equation}
	\tilde{M}_{p,q}^{r_{b}r_{b}}   =\tilde{r}_{bp}^{1}\tilde{r}_{bq}^{1}%
	C_{p,q}(0)+\frac{\gamma_{bl}\gamma_{br}}{\Gamma^{2}}J_{p,q},\tag{E13}
	\end{equation}
		\begin{equation}
	\tilde{M}_{p,q}^{r_{b}l_{b}}   =\tilde{r}_{bp}^{1}\tilde{t}_{bq}^{1}%
	C_{p,q}(0)+\frac{\gamma_{bl}\sqrt{\gamma_{br}\gamma_{bl}}}{\Gamma^{2}}%
	J_{p,q},\tag{E14}
	\end{equation}
		\begin{equation}
	\tilde{M}_{p,q}^{l_{b}r_{b}}   =\tilde{t}_{bp}^{1}\tilde{r}_{bq}^{1}%
	C_{p,q}(0)+\frac{\gamma_{bl}\sqrt{\gamma_{br}\gamma_{bl}}}{\Gamma^{2}}%
	J_{p,q},\tag{E15}
	\end{equation}
		\begin{equation}
	\tilde{M}_{p,q}^{l_{b}l_{b}}   =\tilde{t}_{bp}^{1}\tilde{t}_{bq}^{1}%
	C_{p,q}(0)+\frac{\gamma_{bl}^{2}}{\Gamma^{2}}J_{p,q}.\tag{E16}
\end{equation}
\end{subequations}
Here, $\tilde{M}_{p,q}^{r_{a}r_{a}}$
and $\tilde{M}%
_{p,q}^{l_{a}l_{a}}$
(and correspondingly $\tilde{M}_{p,q}^{r_{b}r_{b}}$
and $\tilde
{M}_{p,q}^{l_{b}l_{b}}$
) denote the reflection and transmission amplitudes of two photons in waveguide $a$ (or $b$), respectively. 
$\tilde
{M}_{p,q}^{r_{a}l_{a}}$
(or $\tilde{M}_{p,q}^{l_{a}r_{a}}$
) describe the amplitude for a photon with wavenumber $p$ (or $q$) being reflected while a photon with wavenumber $q$ (or $p$) is transmitted in waveguide $a$.
$\tilde
{M}_{p,q}^{r_{b}l_{b}}$
(or $\tilde{M}_{p,q}^{l_{b}r_{b}}$
) can be similarly defined in waveguide $b$.
The terms $\tilde
{M}_{p,q}^{r_{a}r_{b}}$
(or $\tilde{M}_{p,q}^{r_{b}r_{a}}$
) give the amplitude for a photon with wavenumber $p$ (or $q$) being reflected in waveguide $a$ and a photon with wavenumber $q$ (or $p$) being reflected in waveguide $b$. 
$\tilde{M}_{p,q}^{l_{a}l_{b}}$
(or $\tilde{M}_{p,q}^{l_{b}l_{a}}$
)] give the amplitude for a photon with wavenumber $p$ (or $q$) being transmitted) in waveguide $a$ and a photon with wavenumber $q$ (or $p$) being transmitted in waveguide $b$. 
$\tilde
{M}_{p,q}^{r_{a}l_{b}}$
(or $\tilde{M}_{p,q}^{l_{b}r_{a}}$
) describe the amplitude for a photon with wavenumber $p$ (or $q$) being reflected in waveguide $a$ and a photon with wavenumber $q$ (or $p$) being transmitted in waveguide $b$.
$\tilde
{M}_{p,q}^{r_{b}l_{a}}$
(or $\tilde{M}_{p,q}^{l_{a}r_{b}}$
) describe the amplitude for a photon with wavenumber $p$ (or $q$) being reflected in waveguide $b$ and a photon with wavenumber $q$ (or $p$) being transmitted in waveguide $a$.

\begin{widetext}
\section*{Appendix F: the scattering state of two photons incident from port 1 and port 3 respectively}
\phantomsection\label{app:F}
\end{widetext}
\[
	\psi(t   \rightarrow\infty)=\int_{0}^{\infty}\int_{0}^{\infty}dpdqN_{p,q}%
	^{r_{a}r_{a}}e^{-i\left(  \Delta_{p}+\Delta_{q}\right)  t}\hat{r}%
	_{ap}^{\dagger}\hat{r}_{aq}^{\dagger}\left\vert 0\right\rangle \]
	\[  +\int_{0}^{\infty}\int_{0}^{\infty}dpdqN_{p,q}^{r_{a}l_{a}}e^{-i\left(
		\Delta_{p}+\Delta_{q}\right)  t}\hat{r}_{ap}^{\dagger}\hat{l}_{aq}^{\dagger
	}\left\vert 0\right\rangle \]
	\[  +\int_{0}^{\infty}\int_{0}^{\infty}dpdqN_{p,q}^{l_{a}r_{a}}e^{-i\left(
		\Delta_{p}+\Delta_{q}\right)  t}\hat{l}_{ap}^{\dagger}\hat{r}_{aq}^{\dagger
	}\left\vert 0\right\rangle \]
	\[  +\int_{0}^{\infty}\int_{0}^{\infty}dpdqN_{p,q}^{l_{a}l_{a}}e^{-i\left(
		\Delta_{p}+\Delta_{q}\right)  t}\hat{l}_{ap}^{\dagger}\hat{l}_{aq}^{\dagger
	}\left\vert 0\right\rangle \]
	\[  +\int_{0}^{\infty}\int_{0}^{\infty}dpdqN_{p,q}^{r_{a}r_{b}}e^{-i\left(
		\Delta_{p}+\Delta_{q}\right)  t}\hat{r}_{ap}^{\dagger}\hat{r}_{bq}^{\dagger
	}\left\vert 0\right\rangle \]
	\[  +\int_{0}^{\infty}\int_{0}^{\infty}dpdqN_{p,q}^{r_{a}l_{b}}e^{-i\left(
		\Delta_{p}+\Delta_{q}\right)  t}\hat{r}_{ap}^{\dagger}\hat{l}_{bq}^{\dagger
	}\left\vert 0\right\rangle \]
	\[  +\int_{0}^{\infty}\int_{0}^{\infty}dpdqN_{p,q}^{l_{a}r_{b}}e^{-i\left(
		\Delta_{p}+\Delta_{q}\right)  t}\hat{l}_{ap}^{\dagger}\hat{r}_{bq}^{\dagger
	}\left\vert 0\right\rangle \]
	\[  +\int_{0}^{\infty}\int_{0}^{\infty}dpdqN_{p,q}^{l_{a}l_{b}}e^{-i\left(
		\Delta_{p}+\Delta_{q}\right)  t}\hat{l}_{ap}^{\dagger}\hat{l}_{bq}^{\dagger
	}\left\vert 0\right\rangle \]
	\[  +\int_{0}^{\infty}\int_{0}^{\infty}dpdqN_{p,q}^{r_{b}r_{a}}e^{-i\left(
		\Delta_{p}+\Delta_{q}\right)  t}\hat{r}_{bp}^{\dagger}\hat{r}_{aq}^{\dagger
	}\left\vert 0\right\rangle \]
	\[  +\int_{0}^{\infty}\int_{0}^{\infty}dpdqN_{p,q}^{r_{b}l_{a}}e^{-i\left(
		\Delta_{p}+\Delta_{q}\right)  t}\hat{r}_{bp}^{\dagger}\hat{l}_{aq}^{\dagger
	}\left\vert 0\right\rangle \]
	\[  +\int_{0}^{\infty}\int_{0}^{\infty}dpdqN_{p,q}^{l_{b}r_{a}}e^{-i\left(
		\Delta_{p}+\Delta_{q}\right)  t}\hat{l}_{bp}^{\dagger}\hat{r}_{aq}^{\dagger
	}\left\vert 0\right\rangle \]
	\[  +\int_{0}^{\infty}\int_{0}^{\infty}dpdqN_{p,q}^{l_{b}l_{a}}e^{-i\left(
		\Delta_{p}+\Delta_{q}\right)  t}\hat{l}_{bp}^{\dagger}\hat{l}_{aq}^{\dagger
	}\left\vert 0\right\rangle \]
	\[  +\int_{0}^{\infty}\int_{0}^{\infty}dpdqN_{p,q}^{r_{b}r_{b}}e^{-i\left(
		\Delta_{p}+\Delta_{q}\right)  t}\hat{r}_{bp}^{\dagger}\hat{r}_{bq}^{\dagger
	}\left\vert 0\right\rangle \]
	\[  +\int_{0}^{\infty}\int_{0}^{\infty}dpdqN_{p,q}^{r_{b}l_{b}}e^{-i\left(
		\Delta_{p}+\Delta_{q}\right)  t}\hat{r}_{bp}^{\dagger}\hat{l}_{bq}^{\dagger
	}\left\vert 0\right\rangle \]
	\[  +\int_{0}^{\infty}\int_{0}^{\infty}dpdqN_{p,q}^{l_{b}r_{b}}e^{-i\left(
		\Delta_{p}+\Delta_{q}\right)  t}\hat{l}_{bp}^{\dagger}\hat{r}_{bq}^{\dagger
	}\left\vert 0\right\rangle \]
	\[  +\int_{0}^{\infty}\int_{0}^{\infty}dpdqN_{p,q}^{l_{b}l_{b}}e^{-i\left(
		\Delta_{p}+\Delta_{q}\right)  t}\hat{l}_{bp}^{\dagger}\hat{l}_{bq}^{\dagger
	}\left\vert 0\right\rangle\]

where
\begin{subequations}
\begin{equation}
N_{p,q}^{r_{a}r_{a}}   =t_{ap}^{0}t_{aq}^{1}C_{p,q}(0)+\frac{\gamma_{ar}%
	\sqrt{\gamma_{ar}\gamma_{br}}}{\Gamma^{2}}J_{p,q},\tag{F1}
	\end{equation}
		\begin{equation}
N_{p,q}^{r_{a}l_{a}}   =t_{ap}^{0}r_{aq}^{1}C_{p,q}(0)+\frac{\gamma_{ar}%
	\sqrt{\gamma_{al}\gamma_{br}}}{\Gamma^{2}}J_{p,q},\tag{F2}
	\end{equation}
		\begin{equation}
N_{p,q}^{l_{a}r_{a}}   =r_{ap}^{0}t_{aq}^{1}C_{p,q}(0)+\frac{\gamma_{ar}%
	\sqrt{\gamma_{al}\gamma_{br}}}{\Gamma^{2}}J_{p,q},\tag{F3}
	\end{equation}
		\begin{equation}
N_{p,q}^{l_{a}l_{a}}   =r_{ap}^{0}r_{aq}^{1}C_{p,q}(0)+\frac{\gamma_{al}%
	\sqrt{\gamma_{ar}\gamma_{br}}}{\Gamma^{2}}J_{p,q},\tag{F4}
	\end{equation}
		\begin{equation}
N_{p,q}^{r_{a}r_{b}}   =t_{ap}^{0}t_{bq}^{1}C_{p,q}(0)+\frac{\gamma
	_{ar}\gamma_{br}}{\Gamma^{2}}J_{p,q},\tag{F5}
	\end{equation}
		\begin{equation}
N_{p,q}^{r_{a}l_{b}}   =t_{ap}^{0}r_{bq}^{1}C_{p,q}(0)+\frac{\gamma_{ar}%
	\sqrt{\gamma_{br}\gamma_{bl}}}{\Gamma^{2}}J_{p,q},\tag{F6}
	\end{equation}
		\begin{equation}
N_{p,q}^{l_{a}r_{b}}   =r_{ap}^{0}t_{bq}^{1}C_{p,q}(0)+\frac{\gamma_{br}%
	\sqrt{\gamma_{ar}\gamma_{al}}}{\Gamma^{2}}J_{p,q},\tag{F7}
	\end{equation}
		\begin{equation}
N_{p,q}^{l_{a}l_{b}}   =r_{ap}^{0}r_{bq}^{1}C_{p,q}(0)+\frac{\sqrt
	{\gamma_{ar}\gamma_{al}\gamma_{br}\gamma_{bl}}}{\Gamma^{2}}J_{p,q},\tag{F8}
	\end{equation}
		\begin{equation}
N_{p,q}^{r_{b}r_{a}}   =t_{bp}^{0}t_{aq}^{1}C_{p,q}(0)+\frac{\gamma
	_{ar}\gamma_{br}}{\Gamma^{2}}J_{p,q},\tag{F9}
	\end{equation}
		\begin{equation}
N_{p,q}^{r_{b}l_{a}}   =t_{bp}^{0}r_{aq}^{1}C_{p,q}(0)+\frac{\gamma_{br}%
	\sqrt{\gamma_{ar}\gamma_{al}}}{\Gamma^{2}}J_{p,q},\tag{F10}
	\end{equation}
		\begin{equation}
N_{p,q}^{l_{b}r_{a}}   =r_{bp}^{0}t_{aq}^{1}C_{p,q}(0)+\frac{\gamma_{ar}%
	\sqrt{\gamma_{br}\gamma_{bl}}}{\Gamma^{2}}J_{p,q},\tag{F11}
	\end{equation}
		\begin{equation}
N_{p,q}^{l_{b}l_{a}}   =r_{bp}^{0}r_{aq}^{1}C_{p,q}(0)+\frac{\sqrt
	{\gamma_{ar}\gamma_{al}\gamma_{br}\gamma_{bl}}}{\Gamma^{2}}J_{p,q},\tag{F12}
	\end{equation}
		\begin{equation}
N_{p,q}^{r_{b}r_{b}}   =t_{bp}^{0}t_{bq}^{1}C_{p,q}(0)+\frac{\gamma_{br}%
	\sqrt{\gamma_{ar}\gamma_{br}}}{\Gamma^{2}}J_{p,q},\tag{F13}
	\end{equation}
		\begin{equation}
N_{p,q}^{r_{b}l_{b}}   =t_{bp}^{0}r_{bq}^{1}C_{p,q}(0)+\frac{\gamma_{br}%
	\sqrt{\gamma_{ar}\gamma_{bl}}}{\Gamma^{2}}J_{p,q},\tag{F14}
	\end{equation}
		\begin{equation}
N_{p,q}^{l_{b}r_{b}}   =r_{bp}^{0}t_{bq}^{1}C_{p,q}(0)+\frac{\gamma_{br}%
	\sqrt{\gamma_{ar}\gamma_{bl}}}{\Gamma^{2}}J_{p,q},\tag{F15}
	\end{equation}
		\begin{equation}
N_{p,q}^{l_{b}l_{b}}   =r_{bp}^{0}r_{bq}^{1}C_{p,q}(0)+\frac{\gamma_{bl}%
	\sqrt{\gamma_{ar}\gamma_{br}}}{\Gamma^{2}}J_{p,q}.\tag{F16}
\end{equation}
\end{subequations}
Here, $N_{p,q}^{r_{a}r_{a}}$
and $N_{p,q}^{l_{a}l_{a}}%
$
(and correspondingly $N_{p,q}^{r_{b}r_{b}}$
and $N_{p,q}^{l_{b}l_{b}}$
) denote the transmission and reflection amplitudes of two photons in waveguide $a$ (or $b$), respectively. 
$N_{p,q}^{r_{a}l_{a}}$
(or $N_{p,q}^{l_{a}r_{a}}$
) describe the amplitude for a photon with wavenumber $p$ (or $q$) being transmitted while a photon with wavenumber $q$ (or $p$) is reflected in waveguide $a$ (or $b$). $N_{p,q}^{r_{b}l_{b}}$
(or $N_{p,q}^{l_{b}r_{b}}$
) can be similarly defined in waveguide $b$.
$N_{p,q}^{r_{a}r_{b}}$
(or $N_{p,q}^{r_{b}r_{a}}$
) give the amplitude for a photon with wavenumber $p$ (or $q$) being transmitted in waveguide $a$ and a photon with wavenumber $q$ (or $p$) being transmitted in waveguide $b$. 
$N_{p,q}^{l_{a}l_{b}}$
(or $N_{p,q}^{l_{b}l_{a}}$
) give the amplitude for a photon with wavenumber $p$ (or $q$) being reflected in waveguide $a$ and a photon with wavenumber $q$ (or $p$) being reflected in waveguide $b$.
$N_{p,q}^{r_{a}l_{b}}$
(or $N_{p,q}^{l_{b}r_{a}}$
) describe the amplitude for a photon with wavenumber $p$ (or $q$) being transmitted in waveguide $a$ and a photon with wavenumber $q$ (or $p$) being reflected in waveguide $b$.
$N_{p,q}^{r_{b}l_{a}}$
(or $N_{p,q}^{l_{a}r_{b}}$
) describe the amplitude for a photon with wavenumber $p$ (or $q$) being transmitted in waveguide $b$ and a photon with wavenumber $q$ (or $p$) being reflected in waveguide $a$.

\begin{widetext}
\section*{Appendix G: the scattering state of two photons incident from port 2 and port 4 respectively}
\phantomsection\label{app:G}
\end{widetext}
\[
	\psi(t   \rightarrow\infty)=\int_{0}^{\infty}\int_{0}^{\infty}dpdq\tilde
	{N}_{p,q}^{r_{a}r_{a}}e^{-i\left(  \Delta_{p}+\Delta_{q}\right)  t}\hat
	{r}_{ap}^{\dagger}\hat{r}_{aq}^{\dagger}\left\vert 0\right\rangle \]
\[  +\int_{0}^{\infty}\int_{0}^{\infty}dpdq\tilde{N}_{p,q}^{r_{a}l_{a}%
	}e^{-i\left(  \Delta_{p}+\Delta_{q}\right)  t}\hat{r}_{ap}^{\dagger}\hat
	{l}_{aq}^{\dagger}\left\vert 0\right\rangle \]
\[ +\int_{0}^{\infty}\int_{0}^{\infty}dpdq\tilde{N}_{p,q}^{l_{a}r_{a}%
	}e^{-i\left(  \Delta_{p}+\Delta_{q}\right)  t}\hat{l}_{ap}^{\dagger}\hat
	{r}_{aq}^{\dagger}\left\vert 0\right\rangle \]
\[ +\int_{0}^{\infty}\int_{0}^{\infty}dpdq\tilde{N}_{p,q}^{l_{a}l_{a}%
	}e^{-i\left(  \Delta_{p}+\Delta_{q}\right)  t}\hat{l}_{ap}^{\dagger}\hat
	{l}_{aq}^{\dagger}\left\vert 0\right\rangle \]
\[ +\int_{0}^{\infty}\int_{0}^{\infty}dpdq\tilde{N}_{p,q}^{r_{a}r_{b}%
	}e^{-i\left(  \Delta_{p}+\Delta_{q}\right)  t}\hat{r}_{ap}^{\dagger}\hat
	{r}_{bq}^{\dagger}\left\vert 0\right\rangle \]
\[  +\int_{0}^{\infty}\int_{0}^{\infty}dpdq\tilde{N}_{p,q}^{r_{a}l_{b}%
	}e^{-i\left(  \Delta_{p}+\Delta_{q}\right)  t}\hat{r}_{ap}^{\dagger}\hat
	{l}_{bq}^{\dagger}\left\vert 0\right\rangle \]
\[  +\int_{0}^{\infty}\int_{0}^{\infty}dpdq\tilde{N}_{p,q}^{l_{a}r_{b}%
	}e^{-i\left(  \Delta_{p}+\Delta_{q}\right)  t}\hat{l}_{ap}^{\dagger}\hat
	{r}_{bq}^{\dagger}\left\vert 0\right\rangle \]
\[ +\int_{0}^{\infty}\int_{0}^{\infty}dpdq\tilde{N}_{p,q}^{l_{a}l_{b}%
	}e^{-i\left(  \Delta_{p}+\Delta_{q}\right)  t}\hat{l}_{ap}^{\dagger}\hat
	{l}_{bq}^{\dagger}\left\vert 0\right\rangle \]
\[ +\int_{0}^{\infty}\int_{0}^{\infty}dpdq\tilde{N}_{p,q}^{r_{b}r_{a}%
	}e^{-i\left(  \Delta_{p}+\Delta_{q}\right)  t}\hat{r}_{bp}^{\dagger}\hat
	{r}_{aq}^{\dagger}\left\vert 0\right\rangle \]
\[ +\int_{0}^{\infty}\int_{0}^{\infty}dpdq\tilde{N}_{p,q}^{r_{b}l_{a}%
	}e^{-i\left(  \Delta_{p}+\Delta_{q}\right)  t}\hat{r}_{bp}^{\dagger}\hat
	{l}_{aq}^{\dagger}\left\vert 0\right\rangle \]
\[  +\int_{0}^{\infty}\int_{0}^{\infty}dpdq\tilde{N}_{p,q}^{l_{b}r_{a}%
	}e^{-i\left(  \Delta_{p}+\Delta_{q}\right)  t}\hat{l}_{bp}^{\dagger}\hat
	{r}_{aq}^{\dagger}\left\vert 0\right\rangle \]
\[ +\int_{0}^{\infty}\int_{0}^{\infty}dpdq\tilde{N}_{p,q}^{l_{b}l_{a}%
	}e^{-i\left(  \Delta_{p}+\Delta_{q}\right)  t}\hat{l}_{bp}^{\dagger}\hat
	{l}_{aq}^{\dagger}\left\vert 0\right\rangle \]
\[ +\int_{0}^{\infty}\int_{0}^{\infty}dpdq\tilde{N}_{p,q}^{r_{b}r_{b}%
	}e^{-i\left(  \Delta_{p}+\Delta_{q}\right)  t}\hat{r}_{bp}^{\dagger}\hat
	{r}_{bq}^{\dagger}\left\vert 0\right\rangle \]
\[ +\int_{0}^{\infty}\int_{0}^{\infty}dpdq\tilde{N}_{p,q}^{r_{b}l_{b}%
	}e^{-i\left(  \Delta_{p}+\Delta_{q}\right)  t}\hat{r}_{bp}^{\dagger}\hat
	{l}_{bq}^{\dagger}\left\vert 0\right\rangle \]
\[ +\int_{0}^{\infty}\int_{0}^{\infty}dpdq\tilde{N}_{p,q}^{l_{b}r_{b}%
	}e^{-i\left(  \Delta_{p}+\Delta_{q}\right)  t}\hat{l}_{bp}^{\dagger}\hat
	{r}_{bq}^{\dagger}\left\vert 0\right\rangle \]
\[ +\int_{0}^{\infty}\int_{0}^{\infty}dpdq\tilde{N}_{p,q}^{l_{b}l_{b}%
	}e^{-i\left(  \Delta_{p}+\Delta_{q}\right)  t}\hat{l}_{bp}^{\dagger}\hat
	{l}_{bq}^{\dagger}\left\vert 0\right\rangle\]

where
\begin{subequations}
\begin{equation}
	\tilde{N}_{p,q}^{r_{a}r_{a}}   =\tilde{r}_{ap}^{0}\tilde{r}_{aq}^{1}%
	C_{p,q}(0)+\frac{\gamma_{ar}\sqrt{\gamma_{al}\gamma_{bl}}}{\Gamma^{2}}%
	J_{p,q},\tag{G1}
	\end{equation}
		\begin{equation}
	\tilde{N}_{p,q}^{r_{a}l_{a}}   =\tilde{r}_{ap}^{0}\tilde{t}_{aq}^{1}%
	C_{p,q}(0)+\frac{\gamma_{al}\sqrt{\gamma_{ar}\gamma_{bl}}}{\Gamma^{2}}%
	J_{p,q},\tag{G2}
	\end{equation}
		\begin{equation}
	\tilde{N}_{p,q}^{l_{a}r_{a}}   =\tilde{t}_{ap}^{0}\tilde{r}_{aq}^{1}%
	C_{p,q}(0)+\frac{\gamma_{al}\sqrt{\gamma_{ar}\gamma_{bl}}}{\Gamma^{2}}%
	J_{p,q},\tag{G3}
	\end{equation}
		\begin{equation}
	\tilde{N}_{p,q}^{l_{a}l_{a}}   =\tilde{t}_{ap}^{0}\tilde{t}_{aq}^{1}%
	C_{p,q}(0)+\frac{\gamma_{al}\sqrt{\gamma_{al}\gamma_{bl}}}{\Gamma^{2}}%
	J_{p,q},\tag{G4}
	\end{equation}
		\begin{equation}
	\tilde{N}_{p,q}^{r_{a}r_{b}}   =\tilde{r}_{ap}^{0}\tilde{r}_{bq}^{1}%
	C_{p,q}(0)+\frac{\sqrt{\gamma_{ar}\gamma_{br}\gamma_{al}\gamma_{bl}}}%
	{\Gamma^{2}}J_{p,q},\tag{G5}
	\end{equation}
		\begin{equation}
	\tilde{N}_{p,q}^{r_{a}l_{b}}   =\tilde{r}_{ap}^{0}\tilde{t}_{bq}^{1}%
	C_{p,q}(0)+\frac{\gamma_{bl}\sqrt{\gamma_{ar}\gamma_{al}}}{\Gamma^{2}}%
	J_{p,q},\tag{G6}
	\end{equation}
		\begin{equation}
	\tilde{N}_{p,q}^{l_{a}r_{b}}   =\tilde{t}_{ap}^{0}\tilde{r}_{bq}^{1}%
	C_{p,q}(0)+\frac{\gamma_{al}\sqrt{\gamma_{br}\gamma_{bl}}}{\Gamma^{2}}%
	J_{p,q},\tag{G7}
	\end{equation}
		\begin{equation}
	\tilde{N}_{p,q}^{l_{a}l_{b}}   =\tilde{t}_{ap}^{0}\tilde{t}_{bq}^{1}%
	C_{p,q}(0)+\frac{\gamma_{al}\gamma_{bl}}{\Gamma^{2}}J_{p,q},\tag{G8}
	\end{equation}
		\begin{equation}
	\tilde{N}_{p,q}^{r_{b}r_{a}}   =\tilde{r}_{bp}^{0}\tilde{r}_{aq}^{1}%
	C_{p,q}(0)+\frac{\sqrt{\gamma_{ar}\gamma_{br}\gamma_{al}\gamma_{bl}}}%
	{\Gamma^{2}}J_{p,q},\tag{G9}
	\end{equation}
		\begin{equation}
	\tilde{N}_{p,q}^{r_{b}l_{a}}   =\tilde{r}_{bp}^{0}\tilde{t}_{aq}^{1}%
	C_{p,q}(0)+\frac{\gamma_{al}\sqrt{\gamma_{br}\gamma_{bl}}}{\Gamma^{2}}%
	J_{p,q},\tag{G10}
	\end{equation}
		\begin{equation}
	\tilde{N}_{p,q}^{l_{b}r_{a}}   =\tilde{t}_{bp}^{0}\tilde{r}_{aq}^{1}%
	C_{p,q}(0)+\frac{\gamma_{bl}\sqrt{\gamma_{ar}\gamma_{al}}}{\Gamma^{2}}%
	J_{p,q},\tag{G11}
	\end{equation}
		\begin{equation}
	\tilde{N}_{p,q}^{l_{b}l_{a}}   =\tilde{t}_{bp}^{0}\tilde{t}_{aq}^{1}%
	C_{p,q}(0)+\frac{\gamma_{al}\gamma_{bl}}{\Gamma^{2}}J_{p,q},\tag{G12}
	\end{equation}
		\begin{equation}
	\tilde{N}_{p,q}^{r_{b}r_{b}}   =\tilde{r}_{bp}^{0}\tilde{r}_{bq}^{1}%
	C_{p,q}(0)+\frac{\gamma_{br}\sqrt{\gamma_{al}\gamma_{bl}}}{\Gamma^{2}}%
	J_{p,q},\tag{G13}
	\end{equation}
		\begin{equation}
	\tilde{N}_{p,q}^{r_{b}l_{b}}   =\tilde{r}_{bp}^{0}\tilde{t}_{bq}^{1}%
	C_{p,q}(0)+\frac{\gamma_{bl}\sqrt{\gamma_{al}\gamma_{br}}}{\Gamma^{2}}%
	J_{p,q},\tag{G14}
	\end{equation}
		\begin{equation}
	\tilde{N}_{p,q}^{l_{b}r_{b}}   =\tilde{t}_{bp}^{0}\tilde{r}_{bq}^{1}%
	C_{p,q}(0)+\frac{\gamma_{bl}\sqrt{\gamma_{br}\gamma_{al}}}{\Gamma^{2}}%
	J_{p,q},\tag{G15}
	\end{equation}
		\begin{equation}
	\tilde{N}_{p,q}^{l_{b}l_{b}}   =\tilde{t}_{bp}^{0}\tilde{t}_{bq}^{1}%
	C_{p,q}(0)+\frac{\gamma_{bl}\sqrt{\gamma_{al}\gamma_{bl}}}{\Gamma^{2}}J_{p,q}.\tag{G16}
\end{equation}
\end{subequations}	 
Here, $\tilde{N}_{p,q}^{r_{a}r_{a}}$
and $\tilde{N}_{p,q}^{l_{a}l_{a}}$
(and correspondingly $\tilde{N}_{p,q}^{r_{b}r_{b}}$
and $\tilde
{N}_{p,q}^{l_{b}l_{b}}$
) denote the reflection and transmission amplitudes of two photons in waveguide $a$ (or $b$), respectively. 
$\tilde
{N}_{p,q}^{r_{a}l_{a}}$
(or $\tilde{N}_{p,q}^{l_{a}r_{a}}$
) [and $\tilde
{N}_{p,q}^{r_{b}l_{b}}$
(or $\tilde{N}_{p,q}^{l_{b}r_{b}}$
)] describe the amplitude for a photon with wavenumber $p$ (or $q$) being reflected while a photon with wavenumber $q$ (or $p$) is transmitted in waveguide $a$ (or $b$). $\tilde
{N}_{p,q}^{r_{b}l_{b}}$
(or $\tilde{N}_{p,q}^{l_{b}r_{b}}$
) can be similarly defined in waveguide $b$.
$\tilde
{N}_{p,q}^{r_{a}r_{b}}$
(or $\tilde{N}_{p,q}^{r_{b}r_{a}}$
) give the amplitude for a photon with wavenumber $p$ (or $q$) being reflected in waveguide $a$ and a photon with wavenumber $q$ (or $p$) being reflected in waveguide $b$. 
$\tilde{N}_{p,q}^{l_{a}l_{b}}$
(or $\tilde{N}_{p,q}^{l_{b}l_{a}}$
) give the amplitude for a photon with wavenumber $p$ (or $q$) being transmitted in waveguide $a$ and a photon with wavenumber $q$ (or $p$) being transmitted in waveguide $b$. 
$\tilde
{N}_{p,q}^{r_{a}l_{b}}$
(or $\tilde{N}_{p,q}^{l_{b}r_{a}}$
) describe the amplitude for a photon with wavenumber $p$ (or $q$) being reflected in waveguide $a$ and a photon with wavenumber $q$ (or $p$) being transmitted in waveguide $b$.
$\tilde
{N}_{p,q}^{r_{b}l_{a}}$
(or $\tilde{N}_{p,q}^{l_{a}r_{b}}$
) describe the amplitude for a photon with wavenumber $p$ (or $q$) being reflected in waveguide $b$ and a photon with wavenumber $q$ (or $p$) being transmitted in waveguide $a$.

\end{document}